\documentclass[12pt]{article}

\usepackage{amsfonts}
\usepackage{eurosym}
\usepackage{graphicx}
\usepackage{booktabs}
\usepackage{float}
\usepackage{subfig}
\usepackage{amssymb}
\usepackage{comment}
\usepackage{amsmath}
\usepackage{booktabs}
\usepackage{multirow}
\usepackage{enumerate}
\usepackage{color}
\usepackage[authoryear]{natbib}
\usepackage{hyperref}
\usepackage{algorithm2e}
\usepackage{empheq}

\usepackage{csvsimple}
\usepackage{bbm}
\usepackage{booktabs}
\usepackage{longtable}
\usepackage{amsthm}
\usepackage[font=small,labelfont=bf]{caption}

\definecolor{shadecolor}{cmyk}{0,0,0,0}
\definecolor{light-blue}{cmyk}{0,0,0,0}
\newsavebox{\mysaveboxM}
\newsavebox{\mysaveboxT}
\newcommand*\Garybox[2][A Nice Box]{%
  \sbox{\mysaveboxM}{#2}%
  \sbox{\mysaveboxT}{\fcolorbox{black}{light-blue}{#1}}%
  \sbox{\mysaveboxM}{%
    \parbox[b][\ht\mysaveboxM+0.5\ht\mysaveboxT+0.5\dp\mysaveboxT][b]{%
      \wd\mysaveboxM}{#2}%
  }%
  \sbox{\mysaveboxM}{%
    \fcolorbox{black}{shadecolor}{%
      \makebox[\linewidth-17.5em]{\usebox{\mysaveboxM}}%
    }%
  }%
  \usebox{\mysaveboxM}%
  \makebox[0pt][r]{%
    \makebox[\wd\mysaveboxM][c]{%
      \raisebox{\ht\mysaveboxM-0.5\ht\mysaveboxT
                +0.5\dp\mysaveboxT-0.5\fboxrule}{\usebox{\mysaveboxT}}%
    }%
  }%
}

\csvset{
  autotabularcenter/.style={
    file=#1,
    after head=\csv@pretable\begin{tabular}{|*{\csv@columncount}{c|}}\csv@tablehead,
    table head=\hline\csvlinetotablerow\\\hline,
    late after line=\\,
    table foot=\\\hline,
    late after last line=\csv@tablefoot\end{tabular}\csv@posttable,
    command=\csvlinetotablerow},
  autobooktabularcenter/.style={
    file=#1,
    after head=\csv@pretable\begin{tabular}{*{\csv@columncount}{c}}\csv@tablehead,
    table head=\toprule\csvlinetotablerow\\\midrule,
    late after line=\\,
    table foot=\\\bottomrule,
    late after last line=\csv@tablefoot\end{tabular}\csv@posttable,
    command=\csvlinetotablerow},
}

\newcommand{\twopartdef}[4]
{
  \left\{
    \begin{array}{ll}
      #1 & \mbox{if } #2 \vspace{0.2cm} \\
      #3 & \mbox{if } #4
    \end{array}
  \color{white}\right\}
}

\hypersetup{colorlinks=true, linkcolor=blue, citecolor=black, urlcolor=black}

\newtheorem{claim}{{\bf \sc Claim}}
\newtheorem{theorem}{{\bf \sc Theorem}}
\newtheorem{lemma}{{\bf \sc Lemma}}

\newtheorem{corollary}{{\bf \sc Corollary}}
\newtheorem{proposition}{{\bf \sc Proposition}}

\newtheorem{observation}{{\bf \sc Observation}}

\def\eproof{\hbox{\hskip3pt\vrule width4pt height8pt depth1.5pt}}

\DeclareMathOperator*{\argmin}{\textrm\upshape{arg\,min}}
\DeclareMathOperator{\cor}{\textrm{Corr}}
\newtheoremstyle{break}{3pt}{3pt}{\upshape}{}{\bfseries}{.}{\newline}{}
\theoremstyle{break}
\newtheorem{algo}{Algorithm}
\begin{document}

\title{Behavioral Communities and the Atomic Structure of Networks}
\author{ Matthew O. Jackson and Evan C. Storms\thanks{%
Department of
Economics, Stanford University, Stanford, California 94305-6072 USA,
Jackson is also an external faculty member at the Santa Fe Institute.
Emails:  and jacksonm@stanford.edu and estorms@stanford.edu.
We gratefully acknowledge support under NSF grants SES-1629446, SES-2018554, and ARO MURI Award No. W911NF-12-1-0509.
We thank Kostas Bimpikis, Ben Golub, Evan Sadler, Omer Tamuz, Ricky Vohra, Yiqing Xing, and participants from RINSE, the Social and Economic Networks conference at the University of Chicago, the NSF Network Science in Economics Conference, and  seminars at Georgetown,
Stanford, and Yale 
for helpful comments and suggestions.
}}
\date{Draft: November 2023}
\maketitle

\begin{abstract}

When people prefer to coordinate their behaviors with their friends---e.g., choosing whether to adopt a new technology, to protest against a government,
to attend university---divisions within a social network can sustain different behaviors in different parts of the network.
We define a society's  `behavioral communities' via its network's `atoms': groups of people who adopt the same behavior in every equilibrium.
We analyze how the atoms change with the intensity of the peer effects, and characterize the atoms in a prominent class of network models.
We show that using knowledge of atoms to seed the diffusion of a behavior significantly increases diffusion compared to seeding based on standard community detection algorithms.
We also show how to use observed behaviors to estimate the intensity of peer effects.


\textsc{JEL Classification Codes:} D85, D13, L14, O12, Z13

\textsc{Keywords:} Social Networks, Networks, Cohesion, Community Detection, Communities, Games on Networks, Coordination, Complementarities, Peer Effects, Peer Influence, Diffusion, Contagion, Atoms
\end{abstract}

\thispagestyle{empty}

\setcounter{page}{0} 

\section{Introduction}


Many behaviors---e.g., attending college, adopting a new product or technology, engaging in criminal behavior, vaccinating one's children, participating in government programs, etc.---are influenced by the choices of one's peers. 
This means that divides in the networks of peer interactions enable different communities to maintain different behaviors, norms, and cultures (for discussion and references see \cite*{jacksonz2014,jacksonrz2017,jackson2019},\cite{chetty2022I}).
Thus, when faced with implementing a policy, for instance to spread a new technology or encourage participation in some program,\footnote{For examples, see \cite*{granovetter1978,duflos2003,centolaem2007,banerjeed2012,aralms2013,banerjeecdj2013},\cite{banerjee2021selecting,alexander2022algorithms,lee2022complex}.  }
it is essential to understand
how the network structure determines that spread.

For example, consider people who prefer to adopt a new behavior (e.g., to participate in an education program, use a new communication technology, or join some social platform,...)
if and only if at least one third of their friends do.
Nobody adopting the new behavior is always a possible equilibrium outcome, and similarly, everybody adopting the technology is always a possible outcome.
More often relevant, however, is that there can also be other equilibria in which some people adopt the behavior and others do not.
If some group of people are sufficiently tightly connected to each other so that each of them at least a third of their links to others in the group,
but are also sufficiently closed so that everybody outside of the group has less than a third of their friends inside the group,  then that group can adopt the behavior while those outside do not.  This divides the society into two different groups:  those who adopt and those who do not.
Whether such split outcomes exist depend on the network structure.
Understanding how a behavior depends on a network structure is essential if a policymaker wants to influence that behavior.

In particular, consider a policymaker who can give incentives to some subset of the population to adopt a behavior---let's call these the initial seeds.
How should the policymaker choose those seeds to maximize the eventual adoption of the behavior?
Optimally choosing the seeds requires taking advantage of network knowledge.  Given the peer effects---e.g., the one third threshold---randomly spreading the seeds around the network is likely
to fail to induce any adoption even if the seeds have high centrality, since most people may only end up with a small fraction of their friends being seeded and thus not adopt the behavior.
Instead, concentrating the seeds carefully in certain parts of the network where people are sufficiently tightly connected is much more effective and can lead to widespread adoption.

To address this problem, one might be tempted to use tools from the enormous ``community detection'' literature (e.g., see the surveys by \cite{fortunato2010,yang2016comparative,moore2017}).
Such algorithms partition the nodes of a network by using variations on a theme of identifying groups of nodes that have relatively high level of connections within the groups and relatively low levels of connections across the groups.
Despite their wide use, community detection techniques are not based on whether the splits in the network
are sufficient to sustain different behaviors or norms across different parts of the partition, but are instead based on optimizing some abstract property of the graph without reference to context (e.g., modularity, cut size, spectrum property...).
Effectively, existing community detection methods ignore that nodes, in many applications, are humans whose interactive behaviors are the point of identifying ``communities''.  What a community is depends on the context, and such methods ignore context.  In fact, it is not even clear how one should interpret what a community is in the detection literature other than a group of nodes that some particular algorithm identified.  This leaves a researcher without any scientific method of selecting a method or being sure it is the right method.   

We provide a new foundation and technique for defining a network's communities that is based upon identifying the patterns of behavior that are sustainable by the network.   Thus, we define the technique directly based on a context of wide interest that not only provides a scientific basis for the technique, but also turns out to produce different results from any existing technique.  
We also show how this technique is useful in optimally seeding a peer-influenced behavior, as well as estimating the level of peer influence underlying a behavior.

Specifically, we consider peer-influence settings in which people prefer to adopt a behavior if and only if at least a given threshold of their friends do.
We define people to be in the same  `atom' of a network if
their behaviors are the same in every (pure strategy) equilibrium of the game.
The atoms that we identify are exactly the building blocks of the equilibria:  every equilibrium is a union of atoms, and every two atoms have some equilibrium that separates them.
Thus, the atomic structure is fundamental to understanding the structure of potential equilibrium behaviors of the society and in influencing those behaviors.

Importantly,  the atomic structure of the network depends on the context/behavior in question.   As the threshold corresponding to adoption of a behavior
changes, so does the atomic structure.  For instance, if people wish to adopt a behavior when only one quarter of their friends adopt the behavior, compared to a different behavior in which they prefer to adopt it
only once at least half of their friends adopt the behavior, then the equilibria and atomic structure can change substantially.
This is another fundamental distinction of our approach from standard community detection techniques.   Instead of
providing just one set of communities, or abstract hierarchy,\footnote{There are some community detection algorithms that return a whole hierarchy of
nested communities, but then the techniques for choosing among those are not tied to behaviors, and often left to the discretion of the researcher.
Moreover, as our behavioral threshold is varied, the atoms are not nested.  Thus, the behavioral atoms that we identify do not correspond to any hierarchy.}
our atomic structure depends on the behavior in question.
We illustrate this difference in the context of several data sets.
This is important, as a policy-maker should be interested in different atomic structures for different programs or behaviors.

We provide results outlining how to identify the different atomic structures as the behavioral threshold is varied.
One theorem characterizes how the atomic structure relates to blocks if the network is generated by
a stochastic block model.  This is the standard random graph model due to \cite*{holland1983stochastic} that has been used extensively to model homophily and communities,
since it is the simplest model that allows linking probabilities to depend on which groups
two nodes belong to, and is thus the minimal extension of a classic Erdos-Renyi random graph that presumes a community structure.
Although there are techniques for discovering the blocks underlying a stochastic block model  (e.g., see \cite{fishkind2013consistent,lei2015consistency,chen2021spectral}),
those methods are not tied to behavior and make it unclear how to group blocks from that perspective.
Our approach provides an intuitive way to identify all of the blocks (even if the number of blocks and the characteristics that matter
are not known by the researcher), by varying the behavioral threshold, and under minimal restrictions on the underlying block model.
In particular, one of our theorems shows that on a large enough network: (i) the atoms are always supersets of the blocks, and (ii) conversely, that any given block can be an atom for some particular behavioral threshold.
Importantly, however, as the behavioral threshold changes, different combinations of blocks comprise the communities in ways that are not always monotone or nested.

We emphasize that which blocks combine into which communities or atoms depends on the behavioral threshold.
Thus, if one wants to understand potential behavioral outcomes on a network, using standard techniques to uncover communities does not provide coherent or accurate predictions about how behavior will actually break
across the graph.  Uncovering the atoms associated with some behavior goes beyond standard techniques by providing the building blocks (which could be combinations of the stochastic blocks) and showing how different combinations comprise the communities as dependent upon the behavior in question.
It is also important to emphasize that our approach and algorithms work well even when the network has an arbitrary structure and is not generated by a stochastic block model.

An additional step enables us to fully understand the potential equilibrium structures.
We further define what we call the ``atomic metagraph.''
This is a directed graph where the nodes of the graph are the collections of atoms and a directed edge from one collection of atoms to another indicates that the second collection of atoms must adopt the behavior whenever all people in the first collection adopt the behavior.
This conceptual object encapsulates the full behavioral implications, effectively delineating all the possible equilibria in one graph.   Moreover, it not only delineates the equilibria, but also captures the directional implications, so that one can predict dynamics of which atoms will adopt a behavior conditional on which others have already adopted it under best response dynamics.
Even though this can become challenging to estimate in some large graphs for some behaviors, its conceptual foundation can be useful for understanding equilibrium structure, and it can still be tractable in larger networks
when the behavior in question only induces a bounded number of atoms (as we illustrate with an application).

We further illustrate the power of estimating the atomic structure of a network by showing how it can be used
to help seed a desired behavior.   We show that an algorithm that uses the atomic structure for seeding a behavior
significantly outperforms not only a random seeding, but also a seeding based on the most popular community-detection algorithm.

In some settings a researcher or policymaker may also be interested in understanding the threshold that characterizes a behavior.
We provide methods of identifying the behavioral threshold by observing an equilibrium.
This can be used as a first step, for example, in a seeding problem.
Having observed some similar past behavior, one can estimate the threshold associated with the behavior and then
identify the atoms, which can then be used to optimize a seeding.
We also show that the combination of first estimating the behavioral threshold from observation of a (noisy) equilibrium and then using it to determine the
atomic structure and seeding a behavioral diffusion performs well relative to knowing the threshold, provided there is not too much noise in behavior.

Finally, we provide a novel observation about the evolution of networks.
In two different applications for which we have time panels of networks, we show that links within
atoms are significantly more likely to be maintained over time than links between atoms.
This is consistent, for instance, with people getting higher payoffs from interacting with others with whom their behaviors coordinate.

An appendix collects a number of further results.\footnote{
One is that we show how the observation of a number of different behaviors can be used to recover information about the network when data about the network is missing. (For
discussion of the importance of inferring networks
without observing links
see \cite{brezaetal2018}.)
In particular, we provide bounds on the number of behaviors that are needed to be observed to recover the complete atomic structure (even without knowing the number of atoms).
These techniques can also be used to discern which types of networks matter in settings with multi-layered networks.
We also provide additional results for situations in which the threshold above which an individual wishes to adopt a behavior depends on the absolute number of friends rather than the fraction of friends.
All of our definitions and techniques have analogs for this case.
In addition, we show how to distinguish whether behavior is driven by a fractional threshold or an
absolute threshold.
We illustrate this by showing that smoking behavior by students in a US high school is better explained by a fractional than an absolute threshold -- and we  estimate the behavioral threshold that best matches the atomic structure to the observed behavior.
We provide further results on how to identify thresholds in situations in which different individuals have different thresholds.}

\paragraph{The Related Literature.}

\

Our results relate to the games-on-networks literature.\footnote{See \cite*{jacksonz2014} for an overview and references.}
A key early contribution to that literature was \cite{morris2000},
whose results addressed the question of which networks allow for two different actions to be
played in equilibrium.\footnote{There is an earlier literature on the majority game or voting game, e.g.,
\cite{cliffords1973,holleyl1975} that is a precursor to Morris's analysis
for a specific threshold.   There is also a broad literature on conventions (e.g., see the discussion in \cite{young1996,young1998}) that examines coordination games
played by populations and discusses issues about supporting multiple conventions.  Our analysis provides a network-explicit foundation for such an analysis (see also  \cite{ellison1993,ely2002,jacksonw2002b} for some discussion of how some specific networks as well as endogenous networks determine the stochastically stable conventions).}
Our analysis builds upon some of Morris's definitions but explores different questions.   A recent (independent) paper by \cite*{leister2022social} uses a global games approach to partition players in a network based on their (hierarchical) thresholds for adoption in a coordination game.  Despite also working with equilibria in coordination games, the approach and concepts are based on incomplete information and ordering of adoption in a unique equilibrium,  rather than having the same equilibrium behaviors across multiple equilibria. Thus their partition has a very different interpretation and structure from the atoms identified here, which are building blocks of multiple equilibria. This also leads to different insights and applications.

There is a literature in which communities emerge from people choosing relationships and/or the groups to which they wish to belong, such as \cite*{currarinijp2009,currarinijp2010,currarinim2012,ketss2016,athey2021theory,zuckerman2022unseen,canen2022social}.
That is a reverse perspective from ours, since it is preferences over partners that determines the network, while our focus here is on how networks determine behaviors.
Some work such as \cite{chenetal2010} endogenizes community structures as equilibria phenomena in which agents choose which communities to join and may join multiple communities.
That is closer to our motivation in terms of examining communities as potential equilibria on a given network, but our interest is in terms of defining communities based on behavioral norms rather than as groups that agents join, and so the communities that we define are quite different from those of \cite{chenetal2010}.
Nonetheless, we provide new empirical observations about how networks evolve based on the atomic structure.  These could be useful in
further development of models of network formation.

Our results on how to identify all of the blocks of
a stochastic block model make use of a recent breakthrough
in random graph theory characterizing a modularity property of random graphs (\cite{mcdiarmid2018ER_modularity}),
that we extend to a more general class of random graphs.

Our approach is not only different from the previous community detection literature in terms of what we identify as a community, but also in terms of
the perspective that drives our definition.   We microfound our definition of community in the patterns of
behavior that a network can support: so our approach is motivated and {\sl directly defined} by
how networks shape behaviors.
This change in focus is fundamental and does a couple of things.
First, it provides a reasoning behind what a community
is: a microfoundation for the definition.    Second,
it often leads us to find more basic `atoms' that differ substantially from the communities identified by standard community detection algorithms.
In particular, our community partitions vary with the behavior in question,
and cut across those produced by other algorithms:  refining some communities and grouping others together based on the contagion of behavior that would occur.

Our results provide a new perspective on the distinction between simple and complex contagion (e.g., see \cite*{centolaem2007,centola2010,centola2018}).
The usual distinction is whether people need only a single contact to become ``infected'' (e.g., one neighbor adopting enough to induce a person to adopt) or more than one contact.  While something like a flu spreads with a single contact, the adoption of a new technology typically depends on the relative fraction of a person's friends who adopt and that makes such a behavior harder to spread.
Nonetheless, some (but not all) complex contagions spread  widely (e.g., see \cite*{ecklesetal2018}).
Our results show that the key distinction is not whether a behavior requires a threshold more than one neighbor adopting, but
on how high the adoption threshold is, and where the initial adopters sit.
An important distinguishing factor is whether the threshold is lower or higher than the relative frequency of links across blocks in the network.   If the threshold is below that frequency, then behavior necessarily spills over
across blocks and the atoms are combinations of blocks and behavior diffuses.  In contrast, when the threshold of behavior exceeds the relative frequency of links across blocks then the atoms coincide with the blocks, and behavior is much harder to spread across blocks.
This is a  different distinction between simple and complex contagion: diffusions that require a high enough {\sl fraction} of an agent's neighbors to be infected before the agent becomes infected, where that fraction is above the frequency of links across cuts in the network, are the ones that are truly `complex' in the sense that they can have limited spread.\footnote{Some of the early experiments distinguishing simple from complex diffusions were on induced networks with small degrees (e.g., \cite{centola2011}), and then there is no real distinction between more than one contact and a nontrivial fraction,
which explains why \cite*{ecklesetal2018} found that some complex diffusions spread just as widely as a simple ones without very high clustering.  Once one has a larger degree, then as we show below, diffusions that require some relatively small number of contacts to be infected before an agent
becomes infected spread much more widely than
those that require larger fractions of contacts to be infected, which can help make sense of the different conclusions of the prior experimental and empirical work examining complex diffusion.}

The seminal work by \cite*{kempekt2003,kempekt2005,mosroch2009} on optimal seeding focused on the special case in which
influence is ``submodular'', so that there is no particular threshold of adoption behavior: the probability of
adopting can increase as more friends adopt, but there are diminishing returns to having more friends adopt.
In contrast, our setting captures applications involving coordination
in behavior, in which influence is decidedly not
submodular.  The case of submodularity is closer in its properties to the spread of a disease or meme, in which infection can happen by contact with a single neighbor and there is a simple probability of actually having contact with the neighbor during an infectious period.
Such simple contagion exhibits diminishing returns and submodularity. Instead, behaviors that involve coordination and preferences over the number of neighbors taking an action fail submodularity.
For example,
adopting a new technology requires that some threshold of one's friends have adopted,
and that threshold depends on how much better the new technology is than the old technology.
This corresponds to a coordination game.  For instance, if the new technology is slightly better
than the old one, and there is a benefit to having the same technology as one's neighbors,
then a threshold of $q$ somewhat lower than 1/2 might apply, say 1/3.
If one has 12 friends, then one is willing to adopt if 4 or more friends do.
Here the fourth friend's adoption is more influential than the first, second or third friend's
adoption, violating submodularity.
Even noising up the threshold to make it probabilistic does not change the fact that some person beyond the first person is more influential.
This sort of failure of submodularity is present in many settings where one needs sufficient convincing or encouragement in order to act, especially in which there are coordination motives.
This difference introduces substantial challenges that completely invalidate previous techniques for approximately optimal seeding, that our approach makes first steps
in addressing.
In particular, the previous literature showed that one
could adapt existing algorithms to do well on the seeding problem under submodularity; but those
results do not hold without that condition.
We develop a new type of algorithm that makes explicit use of the atoms---choosing and seeding atoms that
provide the widest spread of behavior for the minimal number of seeds---recognizing
that one needs to understand the atoms to effectively seed in non-submodular settings.\footnote{There is also variations on the seeding literature that examine offering incentives to people for adopting behavior or influencing friends in settings with networked peer effects (e.g., \cite*{leducjj2016,nora2023exploiting}).  Our approach is less related to those papers both in results and model, but our seeding could be a useful foundation for such an analysis in this setting.}

The observation that concentrating seeds can help in diffusing behavior in the presence of complex contagion has been noted previously
(e.g., \cite{calvoj2004,dodds2011threshold,aralms2013,guilbeault2021topological}), and our approach provides additional structure by showing how that concentration relates to the atomic structure.

\section{A Model}

A network describes relationships and people choose behaviors as a function of their neighbors' choices.

\subsection{Networks}

A \emph{network}, $(N,g)$, consists of a finite set of nodes or agents, $N=\{1,\ldots, n\}$, together with a list of the undirected links or friendships that are present, denoted by $g$.
Specifically,  $ij\in g$ indicates the presence of a link or friendship between nodes $i$ and $j$.\footnote{
Formally, $g\subset G\equiv \{  S\subset N: |S|=2\}$, and the notation $ij$ is shorthand for $\{i,j\}$.
Our focus is on undirected networks (mutual friendships) and so $g$ is a set of unordered pairs (e.g., $ij$ and $ji$ are the same link).
Our definitions extend readily to directed graphs, and the results have analogs for random directed graphs.}
We adopt the convention that $ii\notin g$, so that agents are not friends with themselves.

Agent $i$'s friends or neighbors in $g$ form the set $N_i(g) \subset N\setminus \{i\}$ such that
$N_i(g)\equiv \{j | ij\in g\}$. Agent $i$'s degree is the size of $N_i(g)$,
denoted $d_i(g) = |  N_i(g)|$.
Isolated nodes are not of any interest in our setting.
Thus, in the definitions that follow we presume that a network $g$ is such that each node has at least one neighbor ($d_i(g)\geq 1$ for all $i\in N$), as isolated nodes have no friends to influence their decisions.

\subsection{Conventions and Equilibria}

Each agent chooses one of two actions: to adopt a behavior or not (e.g., adopt a new interactive technology or product, learn a new language, participate in a program, go to university, to commit a crime, etc.)  and cares about how that action matches with the
actions of their friends.
In particular, a behavior is characterized by a threshold $q \in (0,1)$ such that an agent adopts
that behavior if at least a fraction $q$ of their neighbors do, and not otherwise.

A {\sl convention}---equivalently an {\sl (pure strategy) equilibrium}---for a threshold $q$ on a network $(N,g)$ is a set $S\subset N$ of agents
who adopt the behavior such that every agent $i\in S$ has at least a fraction $q$ of their
neighbors adopting behavior and every $i\notin S$ has less than a fraction $q$ of their friends adopting the behavior.
That is an equilibrium is a set $S\subset N$ of agents such that
\begin{itemize}
\item $|N_i(g)\cap S|/d_i(g)\geq q$ for all $i\in S$,  and
\item $|N_i(g)\cap S|/d_i(g)< q$ for all $i\notin S$.
\end{itemize}

As in \cite{morris2000}, an interpretation is that an agent is playing a coordination game and gets a payoff based on
how their behavior matches with each one of their neighbors:
\[
\begin{array}{cccc}
 & \ & Neighbor's \ Choice: & \\
& \  & Adopt \ Behavior  & \ \ \ \ \ Not \ Adopt \ \ \ \ \  \\  \cline{3-4}
Own \ Choice: \ \ \ \ \ \ \ & Adopt \ Behavior  & \multicolumn{1}{|c}{x} & \multicolumn{1}{|c|}{y} \\   \cline{3-4}
& Not \ Adopt &\multicolumn{1}{|c}{w} & \multicolumn{1}{|c|}{z}    \\  \cline{3-4}
\end{array}%
\]
where $x>w$ and $ z>y$.
This coordination game has a corresponding threshold such that if a fraction of at least $q= \frac{z-y}{x-w+z-y}$ of an agent's neighbors adopt the behavior, then the agent's best response is to adopt too.
At exactly this threshold an agent is indifferent, but otherwise has a unique best response.

Generically, the threshold would never be hit exactly.  However, some rational thresholds, such as $q=1/2$, are prominent in the literature,
as, for instance, people might simply follow the crowd and do what the majority of their friends do,\footnote{The `voter game' or `majority game' is well-studied in the statistical physics literature, among others (e.g., see the references in \cite{jacksonz2014}) .}
and so we allow for rational thresholds.
Unless otherwise noted, we break ties so that an agent adopts the behavior if exactly $q$ neighbors do.
If $q$ involves tie-breaking, then the case in which agents do not adopt the behavior exactly at the threshold can be analyzed by looking at a threshold of
$q-\varepsilon$ for some small $\varepsilon<\frac{1}{n-1}$.
If one reverses the tie-breaking rule, then our use of open and closed intervals below reverses.

A convention is thus the set of adopters in a pure strategy Nash equilibrium of the above coordination game.
Mixed strategy equilibria are unstable in that slight perturbations in mixed actions lead best response dynamics away from the mixture, and thus they are of little interest in defining communities and so we concentrate on pure strategy equilibria. 
Pure strategy equilibria always exist---e.g., both the empty set and the whole set of nodes form conventions---and are stable for generic choices of $q$.

More generally, as discussed in \cite{jackson2008}, most applications of technology adoption and choices of norms are such that people can see and react to the choices of others around them.  Thus, pure strategy equilibria are the natural resting points of dynamic processes, while mixed strategy equilibria are not.  Nonetheless, the calculations behind mixed strategy equilibria can be useful and are related to the thresholds that we use here.

In addition, the complete information setting in which people see and react to the choices of their neighbors is a more natural one than incomplete information games in many settings of long-term choices, such as whether to learn a language or whether to adopt a new technology, where people are explicitly reacting to the actions of their neighbors, and this is built in to the dynamics we describe below and single out pure strategy equilibria as resting points. 

\begin{figure}[h!]
\centering
\subfloat[The shaded nodes form a convention]{
\includegraphics[width=0.3\textwidth]{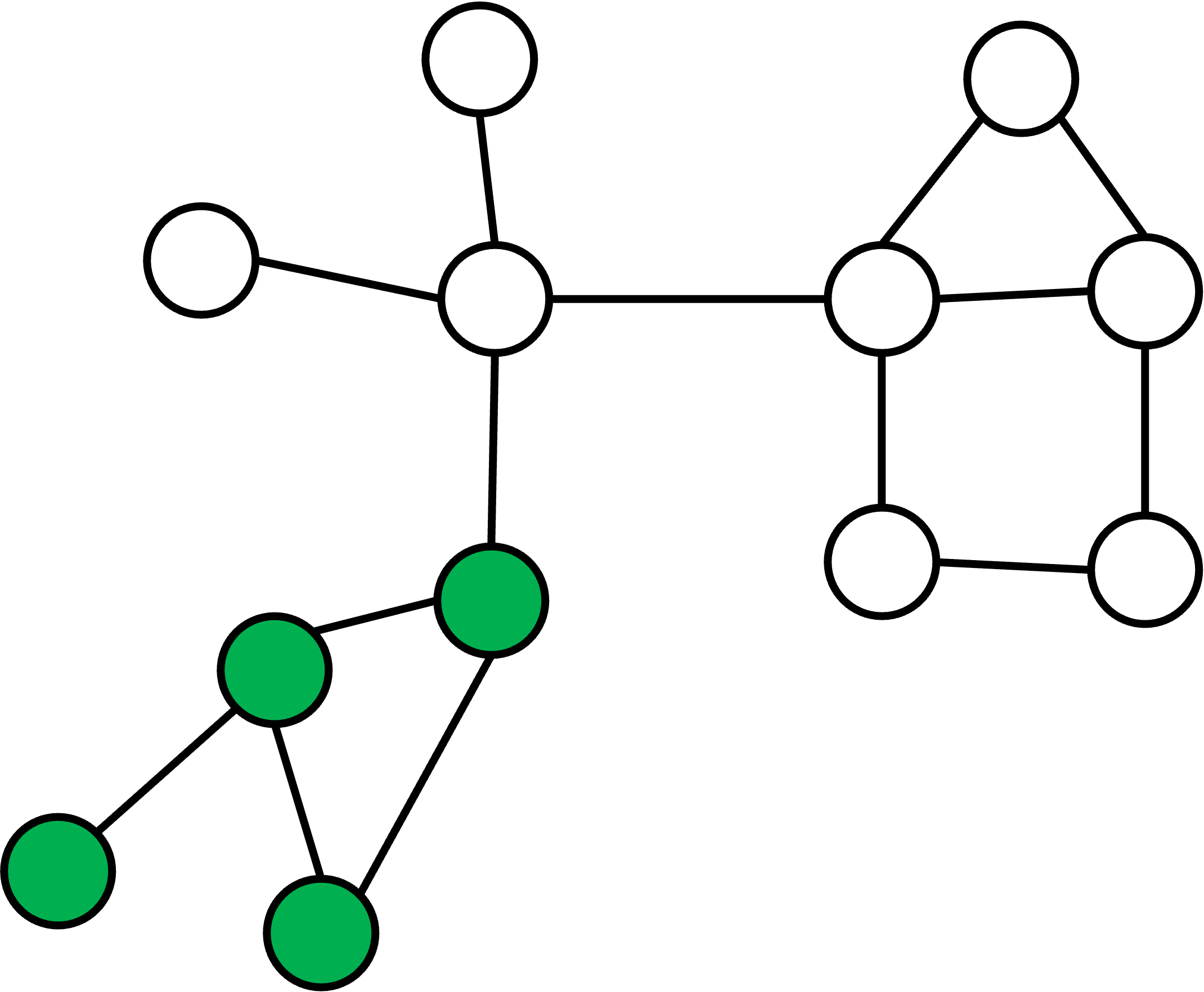}
}
\includegraphics[width=0.05\textwidth]{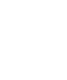}
\subfloat[In this example, the complement is also a convention for the same $q$]{
\includegraphics[width=0.3\textwidth]{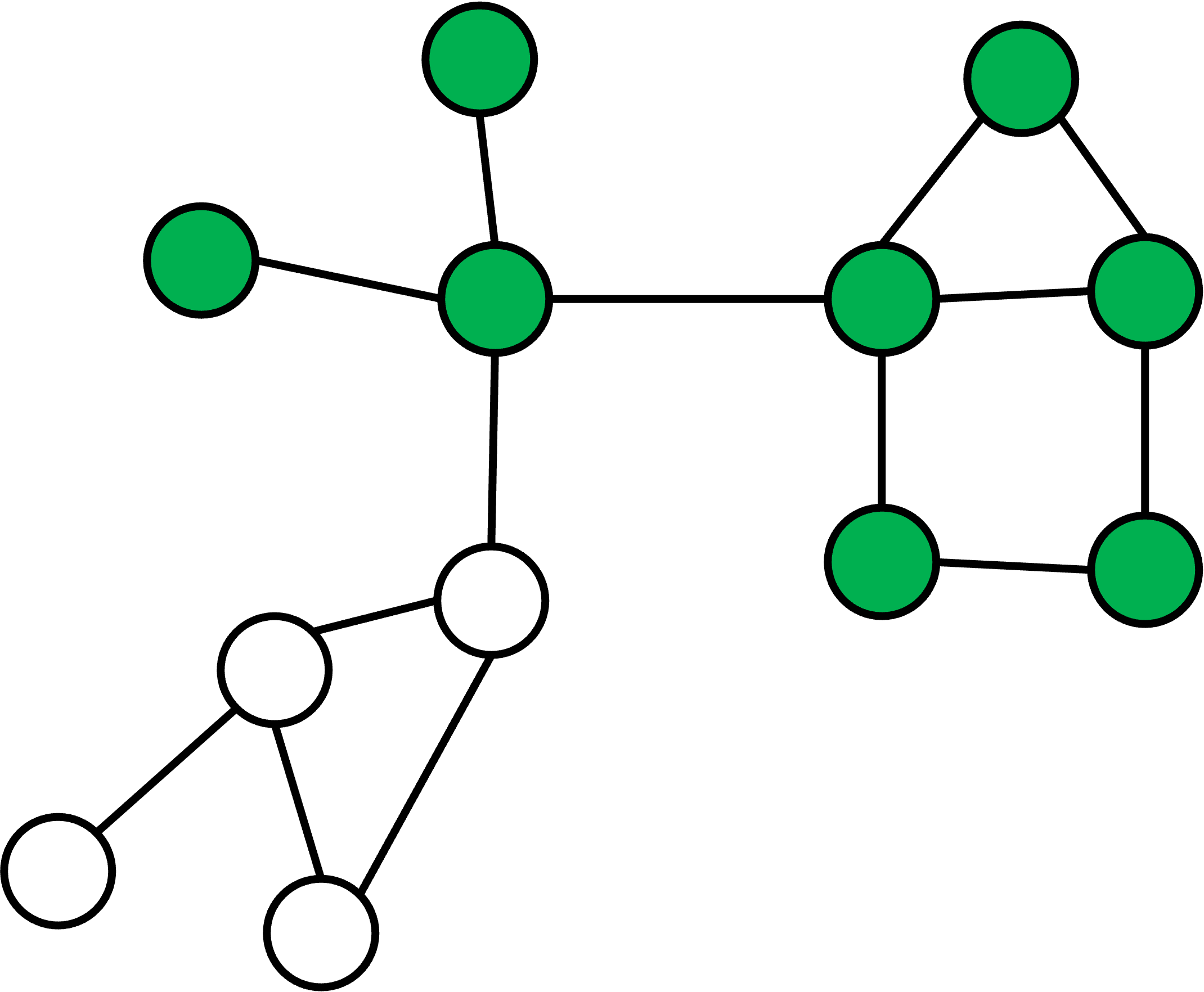}
}
\caption{\label{ExampleConvention}
{\bf Conventions:} Two conventions when the threshold is such that a person adopts if and only if at least forty percent of their
neighbors do ($q=.4$).}
\end{figure}

Two conventions are pictured in Figure \ref{ExampleConvention} for a behavior with a
threshold of $q=0.4$.

Following \cite{morris2000}, define
a group of agents $S\subseteq N$ to be $q$-cohesive if each of its members have a fraction at least $q$ of their neighbors in the group  ($|N_i(g)\cap S|/d_i(g)\geq q$ for all $i \in S$).
We say that a group $S\subseteq N$ is $q$-closed if every individual outside of $S$ (in $N\setminus S$) has
a fraction of their friends in the group that is strictly less than  $q$  ($|N_i(g)\cap S|/d_i(g)< q$
for all $i \notin S$).


A { convention} for threshold $q$ on a network $g$ is a subset of nodes that is both $q$-closed and $q$-cohesive.

\subsection{Absolute Thresholds }

The above definitions are relative to some fraction of at least $q$ of neighbors taking an action.
This applies naturally to coordination problems.
For some other games of strategic complements, it can be natural to adopt a behavior if at least $t$
neighbors do, for some $t\in \{0,1,2,\ldots, n-1\}$.
For instance, one might benefit from learning to play a particular game that requires $k$ agents if and only if at least $k-1$ friends play the game.

In what follows, there are equivalent definitions switching $q$ and $t$ everywhere, so we just present the definitions for the $q$ case.  Some interesting contrasts between the relative and absolute threshold atoms are discussed in an appendix.

\subsection{Atomic/Community Structures as Partitions Generated by Conventions}

We now define the central concept of our analysis:  how community structures are defined from conventions.

Given a network $g$, let $\sigma(q,g)$ denote the $\sigma$-algebra\footnote{Although most of
our exposition presumes finite $N$, the definitions and discussion apply to infinite $N$ as well.}
on $N$ generated by all the sets of agents $S\subset N$ that are conventions on $g$ given
threshold $q$.
The {\sl atoms} of $\sigma(q,g)$ are the minimal nonempty sets in $\sigma(q,g)$, which exist by finiteness.
They form a partition that generates $\sigma(q,g)$ and are denoted by $A(q,g)$.  
The atoms are weakly finer than the conventions themselves, but are the minimum building blocks of those conventions (and hence the name ``atoms''), as we discuss shortly.

\begin{figure}[h!]
  \centering
    \includegraphics[width=0.3\textwidth]{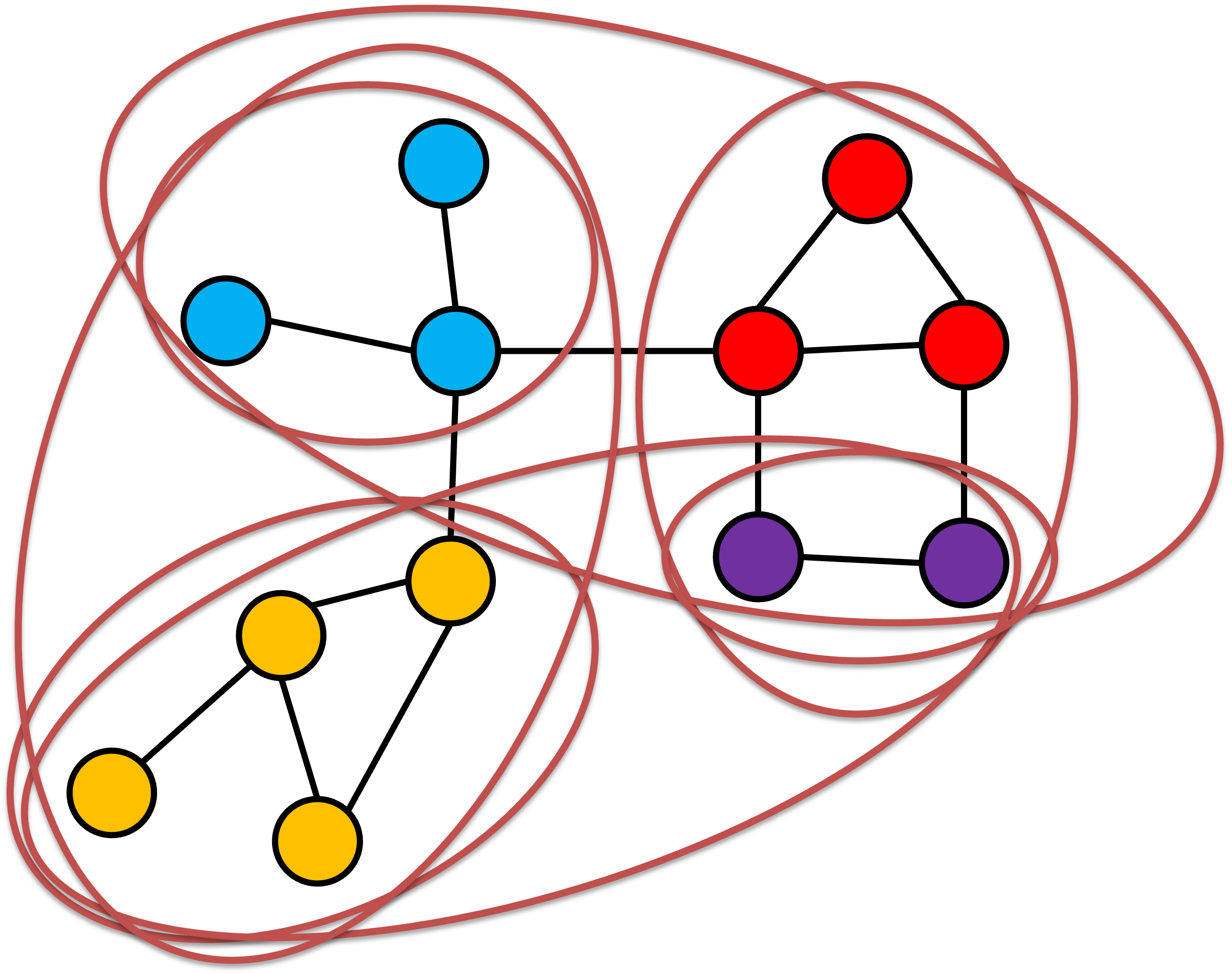}
    \includegraphics[width=0.10\textwidth]{blank.jpg}
    \includegraphics[width=0.23\textwidth]{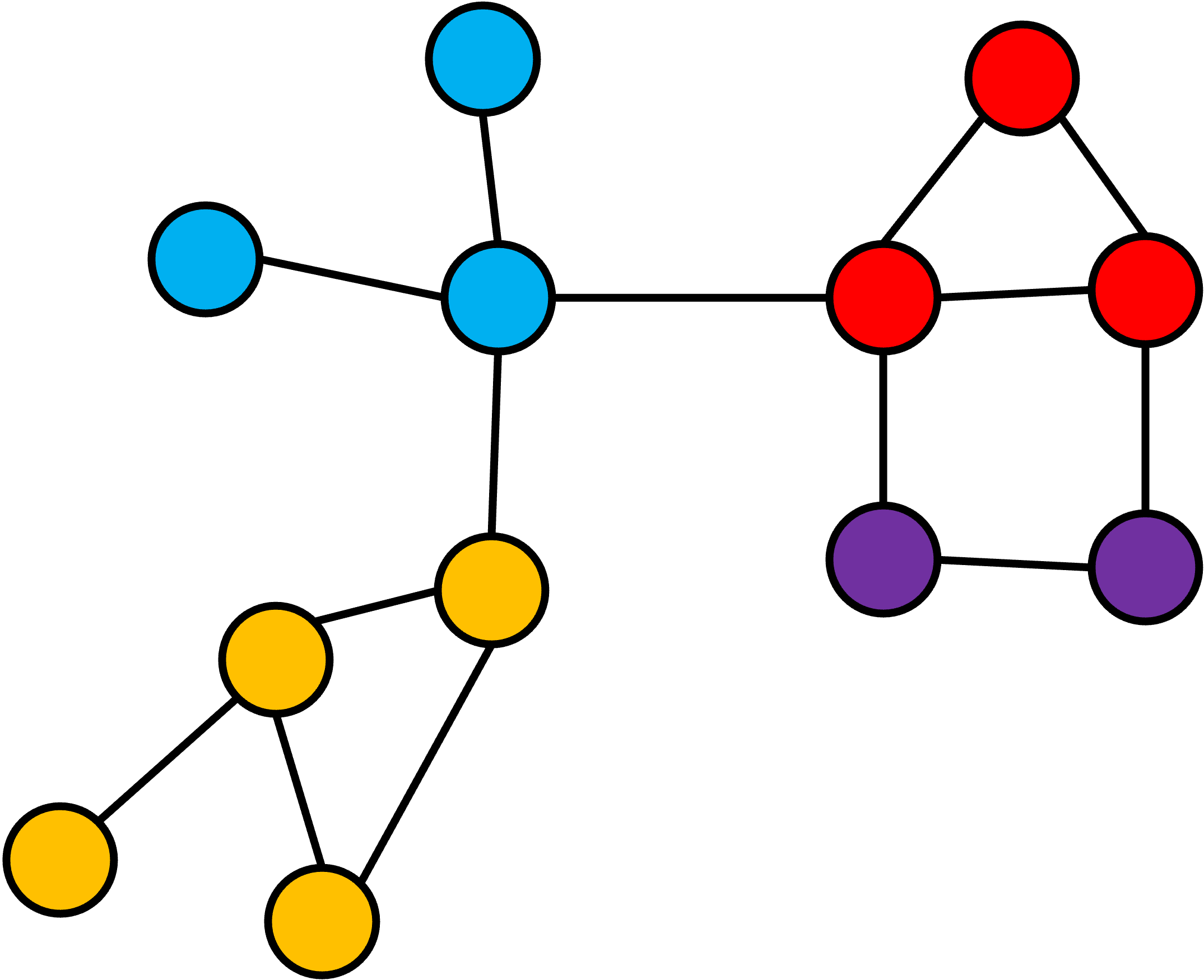}
    \caption{\label{ExampleCommunity}
{\bf The Atoms or Behavioral Communities:} Each oval contains the nodes associated with one of the possible conventions (not drawn are the trivial conventions that include all nodes and no nodes).   Taking the intersections of all of these as well as all of their complements defines which nodes are always in/out of conventions together, which in turn defines the atoms or behavioral communities for $q=.4$; i.e.,  the atoms, $A(.4,g)$, which are color-coded.
These are also the atoms for any $q$ between 1/3 and 1/2.  Moreover, the complement of each equilibrium is an equilibrium for the complementary $q$, and so these are also the atoms for each $q$ between 1/2 and 2/3.
}
\end{figure}

Given a finite $n$, one can find the atoms by partitioning the nodes by successive bisection from
each convention and its complement, as in Figure \ref{ExampleCommunity}.
We use the terms `communities' and `atoms' interchangeably in what follows to describe the elements of $A(q,g)$.

To see how conventions define atoms/communities, let us consider all of the other conventions associated with the network from Figure \ref{ExampleConvention}.

\begin{figure}[h!]
\centering
\includegraphics[width=0.25\textwidth]{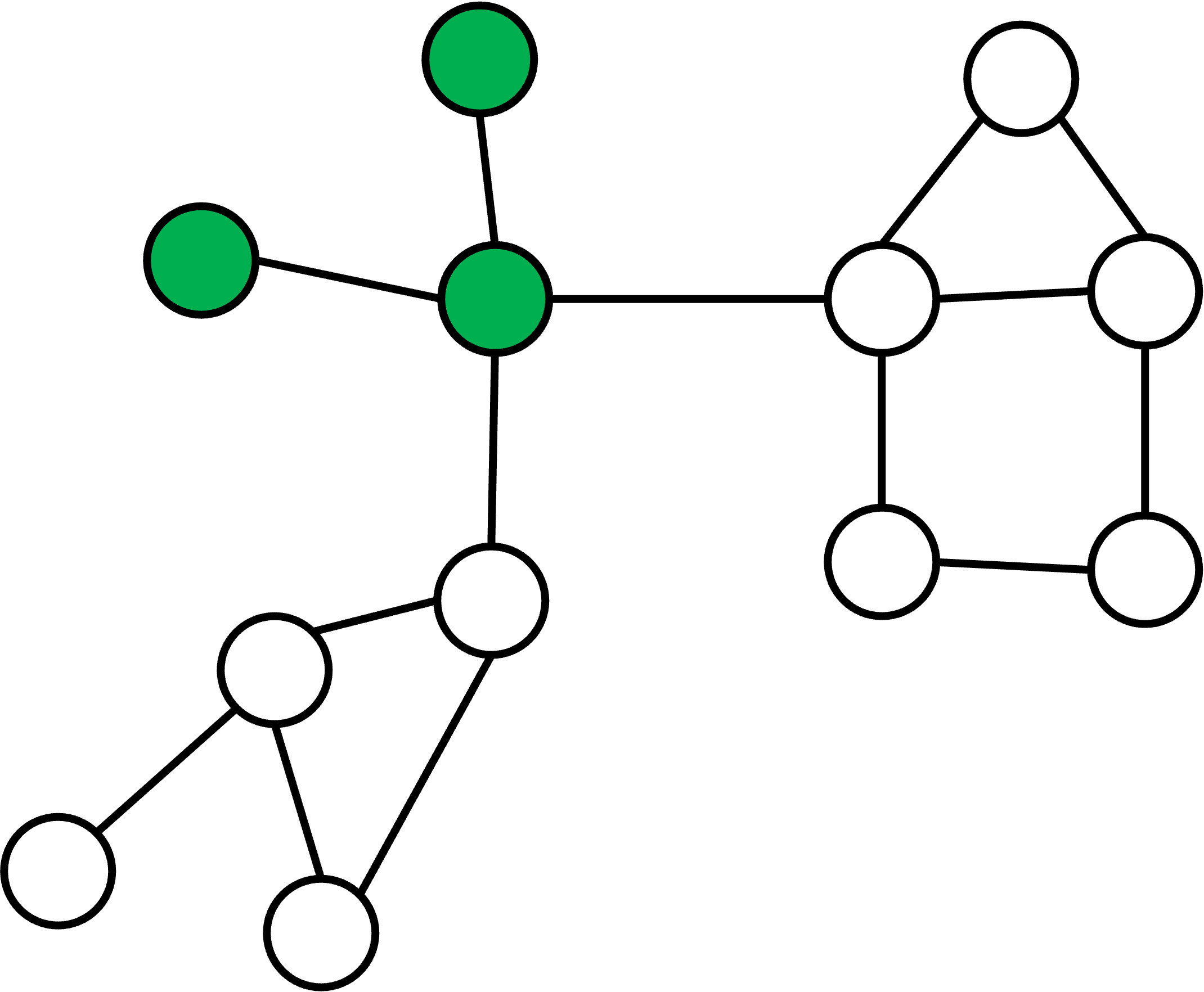}
\includegraphics[width=0.05\textwidth]{blank.jpg}
\includegraphics[width=0.25\textwidth]{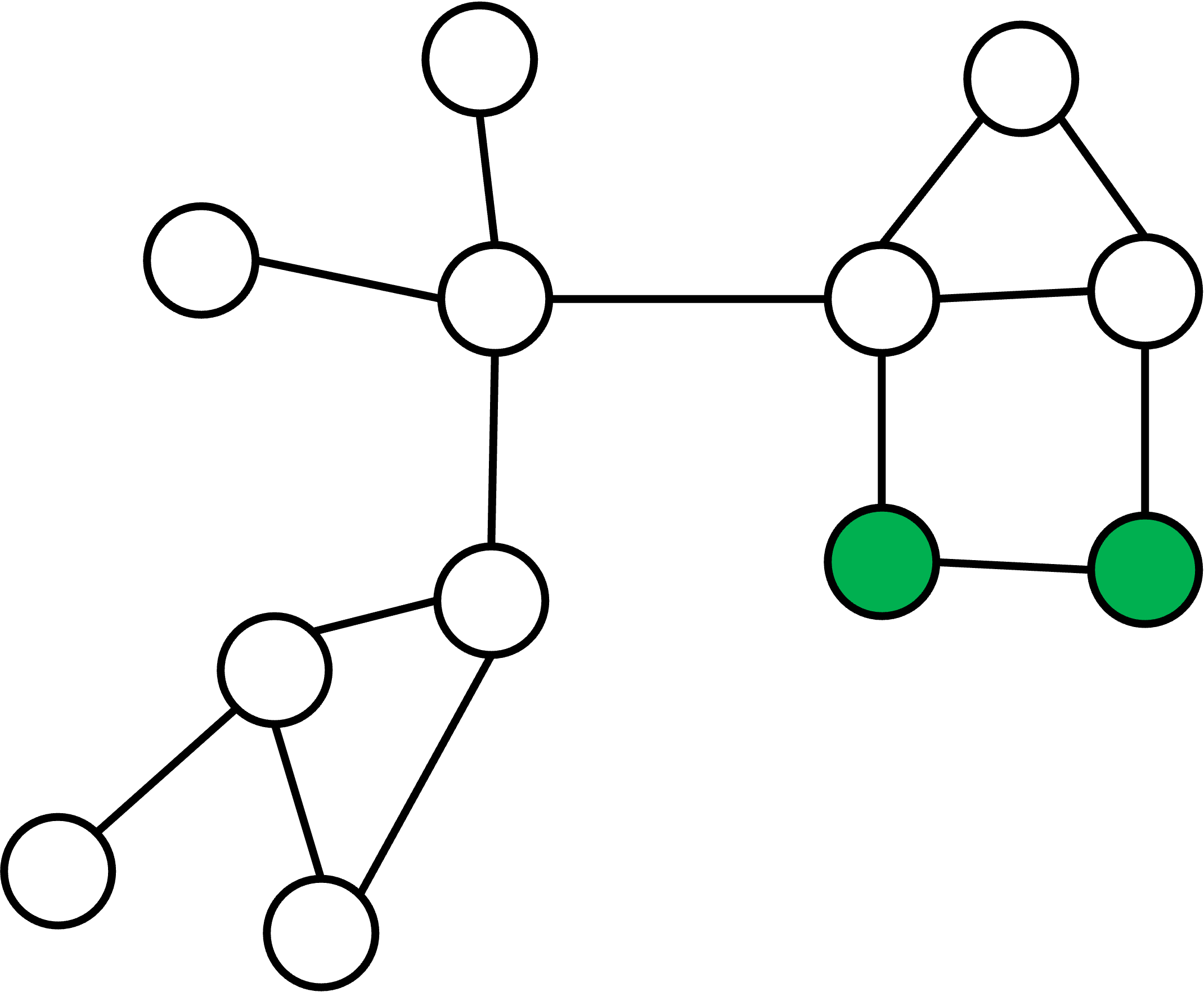}
\includegraphics[width=0.05\textwidth]{blank.jpg}
\includegraphics[width=0.25\textwidth]{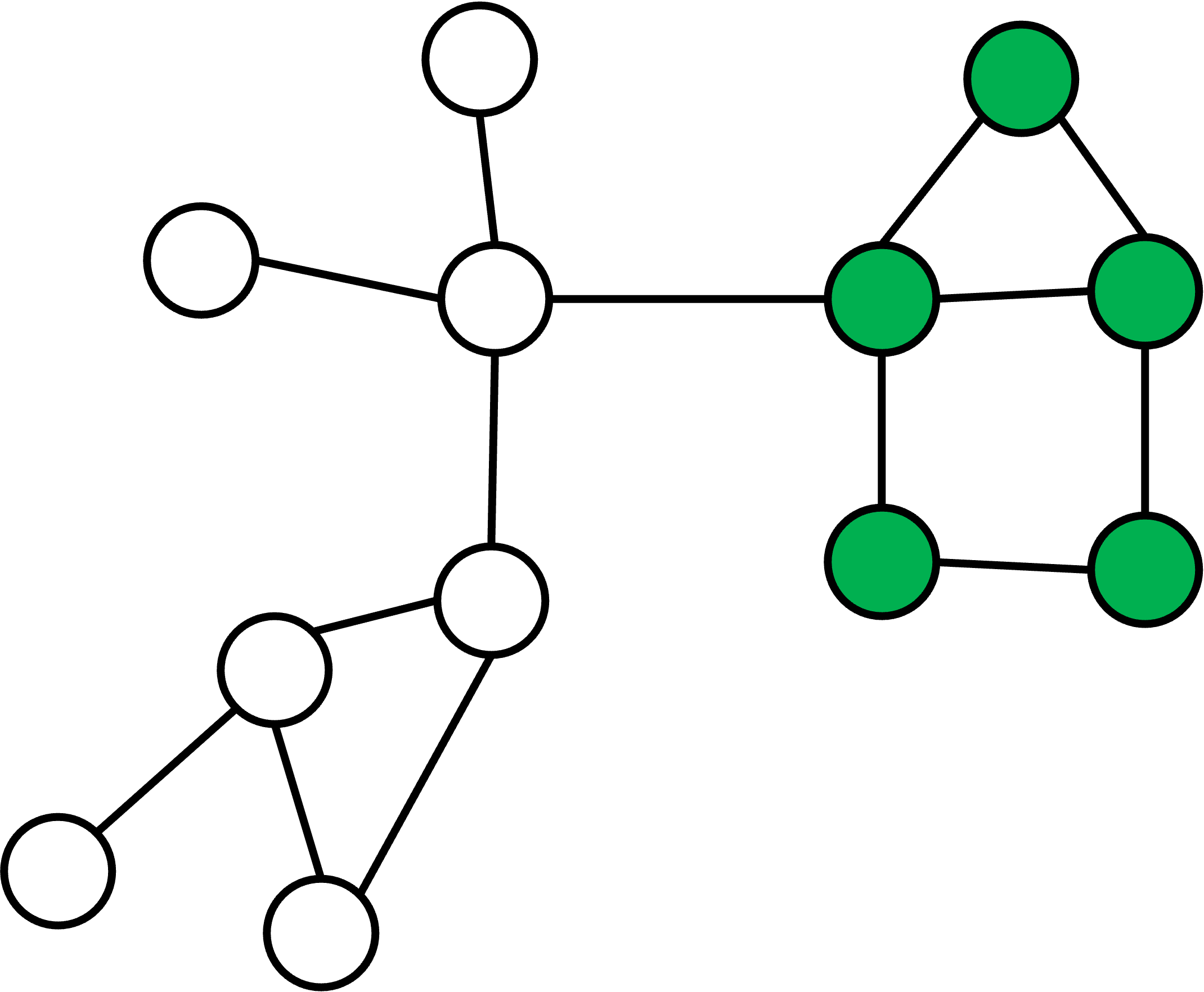}
\includegraphics[width=0.25\textwidth]{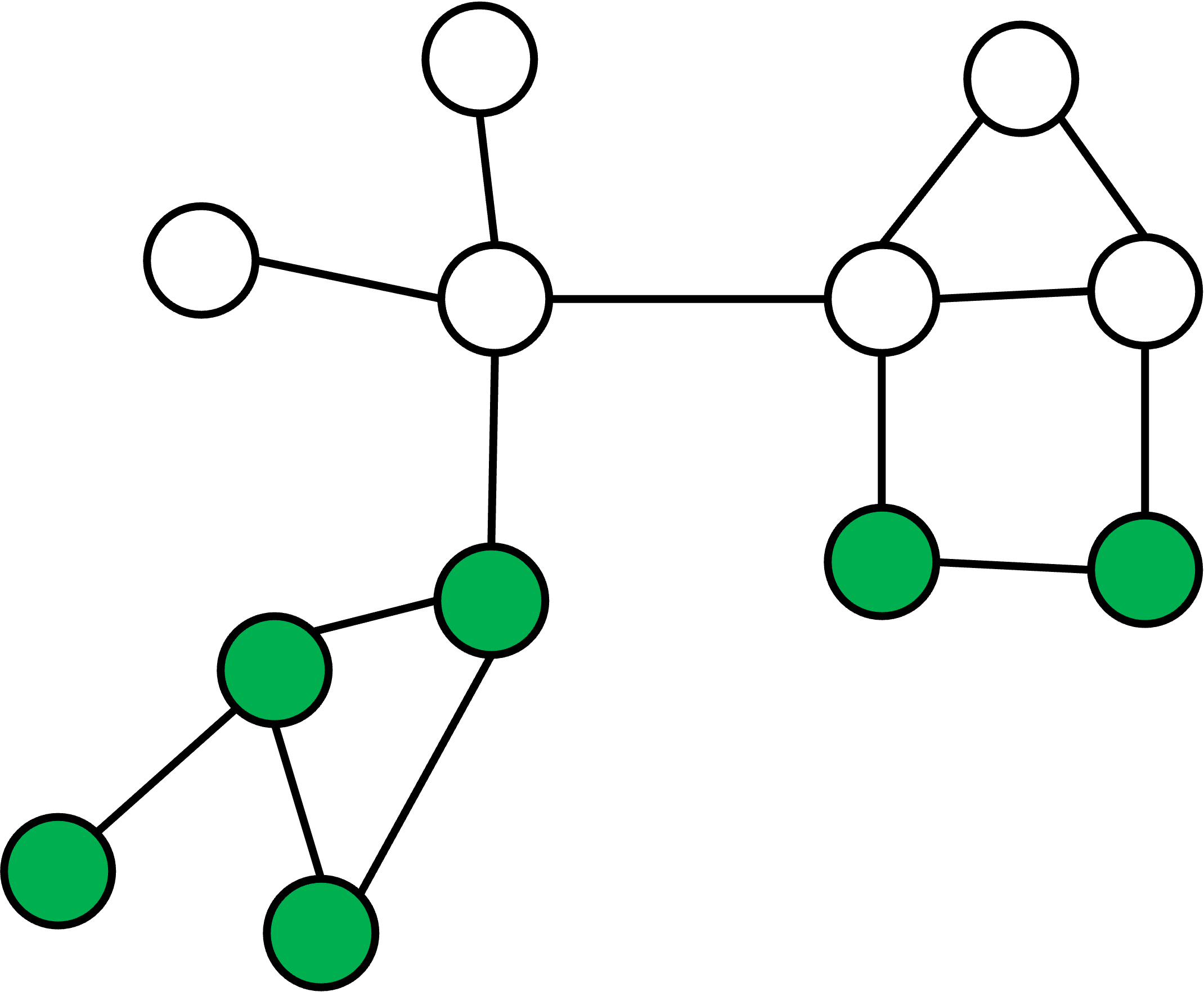}
\includegraphics[width=0.10\textwidth]{blank.jpg}
\includegraphics[width=0.25\textwidth]{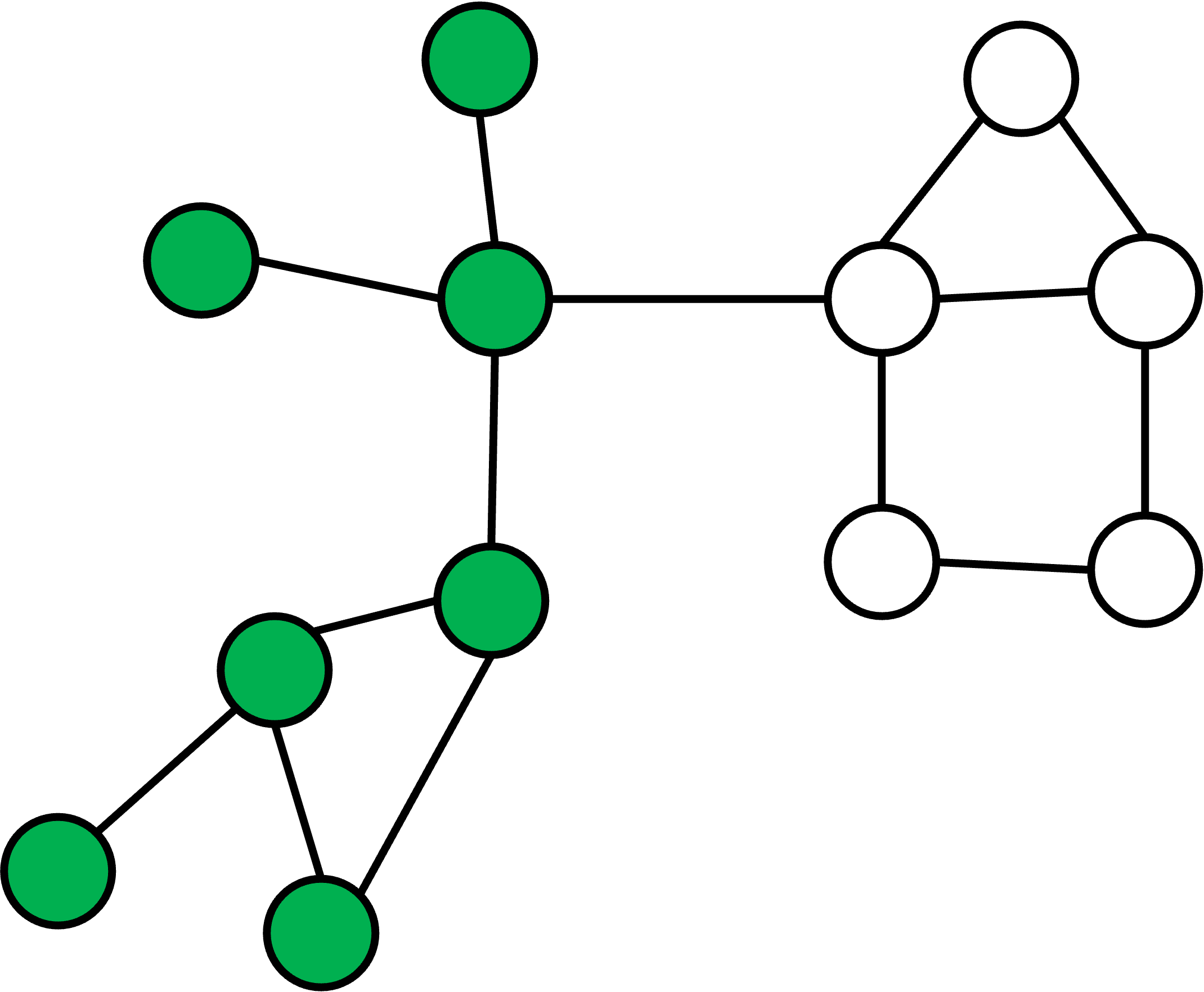}
\caption{\label{ExampleConvention2}
{\bf More Conventions:} The other (non-degenerate) conventions when $q=.4$  (which are also conventions for $q\in (1/3, 1/2)$) .}
\end{figure}

The partition into atoms induced by all of the conventions from Figures \ref{ExampleConvention} and \ref{ExampleConvention2} is pictured in
Figure \ref{ExampleCommunity}.

Note that some atoms may not be conventions by themselves.
For instance, the red nodes in Figure \ref{ExampleCommunity} are never their own convention.  However, every convention is a union of atoms.  
In particular, nodes inside an atom behave the same way in every $q$-convention,
and if two nodes are in different atoms then there is some convention under which they behave differently.
Consequently, conventions are necessarily unions of atoms,
and thus the atoms are the basic building blocks of coordinated behaviors.

\begin{figure}[h!]
  \centering
    \includegraphics[width=0.32\textwidth]{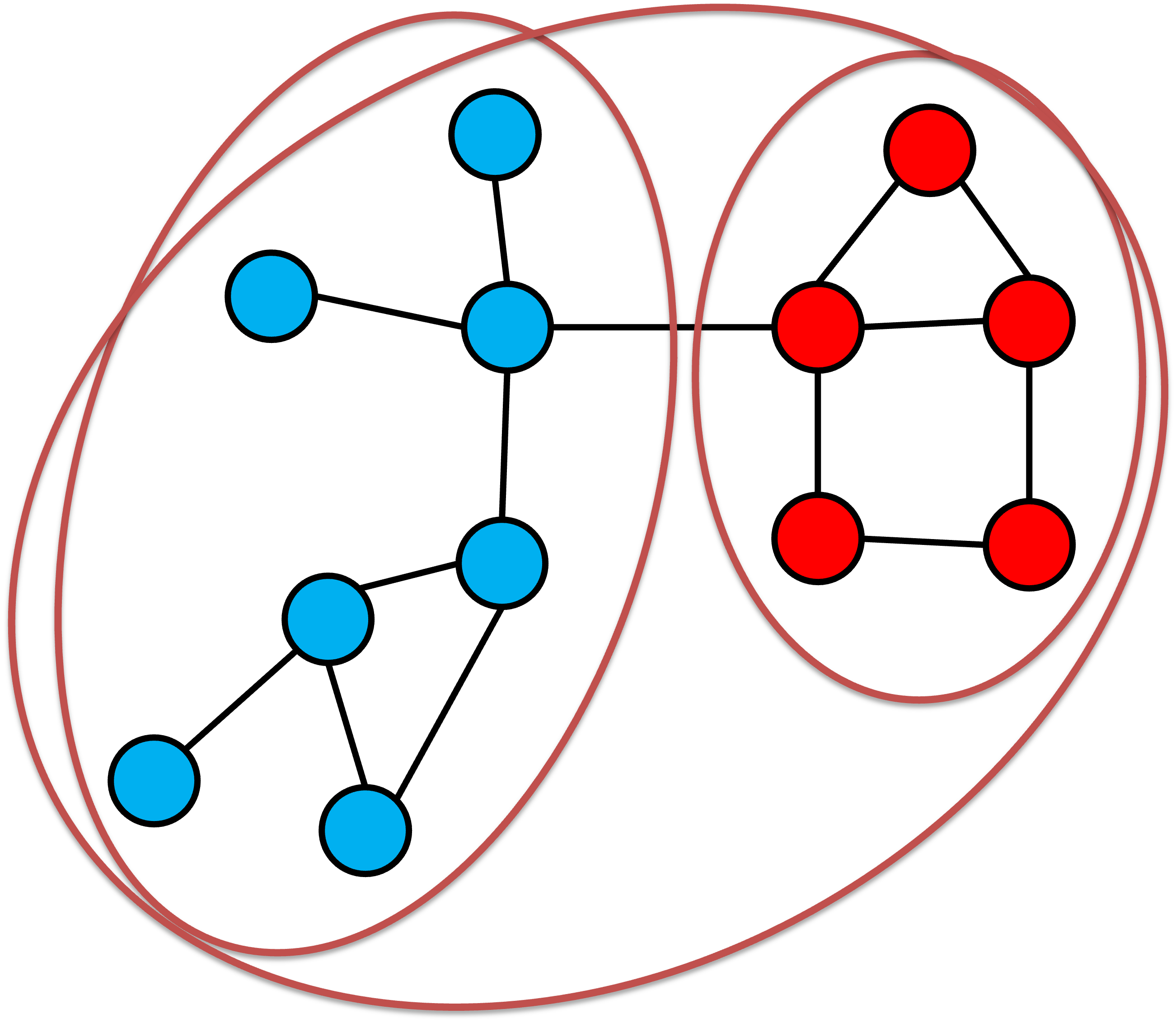}
    \includegraphics[width=0.10\textwidth]{blank.jpg}
    \includegraphics[width=0.32\textwidth]{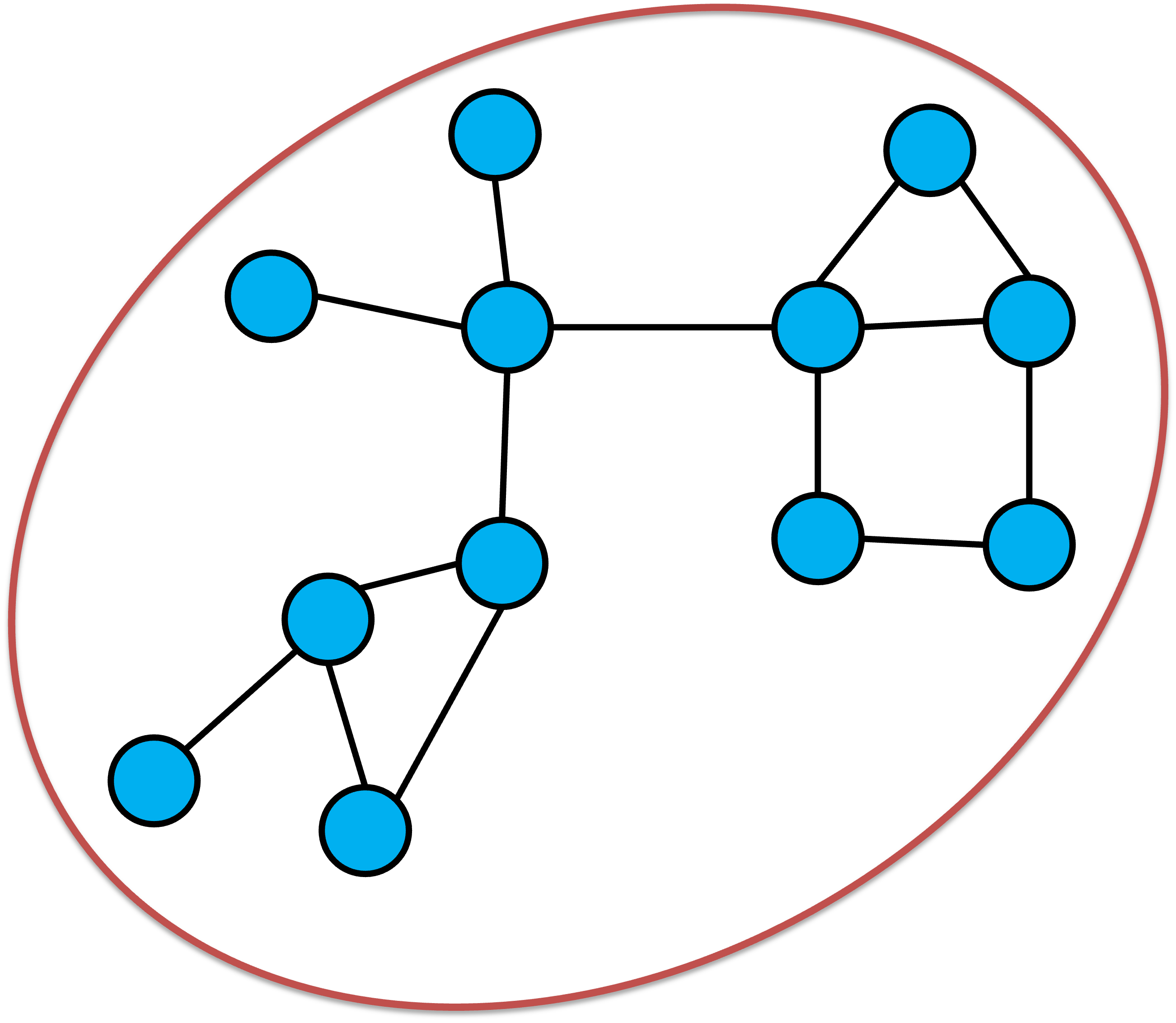}
    \caption{\label{ExampleCommunity2}
{\bf The Atoms for Other $q$s:} The figure on the left pictures the atoms for any $q$ strictly between 1/4 and 1/3 (or between 2/3 and 3/4),
and on the right pictures the atoms for any $q$ less than 1/4 or greater than 3/4.}
\end{figure}

It is important to note that the atoms/communities are not necessarily nested as $q$ varies.
As the threshold $q$ is increased, the cohesiveness requirement for a convention gets {\sl harder}
to satisfy; while in contrast the closure requirement gets {\sl easier} to satisfy.
Thus,  the atoms may change non-monotonically in $q$.
This is evident in Figure \ref{ExampleCommunity2}, which pictures the atoms for other $q$s for the example from Figure \ref{ExampleCommunity}.

This non-nested structure as $q$ varies makes it a challenge to provide simple results on atomic structures, nonetheless, we are able to provide full characterizations in some settings (such as large block models) and to provide algorithms and results that use the atomic structure.

\subsection{Communities Corresponding to Robust Conventions}

Given that conventions and their induced atomic/community structure vary with the threshold $q$, it is useful to define
conventions that are `robust' in the sense that they remain conventions for some range
of $q$s.
There are (at least) four reasons for doing so:
\begin{itemize}
\item One may wish to identify robust communities that remain together for a variety of behaviors.
\item
Individuals may be heterogeneous in their preferences and so behave according to a range of $q$s.
\item The network that is observed may have measurement error in it, so that there may be missing links and/or nodes (or contain extras), and so one
would like to have a convention that is robust to changes in the fractions of neighbors that are undertaking a given action.
\item A network may evolve over time, and so the current network might only be an approximation of what might be in place at some other time.
\end{itemize}
Considering conventions that remain conventions for a range of $q$'s is one way to address these issues.

A {\sl robust convention} relative to some set $Q\subset [0,1]$ is a set $S\subset N$ of agents who form a convention for all $q\in Q$.

As an illustration, the conventions in Figure \ref{ExampleConvention} are both robust conventions for $(1/3, 1/2]$, but not for any additional $q$'s.
Given a network $g$, let $\sigma(Q,g)$ denote the $\sigma$-algebra generated by all robust conventions relative to $Q$,
and $A(Q,g)$ the corresponding atoms (elements of the partition that corresponds to $\sigma(Q,g)$).
Robustness  matters, as it is more stringent to require that a convention hold for a range of $q$'s rather than just a single $q$.
That leads to a coarser convention structure, and correspondingly coarser atoms.

\begin{figure}[h!]
\centering
\subfloat[$q=1/3$]{
\includegraphics[width=0.35\textwidth]{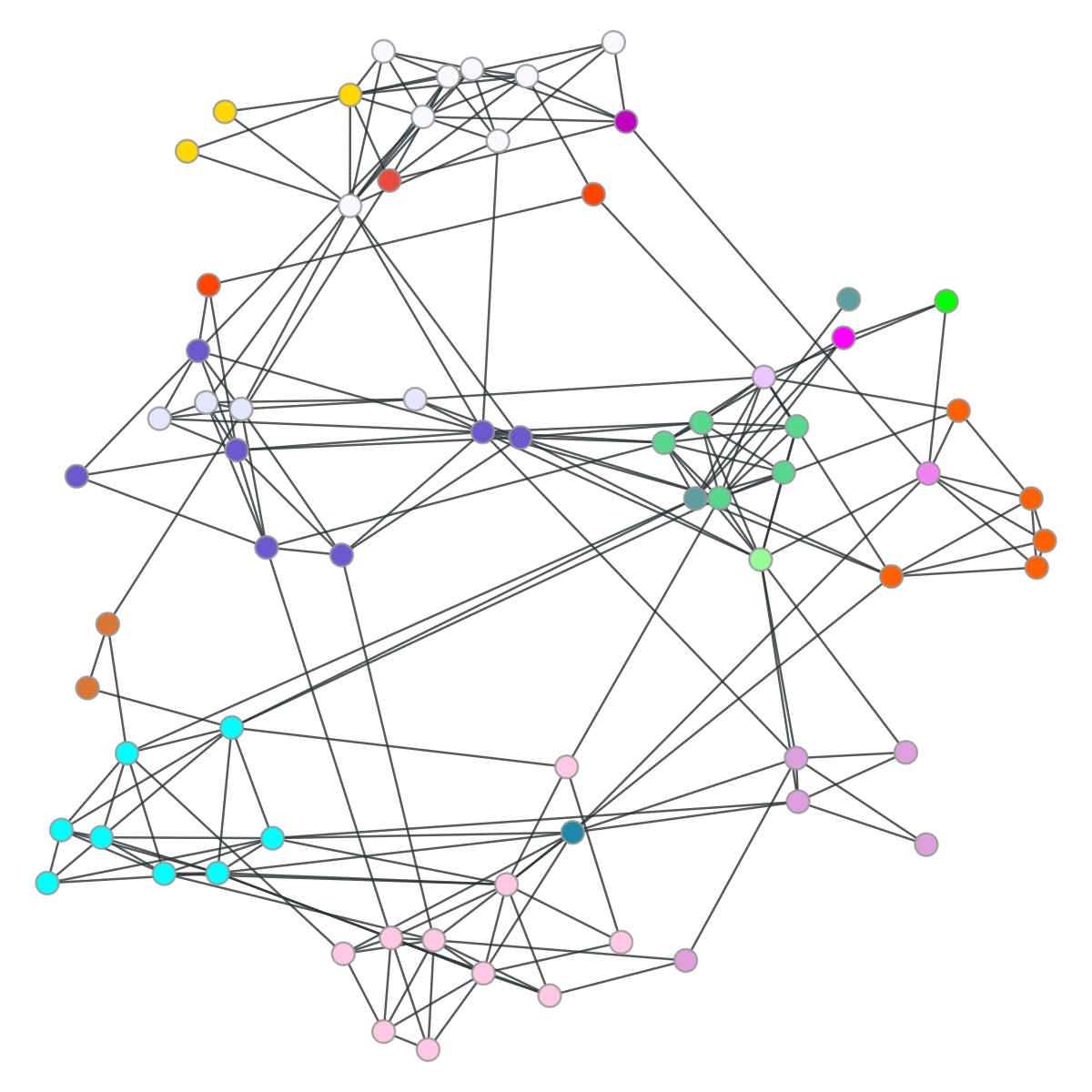}
}
\includegraphics[width=0.02\textwidth]{blank.jpg}
\subfloat[ $Q=\lbrack 1/3-\varepsilon, 1/3+\varepsilon \rbrack $  ]{ 
\includegraphics[width=0.35\textwidth]{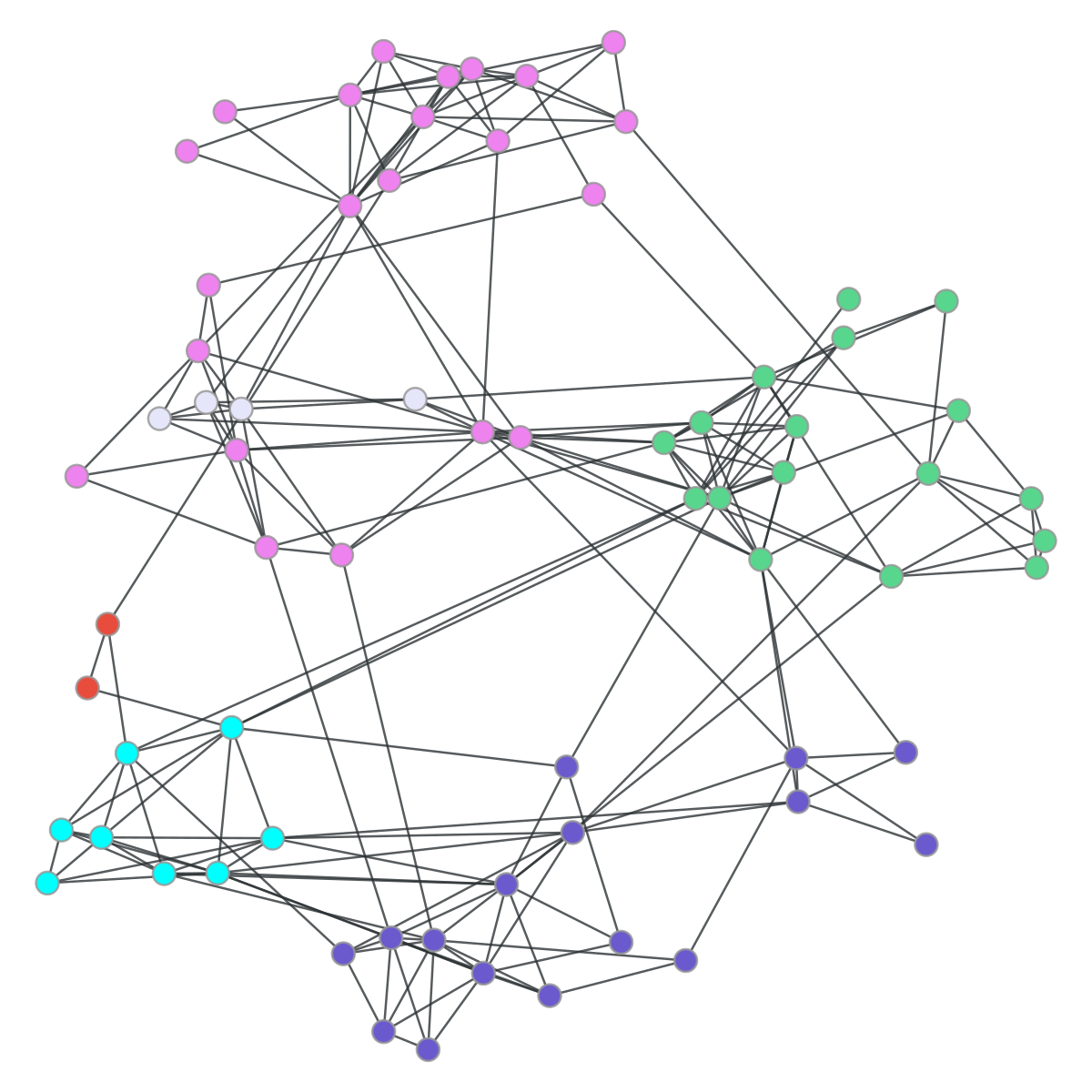}
}
\caption{\label{hs00}
{\bf} Even a small amount of `robustness' coarsens the atomic partition in a high school from the Add Health data set.  The left panel is the atomic structure for $q=1/3$ and the right panel is
the robust atomic structure of $Q=[1/3-\varepsilon, 1/3+\varepsilon]$ (for any $0<\varepsilon<\frac{1}{n}$).
}
\end{figure}

The following example of a social network in an American high school from
The National Longitudinal Study of Adolescent to Adult Health (Add Health) shows that even
a tiny amount of robustness can have a large impact on the atomic structure when dealing with rational $q$'s, as illustrated in Figure \ref{hs00}.

Note  that only the extreme points of $Q$ are needed to check whether some convention is robust with respect to $Q$.
This follows since cohesiveness is most demanding at the threshold of $\sup(Q)$, while closure is most demanding at the threshold of $\inf(Q)$,
and so requiring something be a convention for all of $Q$ then just requires examining the corresponding extremities.

This also means that
if $S$ is a robust convention relative to $Q$, then it is a convention for all $q \in (\inf(Q),\sup(Q)]$.\footnote{If $\inf(Q)\notin Q$  and $\inf(Q)$ is rational, then $S$ may fail to be a convention
at exactly $\inf(Q)$.}
Thus, without loss of generality, we can restrict attention to $Q$s that are intervals.

This implies, for instance, that $S$ remains a convention if people have heterogeneous $q_i$s lying within $Q$.
Note also that if $S$ is a robust convention relative to $Q$ and $Q'\subset Q$, then $S$ is a robust convention for $Q'$.
Thus, if $Q'\subset Q\subset [0,1]$, then $A(Q,g)$ is coarser than $A(Q',g)$ for any $g$.

\section{Using the Atomic Structure to Discover the Blocks in a Stochastic Block Network}
\label{recover}

We now show that the atomic structure of a network can be used to uncover patterns of homophily in a network.
In particular, we consider a network that was generated by some unobserved stochastic block model.
These models are standard ones for modeling and analyzing homophily and communities (e.g., see \cite{golubj2012,lee2019review}).

People have many different characteristics---observed and unobserved---that could potentially define the blocks.
Importantly, people could form ties based on traits that are not observed by the researcher.
As we show here, one can still discover and identify all of the blocks by finding the atoms as $q$ is varied.

This has many applications.  For instance, consider a policy-maker who is going to introduce a new program in which there are peer effects in participation.
By examining the atomic structure of the network as $q$ is varied, the policy-maker can discover all of the blocks.
Moreover, as we show below, the atoms will always be blocks or combinations of blocks.
Then even if the policy-maker does not know
the $q$ associated with participation in the new program, the policy maker knows the blocks which tell them the potential atoms that could result, and thus the potential conventions.
The discovered blocks can then be used to target policies that encourage participation (more on this appears in the seeding discussion below).

\subsection{Stochastic Block Models}\label{blocks}

A stochastic block model is a random graph model in which nodes are separated into different `blocks' and the frequency of links depends on (and only on) which block each of the nodes belongs to \citep{holland1983stochastic,lee2019review}.
So for instance, if there are two characteristics of individuals that turn out to matter in tie formation: ethnicity and gender,  then there would be a probability that any two given female Asians are friends with each other, a (possibly) different probability that any two given male Asians are friends with each other, and a possibly different probability that any given female Asian is friends with any given male Hispanic, and so forth.
Some characteristics that matter are often unobserved
(e.g., \cite{jacksonnsy2018} find homophily on personality traits).
These have become one of the most widely used models in applied work.

To establish results about the atomic structure of a random network generated by a stochastic block model,
we consider growing sequences of stochastic block models (as is standard in random graph theory).
The large numbers of nodes allows us to deduce properties of a typical realized network.  With small
numbers of nodes, there can be nontrivial probabilities of any network arising by chance.  However, as the number of nodes grows,
the probabilities of the graph having certain properties, like the blocks being identifiable and having some particular relationship to the atoms, also grows and goes to one.

Let $n$ index a sequence of random graph models, tracking the number of nodes in the society,
and let $g^n$ denote a random network generated on these $n$ nodes.

The society is partitioned into different blocks of people or nodes, with the partition denoted $B(n)$,
with generic blocks of nodes $b\in B(n),b'\in B(n)$, and associated cardinalities $b(n)$.
The probability that any pair of nodes from blocks $b\in B(n)$ and $b'\in B(n)$ are linked is denoted $p_{bb'}(n)$.
Given the undirected nature of links, it follows that $p_{bb'}(n)=p_{b'b}(n)$.
Links are independent across all pairs of nodes.

Let $d_{bb}(n)=p_{bb}(n)(b(n)-1)$ and
$d_{bb'}(n)=p_{bb'}b'(n)$ denote the expected number of links of a node in block $b\in B(n)$ has to nodes in blocks $b$ and $b'$, respectively.
Let
$d_b(n)=\sum_{b'\in B(n)} d_{bb'}(n)$ be the overall expected degree of a node in block $b\in B(n)$.

We assume that
there exists $f(n)\rightarrow \infty$ such that  $d_{bb'}(n)> f(n)\log(n)$ for all $b,b'$.
This condition
is the familiar one from random graph theory: $\log(n)$ is the threshold that ensures that the network is path-connected.
The required rate of growth in degree is slow (only required to be larger than $log(n)$).

\subsection{The Atomic Structure of Stochastic Block Models}

We say that a stochastic block model is
{\sl weakly homophilous}
if there exists $\varepsilon>0$ such that for large enough $n$
$$d_{bb}(n)/d_b(n)> d_{b'b}(n)/d_{b'}(n)+ \varepsilon {\rm  \ \  for \ \ every \ \ }b, b'\neq b.$$

Weakly homophilous stochastic block models are such that nodes from a block $b$ expect to have a relatively higher fraction
of their friends within their own block $b$ compared to the fraction
of friends that nodes from other blocks expect to have inside block $b$.  The ``weak'' part is that this does \emph{not} require that people
 have higher linking probabilities to their own types than their representation in the population;  it only requires that they are
\emph{relatively} more biased towards own type than other people are to that type.
For example, consider a society with two blocks, blues and greens, each being fifty percent of the population.
It could be that blues expect to form sixty percent of their friendships with greens while greens expect to form seventy percent of their friendships with greens.   Both types are biased towards greens, but greens are relatively more biased towards greens and yet the condition is still satisfied.
Thus, this is weaker than usual homophily; but is precisely the condition that
allows blues and greens to be separated by conventions and thus be in different atoms.  For instance, for a threshold of $q=.35$ blues adopting
and greens not adopting is a convention (in a large enough network), and correspondingly for a threshold of $q=.65$ greens adopting and blues not adopting is a convention.

We say that a sequence of stochastic block models is {\sl convergent} if (i) $|B(n)|$ is constant for all large enough $n$,
indexed as $b_1, \ldots,  b_{|B(n)|}$, and (ii)
the vector of corresponding block connectedness measures, $d_{b_k b_{k'}}(n)/d_{b_k}(n)$, converge for all ${b_k b_{k'}}\in B(n)$.
The theorem extends to allow $|B(n)|$ to grow with $n$, but we work with a finite $|B(n)|$ to keep the notation uncluttered.\footnote{\label{fK} For instance, apply the theorem for any given $|B(n)|$, and it requires a large enough $n$.   For a sequence of $|B(n)|$'s it requires a growing
sequence of $n$'s, and so that determines a bound on how quickly $|B(n)|$ can grow with $n$.}

The following theorem shows that with a probability tending to one, the atoms are supersets of the blocks.
Moreover, if the block model is weakly homophilous and convergent, then for any given block there exists some $q$ for which it is a convention itself
and hence an atom.
Thus, even if a researcher does not know what the blocks are or what characteristics generate blocks, they can still recover all the blocks
by identifying all of the atoms as $q$ is varied, and the resulting recovery of the blocks is semi-parametric in the sense the number of blocks and details
of their statistical relationships is not necessary to recover them.

\begin{theorem}
\label{random}
Consider a growing sequence of stochastic block networks, with corresponding $B(n), \{d_{bb'}(n)\}_{b,b'\in B(n)}$,
and any compact set of thresholds $Q \subset [0,1]$.  Then
\begin{itemize}
\item If $Q$ has a nonempty interior (i.e., $Q$ is a robust convention),
then the atoms, $A(Q,g^n)$, are a coarsening (supersets) of the blocks $B(n)$ with a probability going to 1.
\item If $Q$ is extreme in that there exists $\varepsilon>0$ such that either $\max(Q) > \max [\max_{b\in B(n)} d_{bb}(n)/d_b(n),  1- \min_{b,b'\in B(n)} d_{bb'}(n)/d_b(n) ]+ \varepsilon$ or
$\min(Q)< \min [1-\max_{b\in B(n)} d_{bb}(n)/d_b(n),  \min_{b,b'\in B(n)} d_{bb'}(n)/d_b(n) ]-\varepsilon$ for all large enough $n$, then the atomic structure is degenerate (the trivial partition of all $N$) with a probability going to 1.
\item If the sequence of block models is convergent and weakly homophilous,  then there exists $\varepsilon>0$ such that
any given block $b\in B(n)$
is an atom (and a convention) for  $Q=[d_{bb}(n)/d_b(n) - \varepsilon,  d_{bb}(n)/d_b(n)+\varepsilon ]$ with probability going to 1.
\end{itemize}
\end{theorem}

We prove Theorem \ref{random} using two main lemmas.  First, we use Chernoff bounds to show that with a growing probability, all nodes have relative degrees across the different blocks that do not differ too much from the expected values.
This implies that the blocks are distinguished from each other, which is important in deriving the final conclusion of the theorem.
Second, we extend a theorem on the modularity of random networks to show that no block can fracture into different atoms - so there are no discernable splits within a block.\footnote{A corollary of Theorem \ref{random} is that, in the case of a degenerate block model with a single block, for a sufficiently large network the only robust atom is the entire network.  This aligns with the common intuition in the community detection literature that an Erdos-Renyi graph should serve as a null case of no communities for a sensible community detection approach.
}
Specifically, we extend a powerful theorem of
\cite{mcdiarmid2018ER_modularity} which shows that the modularity of a sequence of Erdos-Renyi random networks tends to 0 in probability.  This implies that no robust convention can splinter a large enough
Erdos-Renyi random network.  We show that the nontrivial modularity that results in a stochastic block model must occur along
block boundaries and cannot fracture any of the blocks.\footnote{This is not a corollary of their result because modularity depends on links formed outside of blocks and not just inside a block.  For instance two nodes in the same block might have different patterns of links across other blocks.  We show that the probability of such occurrences tends to 0.}

Theorem \ref{random} shows several things.   First, it shows that atoms are supersets of the blocks (with probability tending to 1 in the size of the network).   Given the combinatorically large numbers of possible partitionings of an given block, the fact that none happen by a chance  to have a sufficient level of modularity that it allows it to be split by some convention is not obvious.  In fact, in any early version of this paper we conjectured that it was possible that some atom might cut across a block (in large networks with nontrivial probability).  The power of the 
application of the \cite{mcdiarmid2018ER_modularity} theorem is that we can show that all blocks end up being subsets of atoms (with probability tending to 1 in the size of the network).   

The second part of the theorem deals with high or low enough thresholds, and shows that in those cases the atoms degenerate into the whole set of nodes.   Thus, the interesting distinction in communities happens for intermediate thresholds.   This is intuitive as with low enough or high enough thresholds the only equilibria become the entire population either adopting or not adopting a behavior.   

The third part of 
Theorem \ref{random} implies that in a weakly homophilous network as described by a block model, the behavioral atoms can be used to completely uncover
the blocks corresponding to that homophily.
In particular,
Theorem \ref{random} shows that behavior-based atoms can be used to recover (all) the blocks in a stochastic block model.
This is useful in practice, since as mentioned  above, the researcher will not a priori know which characteristics (observed or not) are actually significant in defining the blocks. 
Moreover, the researcher may not know which $Q$ they are interested in, and thus knowing all the possible atoms can be of substantial interest.  

We remark that Theorem \ref{random} implies that robust conventions are immune to rewirings of links (or mismeasurement)---up to some limit.
There are two aspects to this.  First, the theorem applies to any realized network (with high probability) and so does not depend on specific link realizations.
But second, with weak homophily and convergence, there are nontrivial differences in the connectivity within and across atoms.  This means that nontrivial fractions (and
a large absolute number) of links can be rewired before one disturbs an atomic structure for any threshold that does not coincide with a limit connectivity rate.

\subsection{A Comparison of Behavioral Communities and Modularity-Based Community Detection Algorithms}

Theorem \ref{random} provides
a distinction between our approach and other community detection algorithms.
Whether a particular block is identified as an atom depends on the threshold(s) $Q$ in question.
If there is heterogeneity in the relative cohesiveness of different blocks, then different blocks can be identified as atoms for different thresholds,
with some blending together for some and not other thresholds.
It can be that there is no threshold that identifies all the blocks simultaneously.
This is particularly important, since it means that the blocks themselves might not be the right atomic partition for any given $Q$.

Many community detection algorithms simply return one partition, and have no analog to a behavioral threshold.
In particular, it can be that the partition generated by, for instance the Louvain or other modularity based methods, 
never correspond to the partition corresponding to any $Q$.   So it is not that they are identifying some robust partition, but instead that they are identifying something that never corresponds to something behaviorally generated, robust or not.   For instance, if different blocks have different relative densities of internal vs external linking probabilities,  it will never be the case that the partition of all blocks corresponds to a robust partition.   If one includes all the $q$'s that isolate different blocks in one $Q$ then the robustness ends up with a degenerate partition being the unique robust equilibrium.        
Other detection algorithms deliver a nested hierarchy of partitions, which the researcher can then choose between by selecting some way to value the fineness of the partition.   However, those hierarchies can subdivide the blocks as they become fine, and yet not identify the blocks or actual atoms if they are coarse.  Again, they can never correspond to the atoms, except at the extremes where the partition is the degenerate set of all nodes.
In our approach, the partition varies in an interpretable way with the behavioral threshold.

\begin{figure}[h!]
\centering
\subfloat[The communities found via modularity minimization (colored and outlined). ]{
\includegraphics[width=0.35\textwidth]{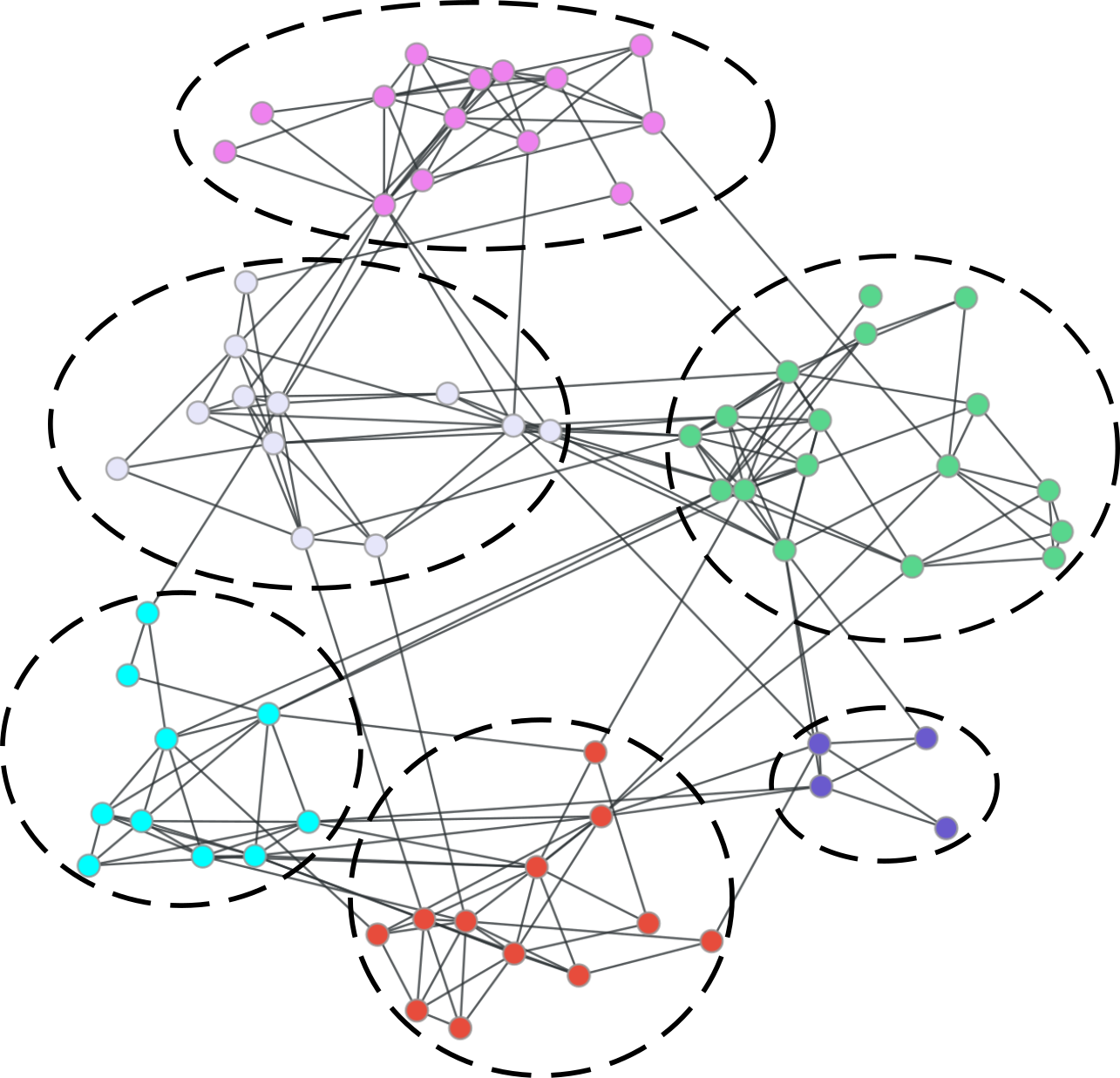}
}
\includegraphics[width=0.02\textwidth]{blank.jpg}
\subfloat[The $Q=1/2 \pm \varepsilon$ robust behavioral atoms (color) compared with the modular communities (outline). 8 out of 69 = 11.6\% of the nodes have to be reclassified to go from one partition to the other.]{
\includegraphics[width=0.35\textwidth]{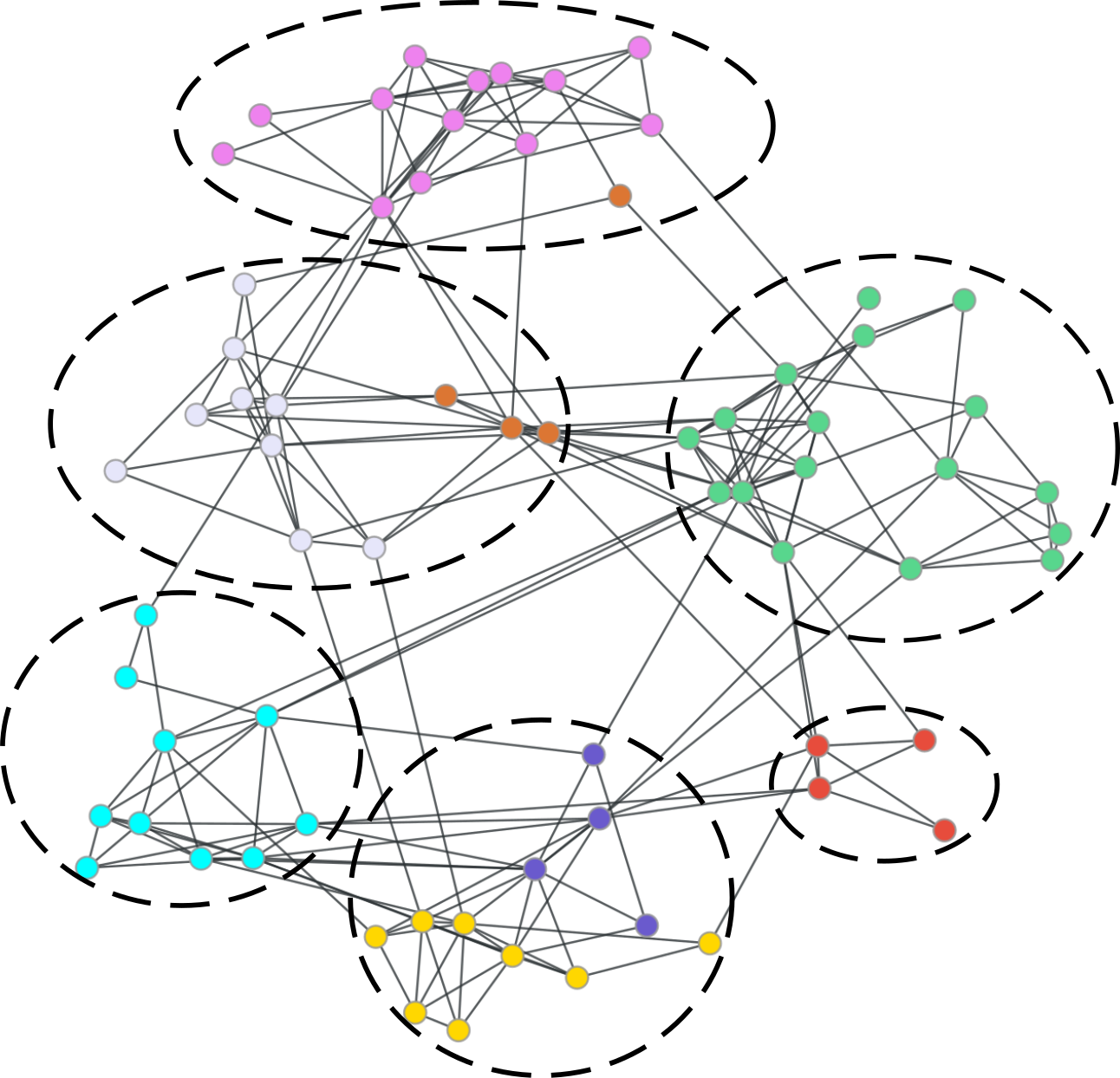}
}

\includegraphics[width=0.02\textwidth]{blank.jpg}
\subfloat[The $Q=1/3\pm \varepsilon$ behavioral atoms (color) compared with the modular communities (outline). 14 out of 69 = 20.3\% of the nodes have to be reclassified to go from one partition to the other.]{
\includegraphics[width=0.35\textwidth]{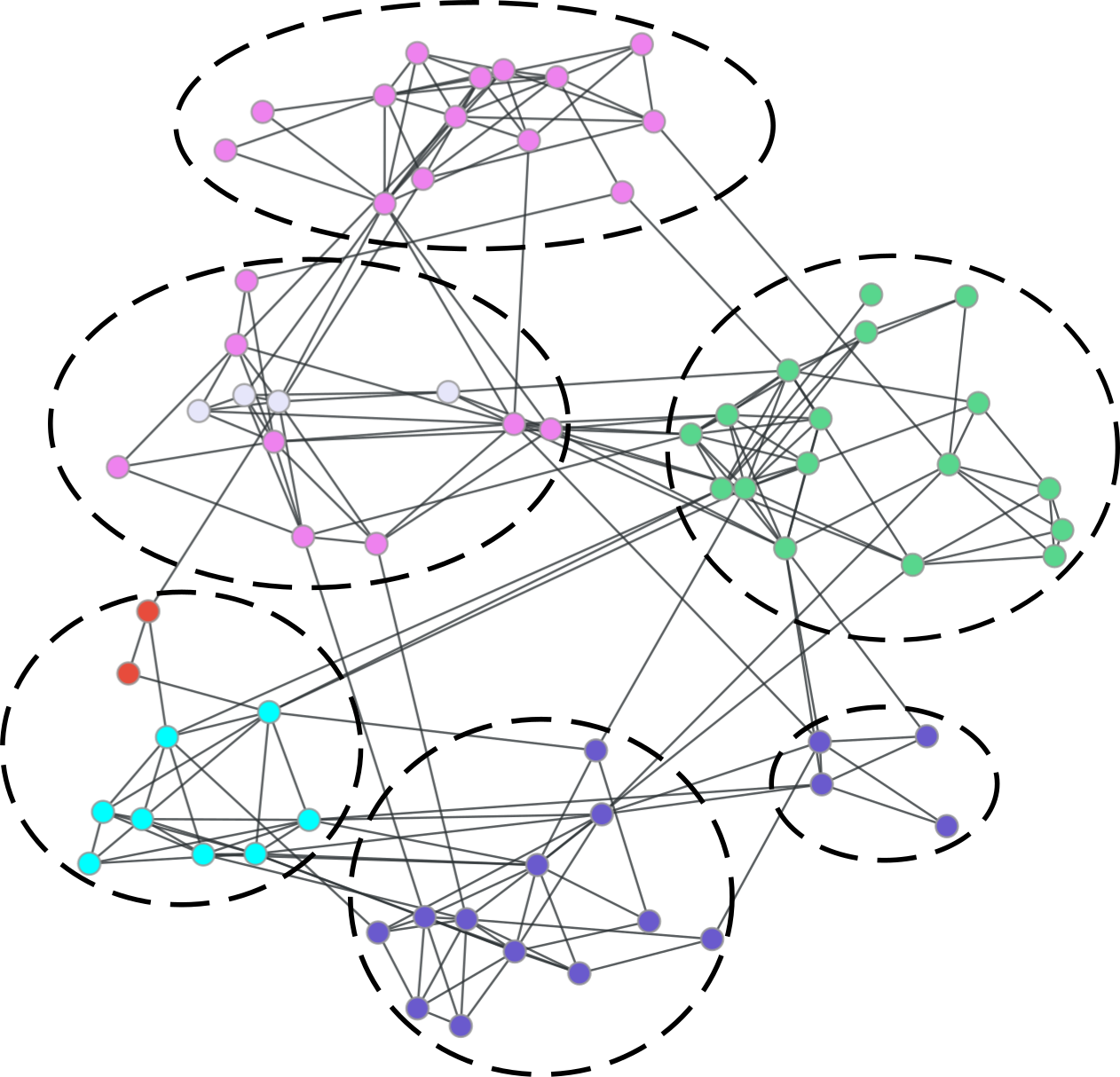}
}
\includegraphics[width=0.02\textwidth]{blank.jpg}
\subfloat[The $Q=1/4 \pm \varepsilon$ robust behavioral atoms (color) compared with the modular communities (outline). 52 out of 69 = 71.1\% of the nodes have to be reclassified to go from one partition to the other.]{
\includegraphics[width=0.35\textwidth]{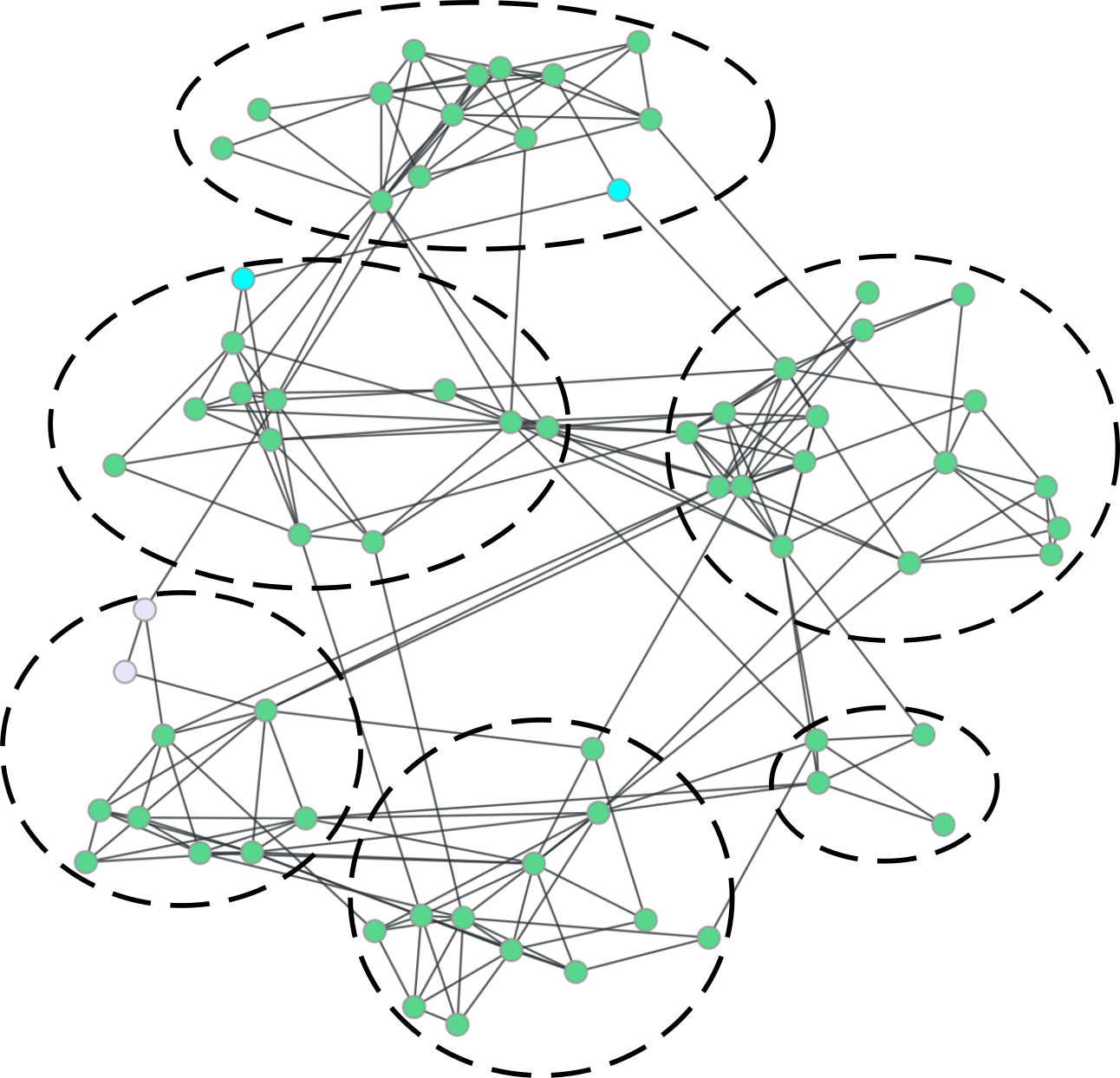}
}
\caption{\label{modular}{}
{\bf :} Comparing the behavioral atoms from various thresholds to the communities found via the Louvain method of modularity minimization \citep{blondeletal2008}.
Dotted outlines and color in subfigure (a) are the Louvain method communities, while the colors in the other panels represent the behavioral atoms for various choices of thresholds.
Nodes with no connections are omitted from the figures.
}
\end{figure}

To illustrate this contrast,  we compare the
 communities that are found via a most popular modular method---the Louvain method---to our atoms.  
That method is based on a modularity measure, and looks to identify sets of nodes that have a higher likelihood of links between them than to outside nodes \citep*{blondeletal2008}.  The issue is that different $q$s can identify different sets of nodes, and if one then tries to include the full range of all those $q$s, the robust atoms become degenerate. 
In particular, we show how for some $q$'s our atoms are strictly finer than those modules, while
for others ours are a coarsening, and for other $q$s they are non-nested.\footnote{Again, this captures the distinction that
 behavioral communities change with the threshold and $Q$, while modular methods are just looking for higher internal than external density, rather than requiring specific conditions on those densities.  Hierarchical community detection methods can be made finer based on some cutoff parameter, but that cutoff is the just a number steps in the dividing algorithm rather than a parameter tied to behavior.}
This is pictured in
Figure \ref{modular}, which shows that behavioral atoms differ from the communities defined by standard community detection algorithms on a prominent data set (Add Health).
We use the high school social network from Figure \ref{hs00}.

Panel (a) of Figure \ref{modular}
 depicts the communities identified by a modularity-minimizing algorithm (specifically, the Louvain method for community detection, \cite{blondeletal2008}) via the dotted outlines and color in subfigure (a), while the colors in the other panels represent the behavioral atoms.

First, note that in comparing panels (a) and (b),  we see that for the $Q=1/2 \pm \varepsilon$ robust neighborhood of $q=1/2$,
the atoms are finer that the communities obtained from the Louvain method.
The additional fineness is capturing real behavioral divides and thus are not defects in terms of defining atoms but real information that behavioral atoms convey that other methods do not.  In particular, it is possible to have conventions that would cut across the Louvain communities.

Next, note that in comparing panels (a) and (c), we see that for the $Q=1/3 \pm \varepsilon$ robust neighborhood of $q=1/3$,
the atoms sometimes subdivide the communities from the Louvain method, and sometimes cut across them, and sometimes doing both at the same time;  e.g., the pink and white nodes in the two communities in the upper left hand corner.

For behaviors with low thresholds, contagion occurs across blocks.   For this network, a behavior that has a $q=1/4$ threshold cannot be contained by any blocks, and the atomic structure reflects that as we see in panel (d).

An interesting feature of Figure \ref{modular} is the orange atom in panel (b) involves nodes that are not
directly connected to each other. Here, the orange nodes nonetheless form an atom because they have similar shares of their neighbors in the white, pink, and green atoms, so that any conventions involving these larger atoms includes either all or none of the orange nodes.
The Louvain method ends up splitting them because it is based only on modularity and not how nodes would behave, and by contrast our method identifies them as an atom that will always act the same in any convention.
Again, we emphasize an important feature of our approach:  the community structure varies with the behavior,
and we can see different atoms and blocks emerge at different thresholds,  some of which are not discovered at all by standard methods.
Modular methods can thus be poor predictors of the possible norms of behavior.

\section{Using Knowledge of the Atoms to Seed a Diffusion Process}
\label{seeding}

Next, we show that knowledge of the atoms helps in optimally seeding a diffusion.

\subsection{Seeds and Dynamics}
\label{dynamics}

Given a social network $g$, a behavioral threshold $q$
and an budget of $s$ nodes to be chosen as ``seeds,'' the objective is to find a set $S$ of no more than $s$ seeds that leads to the largest number of nodes in $g$ eventually adopting the behavior/technology.
The set of nodes eventually adopting the behavior from $s$ seeds is found by iteratively updating behaviors, formally defined as follows:
\begin{itemize}
\item[1.]  The seeds in $S$ are given the new technology/behavior and induced to (tentatively) adopt it.
\item[2.] Iteratively, (for up to $n-|S|$ iterations), nodes who have not yet adopted the new technology/behavior adopt if at least a fraction of $q$ 
of their neighbors in $g$ have adopted.  Once an iteration happens in which there is no change new adoption, proceed to step 3.
\item[3.]  Seeds who have a fraction of less than $q$ 
of their neighbors who have adopted the new technology/behavior by the end point of step 2, drop the new technology/behavior.
\item[4.]  If any seeds drop the technology/behavior, then the technology/behavior is also dropped by any of their neighbors (including other seeds) who
have adopted the new technology/behavior but now have less than a fraction of $q$ 
of their friends still adopting.  This dropping process iterates until there are no further droppings.
\end{itemize}

Given an initial seed set  $S$, we call the resulting convention the \textit{$q$-convention generated by $S$}.
The behavior grows outwards from the seeds via iterative adoptions.  However, in order for what is eventually reached to be an equilibrium convention, the initial seeds
must be willing to maintain the behavior given the eventual set of adopters.  If some of the original seeds wish to drop the behavior given the level of adoption that was reached during the growth phase, then we iteratively follow droppings of the behavior until an equilibrium is reached.
Therefore, this process presumes that the seeds stick with the technology or behavior in the long run only if it eventually reaches enough of their neighbors to make it worthwhile for them to maintain the behavior.
For example, if too few of a seed's friends ever adopt the technology/behavior to have the seed prefer to maintain the behavior, then that seed would eventually drop the technology/behavior themselves.
This can unravel around some of the seeds.
The complementarities in behaviors mean that this process is well-defined, ends in a finite number of steps, and that the ending set of adopters is a (possibly empty) $q$-convention.\footnote{If seeds never drop the behavior,
which might apply in some applications where behavior is nonreversable (e.g., enlisting in the military) then the only change in Theorem \ref{optseed} is that the random seeding results in only the seeds adopting the behavior with a probability converging to 1 as $n$ grows.}

Note that if all seeds are still adopting at the ending point, then the convention that is found via this simple process of iterative adoption (corresponding to iterative best replies in an associated coordination game) is \emph{the} minimum
convention that contains the initial seeds:  any convention that contains the seeds must also contain this convention.\footnote{ By contrast, somewhat surprising things can happen when
some seeds are not willing to still adopt the behavior at the end of the process.  For instance, it is easy to come up with examples in which eventually all of the seeds drop the behavior, but that some other nodes end up maintaining the behavior.  For example, it can be that the adoption by the seeds incentivizes some clique of nodes to all adopt the behavior, and that clique then maintains the behavior, but the seeds
have too many connections outside of the clique and end up dropping the behavior as they are not
relatively connected enough to the clique to pass their threshold.}

\subsection{The Seeding Algorithm}\label{seedalgo}

We provide a new seeding algorithm that seeds based on the atomic structure.

The algorithm for identifying the atoms is provided in detail in Section \ref{algorithm} of the Appendix,
and a brief description is as follows.  Finding the atoms is an NP-hard problem (see Section \ref{Compute}), and so we approach it from two different angles.
First, we compute all the conventions that can be generated
by some number $s$ of nodes for which $\binom{n}{s}$ computations are feasible.  That is, we begin with some number of nodes adopting the behavior of up to $s$ and then see what convention emerges from iterated best responses of neighbors of current adopters of the behavior (with possible reversions).  Together, all such conventions generate a first pass at the atoms, potentially a coarser partition.   Next, we examine pairs of nodes that are in the same atom based on this set of atoms.   We then use an Integer Linear Program to see if we can find a convention that splits those two nodes apart.   If we can find such a partition, then we add it to the list of conventions and update the atoms.  Given the success of Integer Linear Programs on other NP-hard partitioning problems (up to tens of thousands of inputs), this approach seems to work very well.
This may uncover a partition that is coarser than the true partition for some large networks.

Note that if the set of seeds in the seeding problem is limited to some $s$ for which $\binom{n}{s}$ computations are feasible, then this algorithm is fully sufficient, since those are the conventions which then can be reached in any seeding below.

Let us now describe the seeding portion of algorithm.

\begin{enumerate}
\raggedright
\item{} Find the atomic partition $A(q,g)$ as described above and label the atoms $A_1, A_2,...A_m$.
\item{} For each atom, $A_i$, identify the minimum number of seeds up to $s$ within $A_i$ required to result in all nodes in $A_i$ (and possibly others) adopting the behavior under the process defined above. Call this the cost of the atom,
$C_i$.\footnote{The computations required to calculate the seeding costs potentially grow at rate $O(n^s)$.
In particular, process of determining which convention is generated by a given set of any number of seeds is linear in n, and thus relatively fast, regardless of the density of the network or its structure.   Thus, the complexity comes down to searching over sets of no more than $s$ seeds, regardless of the network.   When $k$ is bounded, then $n$ choose $s$ is polynomial.  Moreover, in many applications the atoms are fairly small and then if one additionally limits the search for seeds to be within the atom in question, then that further reduces the number of computations to be of order no more than $m^s$, where $m$ is the size of the atom.  In Section \ref{algorithm} we discuss a condition satisfied by many real-world networks (e.g., those in our sample) that ensures the atoms can be found efficiently and gives upper bounds on the complexity of finding an atom's seeding cost even when the number of seeds is increasing in $n$.}
If no selection of no more than $s$ seeds placed within $A_i$ can result in all nodes in the atom adopting the behavior,
then set $C_i=\infty$.
  \item{} For each atom $A_i$, compute the largest $q$-convention that is generated by some $C_i$ nodes in $A_i$, and let
  the benefit of atom $i$, $B_i$, be the size of this convention.
\item{} Greedily seed the atoms in decreasing order of the size-to-cost ratio $B_i/C_i$ until all $k$ seeds are used or no remaining atoms can be seeded with the remaining number of seeds.
\item{} If there are seeds left over, place the seeds uniformly at random among nodes that are not in the convention generated by the seeds already selected.
\end{enumerate}

As mentioned in the introduction, the computer science literature has looked at optimal seeding problems
that superficially might seem similar to ours, originating with \cite{kempekt2003,kempekt2005} and given a
general treatment in \cite{mosroch2009}.  Like ours, these papers ask which nodes should be seeded to
maximize the spread of a contagion.  However the types of contagion that are considered are fundamentally different.
Their papers assume that the function mapping a node's neighbors' behaviors into that node's behavior
(what they call the activation functions) are submodular---so that the more of a node's neighbors are
already adopting, the smaller the increase in the likelihood of adoption from having an additional neighbor
adopt. That assumption clearly fails in our setting.
While this local submodularity assumption may fit certain cases such as the spread of disease, it does not hold in contexts where agents are deliberately coordinating with their neighbors, since coordination induces a natural complementarity in the effects of neighbors' adoption decisions.  For example, if one only wishes to adopt a technology if a given fraction of one's friends do and one has a nontrivial number of friends, then it is some friend well beyond the first who has the most influence, even if noised up probabilistically,
clearly violating submodularity.\footnote{Some recent work (e.g. \cite{Shoenbeck_2019}, which appeared after the first version of this paper) has considered the absolute threshold setting for the special case of stochastic hierarchical block models where the realized network is unknown. Our algorithm below yields the same seeding recommendations as theirs in that setting. }

This difference is important
since the algorithms from the previous literature are adaptations of well-known algorithms for
submodular optimization, and fail to work in our setting.
  This is the key difference from submodularity, where the largest marginal gain is always
from the incremental seed, and so one just searches for which seed has the largest marginal
impact at the moment.  Instead, in our setting, a group of seeds may have no marginal impact until all of
them are seeded, and then they have a large impact.  In particular, placing them within an atom,
and often in close proximity, can have much more impact then spreading them out.
This contrast shows how useful the atoms can be.

The loss of local submodularity does not make the optimal seeding problem less complex:
just as in the submodular case (e.g., \cite{kempekt2003}), our optimal seeding problem is NP-hard.
Moreover, in our setting even the constant-factor polynomial-time approximations that are
the focus of \cite{kempekt2003} do not apply.
Thus, although we cannot offer an efficient, fully optimal solution to the seeding problem,
we can show that the behavioral atoms contain vital information and inform the above intuitive heuristic for the seeding
problem, and that this approach offers significant improvements over random seeding
as well as seedings that are based on standard community-detection algorithms.

The following theorem shows that in stochastic block models, if there are enough seeds to
optimally seed behavior in part but not all of the network, then the atom-based algorithm defined above leads to spread
of the behavior across some number of
blocks, while random seeding will not lead to the adoption of the
behavior by any nodes.

Let us say that a sequence of convergent and weakly homophilistic
stochastic block models $B(n)$ has blocks of similar sizes if
$|B(n)| b_k(n)/n\rightarrow 1$ for each $k\in \{1, \ldots, |B(n)|\}$,
and there exists $0< \overline{q}<1$ such that
$d_{b_k b_{k}}/d_{b_k} >\overline{q}$ for each $k$.
Again, this extends to allow $|B(n)|$ to grown with $n$ (see Footnote \ref{fK}).

\begin{theorem}\label{optseed}
Consider a sequence of convergent and weakly homophilous stochastic block models with similar sized blocks
with associated parameters $|B(n)|,\overline{q}$ and any behavioral threshold $q$ such that $0< q < \overline{q}$.
For any given $\varepsilon>0$, if the number of seeds, $s(n)$ is such that  $ (1+\varepsilon) \frac{qn}{K\overline{q}}  < s(n) < (1-\varepsilon)q n$ for each $n$,
then atom-based seeding results in at least  $\lfloor \frac{ |B(n)|s(n)\overline{q}}{n(1+\varepsilon)q} \rfloor$
blocks of nodes adopting the behavior with a probability converging to 1
as $n$ grows, while  a random seeding results in no nodes adopting the behavior with a probability converging to 1 as $n$ grows.
\end{theorem}

Theorem \ref{optseed} considers a setting with enough seeds to get behavior to spread to at
least one block ($s(n)>(1+\varepsilon) q \frac{n}{K\overline{q}}$),
but few enough seeds so that less than full infection is possible ($s(n)<(1-\varepsilon)q n$).
Here,  the atom-based seeding infects a significantly larger number of nodes compared to the number of seeds used.
In contrast, a random seeding spreads the seeds across atoms, without concentrating enough seeds
in any atom to get behavior to spread at all.  The ratio of the resulting behavior adoption from atom-based over random seeding heads to infinity (involving a division by 0)
with a probability going to 1 in $n$.

More generally, with a random seeding in such a setting, the fraction of all nodes that have to be seeded would need to be at least $q$, as otherwise the fraction of nodes that would have at least $q$ of their neighbors adopting would be 0.\footnote{This can be proven with Chernoff bounds.}
Thus, with  non-trivial thresholds for complex contagion,  seeding must be done judiciously based on network structure in order to be cost-effective.
In fact, although the comparison in the theorem is with random seeding, it extends to any method of seeding that does not place sufficiently many seeds in particular places in the atoms will do as poorly as the random seeding.

Comparing this to the results of \cite{akbarpourms2017}, we see that when contagion is ``complex'',
so that contact with more than one infected node is needed, then the reach attained by random seeding
can converge to $0$ as a fraction of the reach via atom-based seeding.\footnote{An alternative implication of the result is that one would need at least $|B(n)|qd(n)$ random seeds before
one gets to the same infection rate as the optimal seeding, where $|B(n)|qd(n)/s(n)$ can then be bounded away from 1, and so this contrasts with the results of
\cite{akbarpourms2017}.}

While we state the result in the context of a block model, the result extends beyond such settings and the intuition should be clear:
seeding complex behavior requires concentrating enough seeds in neighborhoods of other nodes to exceed the behavioral threshold,
and thus specific patterns of seeding are needed to achieve an optimum and the atoms provide the critical information about the necessary patterns.  Random seedings are much more wasteful and either
fail to achieve contagion, or else require a seeding that saturates the network at a much higher level.  This is quite different from simple contagions, like a flu.

\begin{figure}[H]
\centering
\includegraphics[width=0.65\textwidth]{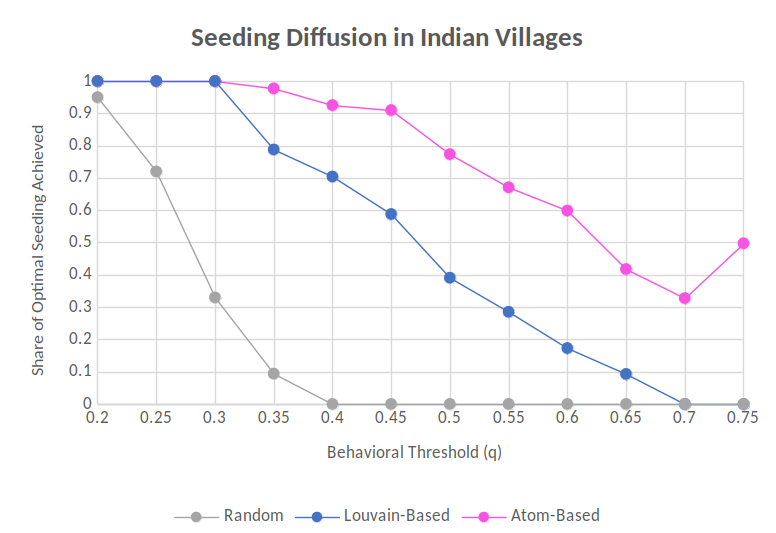}
\caption{\label{heurfig2} Comparing the overall fraction of the network adopting
the behavior under our atom-based vs. random seeding and
Louvain-based seeding as we vary $q$.   We average over 35 Indian villages
each with approximately 200 households (nodes) and using 8 seeds.
The results are listed as a fraction of the fully optimal seeding,
which is computed via an exhaustive (time-expensive) search.  We start at .2 since below this level behavior
spreads easily and widely regardless of the methods.
We stop at $q=0.75$ since beyond this point it becomes impossible to sustain
any adoption in most of the villages even with optimal seeding.}
\end{figure}

To check how different seeding techniques work in empirically observed networks rather than a block model, we compare seeding techniques on a
sample of thirty-five household favor-sharing networks in Indian villages
from the \cite*{banerjeecdj2013} data set. \footnote{To ensure we can compute the fully optimal seeding as a benchmark, and so that we can use the same number of initial seeds for all of the villages, we restrict our testing to the the 35 villages in the dataset with main component size closest to 200. This gives us networks ranging in size from roughly 170 to 220 nodes, with a mean of about 205. }
This gives a fuller impression of the advantages of atom-based seedings, beyond the lower bounds from Theorem \ref{optseed}.
In addition, we not only compare our atom-based seeding
to a random seeding, but also to a seeding technique that is otherwise similar to ours
but instead of using atoms it uses a standard community detection algorithm to define the units that are used in the algorithm.  Specifically, we use the Louvain method of identifying communities, as it is the state-of-the-art modular method.
We show how the comparison varies with $q$ in Figure \ref{heurfig2}.

As we see in Figure \ref{heurfig2},
for very low values of $q$ the three methods are all similar,
as concentrating seeds makes little difference and behavior spreads widely regardless
of the network structure.
As we increase $q$, random seeding quickly fails to concentrate seeds closely enough to attain any adoption.
The Louvain-based method does nearly as well as the atom-based method up to around $q=.3$,
but then
above that level, its communities no longer correspond to how people influence each other in the
network and it performs significantly worse than our atom-based seeding,  getting only about half as much activation by $q=.5$.
Above $q=.65$ the Louvain method converges
to be as bad as random seeding, while the atom-based seeding still yields a nontrivial fraction of the
optimum.%
\footnote{The resurgence of our method at $0.75$ (the `v-shape' coming back up between .7 and .75) arises because at that point even optimal seeding drops dramatically to yielding only a small amount
of participation and our atoms find half of it, and so the gap between atom-based and optimal seeding shrinks.}

This makes clear the
advantage our atom-based approach has over
a standard community-based approach that does not adjust on behavior.

\subsection{Adaptive Seeding and the Atomic Metagraph}

The difference between optimal and our greedy atom-based seeding is that there can be some networks for which seeding several atoms that collectively have many connections to other atoms can induce people to adopt the behavior in those other atoms.   Thus, the fully optimal algorithm would not greedily choose the atoms, but would consider all possible combinations of the atoms to see if some of them would result in additional atoms being induced to adopt the behavior.    This enhanced algorithm is also based on atoms, but would consider benefits of seeding various combinations of them, whereas our simpler algorithm above chooses atoms greedily.
Thus, here we describe a method of tracking the interrelations between atoms and how to use it to improve the algorithm.
The main idea is to generate an ``atomic metagraph'' that records which collections of atoms generate which conventions.

Formally, we define the atomic metagraph $G^{AS}(Q,g)$ of the network $g$ with $Q-atoms$ partition $A(Q,g)=\{ A_1,A_2,...A_K \}$ as a directed graph with the following nodes and edges:

\begin{enumerate}
\item{} The nodes of $G^{AS}(Q,g)$ are all the (nonempty) collections $C\subset A(Q,g)$ of atoms.
\item{} There is a directed edge from $C$ to another  $C'$ in $G^{AS}(Q,g)$ if all the atoms in $C'$ adopt
the behavior in a best response to all of the nodes in $C$ adopting the behavior.
\end{enumerate}

$G^{AS}(Q,g)$  contains self-loops when a collection of atoms $C$ is a convention; hence self-loops track whether a collection of atoms can sustain adoption of the behavior within itself.  More generally, an edge running from $C$ to $C'$ tells us that seeding $C$ also activates all of $C'$.
In Figure \ref{metagraph_fig} we illustrate the atomic metagraph for the example network from Figure \ref{ExampleCommunity} introduction with $Q=[0.4]$ (or any $Q\subset (1/3, 1/2]$:

\begin{figure}[H]
\centering
\includegraphics[width=0.45\textwidth]{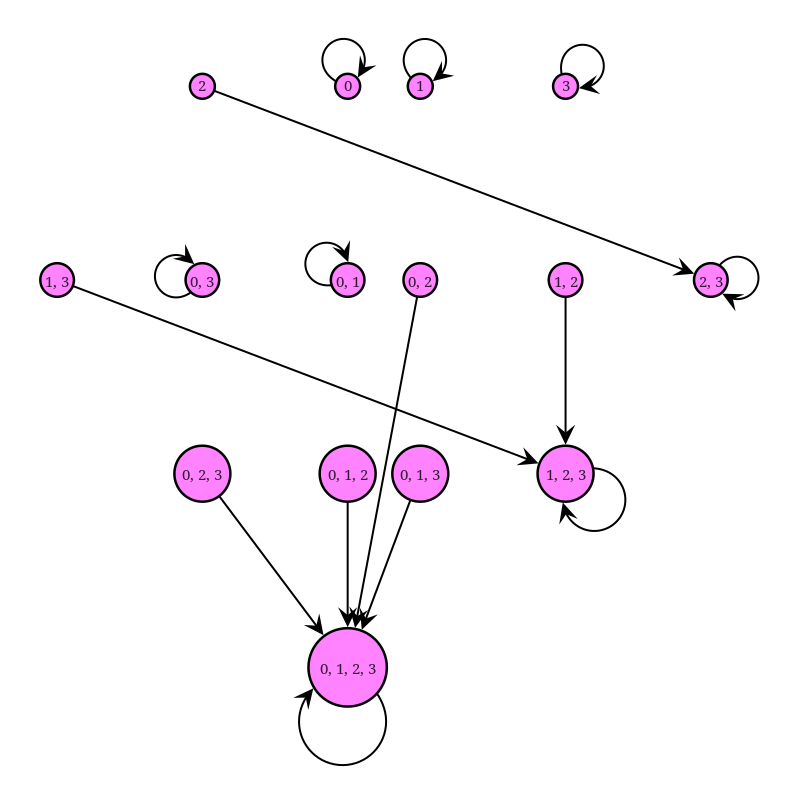}
\includegraphics[width=0.03\textwidth]{blank.jpg}
\includegraphics[width=0.35\textwidth]{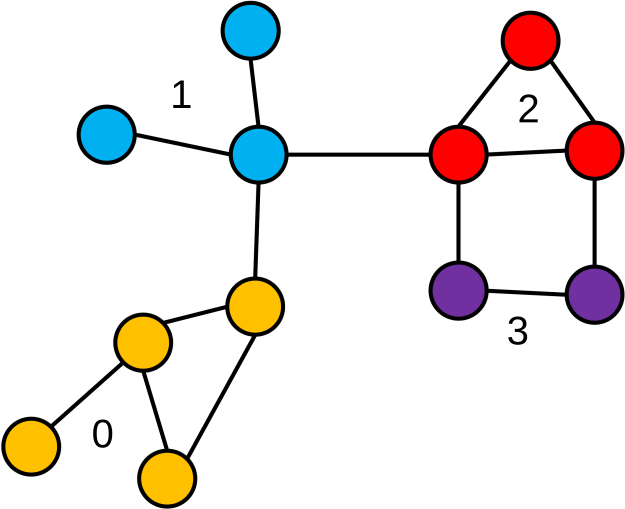}
\caption{\label{metagraph_fig} The atomic metagraph for $q=0.4$ (or any $Q\subset (1/3, 1/2]$) and the network from Figure \ref{ExampleCommunity} which is shown on the right with the atoms labeled by number.}
\end{figure}

The atomic metagraph is a useful object as it provides a complete characterization of all of the potential equilibria in a network, as well as potential dynamics.
In particular, equilibria are the nodes which have self arrows, indicating that if all the atoms in that node are activated (everyone in the atoms adopts the behavior), then they stay activated and do not lead to any further activations.
The atomic metagraph is thus particularly useful in the
seeding problem: we see that there is a way to activate the entire network (induce all nodes to adopt the behavior) by seeding particular pairs of atoms, specifically atoms $0$ and $3$ or atoms $1$ and $2$.
Combined with the knowledge of how many seeds are required to activate each atom, we can see that the entire network can in fact be activated with just two seeds.

To incorporate the atomic metagraph systematically in a seeding algorithm, we take an adaptive approach to calculating the benefit $B_i$ from seeding an atom: at each step after selecting an atom to seed, we recalculate the benefit of each remaining atom as the size of the convention it generates when added to the convention generated by the already-selected atoms.
Given the atomic metagraph as input, finding the nodewise-largest outneighbor is an $O(K)$ operation (where $K$ is the number of atoms), so this does not increase the asymptotic complexity of the greedy algorithm apart from the calculation of the metagraph.\footnote{The atomic metagraph is of exponential size in the number of atoms, so its computation can
become infeasible when a network has a large number of atoms. In practice, one can also set a size threshold so that the atomic metagraph is only computed for atoms respecting that threshold size.}

Figure \ref{heurfig3} shows how using the atomic metagraph in this manner improves the greedy algorithm for the thirty-five Indian village networks of Figure \ref{heurfig2}:

\begin{figure}[h!]
\centering
\includegraphics[width=0.6\textwidth]{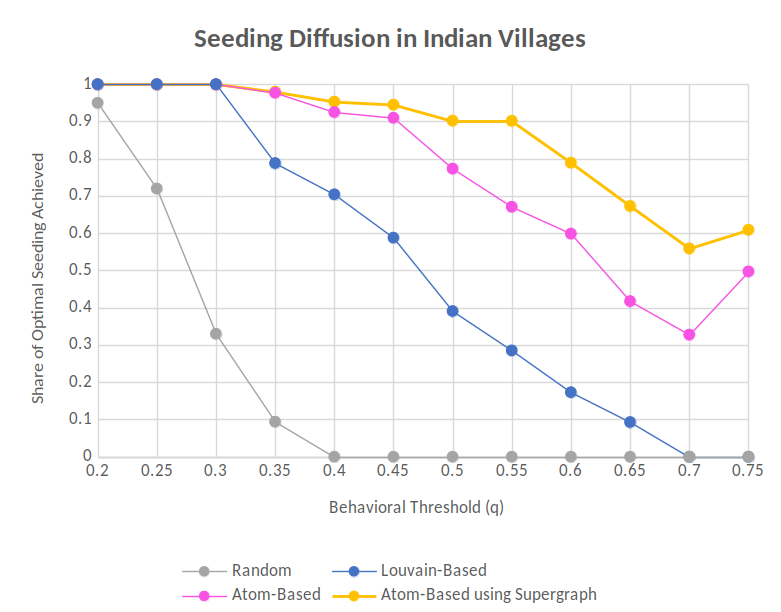}
\caption{\label{heurfig3} Adding the performance of the greedy algorithm using the atomic metagraph for setting of Figure \ref{heurfig2}.}
\end{figure}

Using the metagraph extends the range of $q$ for which our heuristic is close to optimal, as well as now delivering on average more than half the optimal spread over the full range of $q$ values for which seeding is feasible.\footnote{The primary reason the metagraph heuristic still cannot match optimal seeding for higher values of $q$ is that it does not account for how the already seeded atoms change the costs of seeding the remaining atoms.}

\section{Estimating Behavioral Thresholds and Other Extensions}

In some applications, one may not know the threshold associated with a behavior and need to estimate it.
For instance, if a government rolls out a program in some areas before others, then observing adoption in the early areas
allows one to estimate the related $q$ and then use that for subsequent seeding.
One might alternatively see the conventions associated with similar behaviors, and use those as proxies.
In this section, we discuss how to infer the behavior threshold $q$ from observation of a network and
agents' adoption decisions, and evaluate how our estimation procedure performs for informing seeding.\footnote{
Here we work under the assumptions of the model.
As is well-known (e.g., see the discussion in \cite{aralms2009,bramoulledf2009,goldsmith-pinkhami2013,jacksonrz2017,badev2017}), homophily and other unobserved characteristics can confound behavior, and so working from an empirically observed convention might confound a behavior.
This does not impact any of the definitions up to this point in the paper, since our atomic analysis depends only on the network and not on any observed behaviors.
However, when estimating $q$ from observed behavior, then what is driving behavior matters. In a supplementary appendix we discuss the challenges of distinguishing threshold behavior from homophily and offer some suggestions for more general estimation in cases where the $q$s may depend on unobserved characteristics.
The techniques
presented in this section could augment experiments (e.g., \cite{centola2011}) or suitable
instruments (e.g., \cite{araln2017}), to estimate incentives causally.}
(We discuss additional approaches based on different models of errors and heterogeneity in Supplementary Appendix \ref{testqs}.)
\footnote{Since we work with threshold behavior rather than a continuous
influence (such as a linear-in-means model as in \cite*{manski1993,bramoulledf2009} or
a nonlinear, but constant
marginal influence models such as in \cite{brockd2001}), our approach is different from
existing methods.
 \cite{gonzalez2017} estimates peer effects in a threshold model by using regression
 fits of dummy variables for various ranges of influence, which is an indirect or reduced form
 way of fitting a variation of the model we examine.  We instead estimate parameters directly from
 modeling the error structure on behavior and finding parameters that minimize a function of those errors. }

\subsection{Estimating a Threshold $q$}\label{estq}

To fix ideas, we begin with the case in which there is no noise:
we observe a network and a convention and wish to estimate a $q$.

Let $N_{on} = \lbrace i \: | \: $i$ \text{ adopts} \rbrace$ be the set of adopting nodes and $N_{off}$
be its complement.
For each agent $i$, let $s_i$ be the share of $i$'s neighbors who adopt the behavior.
Let $S_{on} = \lbrace s_i \: | \: $i$ \text{ adopts} \rbrace$ be the realized frequency distribution of shares $s_i$ for the agents adopting the behavior, and let $S_{off}= \lbrace s_i \: | \: $i$ \text{not adopt} \rbrace$ be the analog for the agents not adopting.

\begin{figure}[h!]
  \centering
    \includegraphics[width=0.4\textwidth]{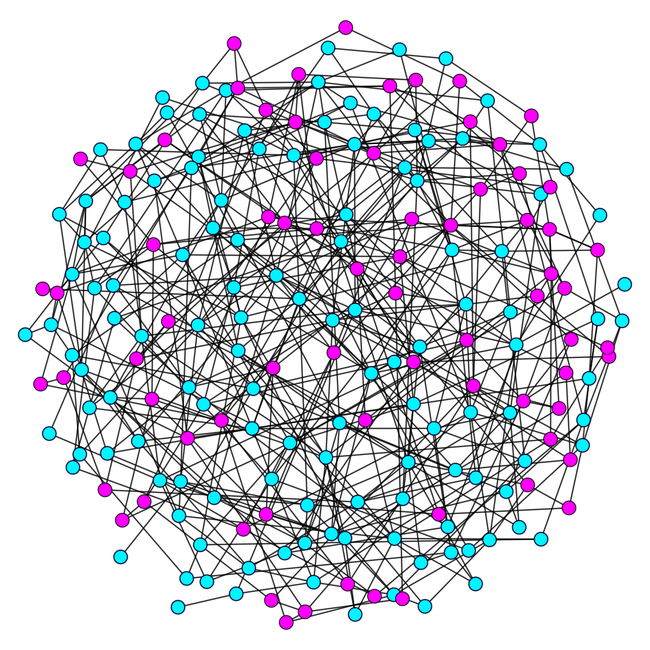}
    \caption{\label{Example_Random_Conventions}
A convention for $q=.4$ on an Erdos-Renyi Network with 200 nodes}
\end{figure}

For an equilibrium convention, there will be perfect separation of the observed $s_i$'s between adopters and non-adopters, with $q$ satisfying:
$$\max S_{off} < q \leq \min S_{on}.$$
To illustrate, Figure \ref{Example_Random_Conventions} depicts an Erdos-Renyi network with $200$ nodes
and an associated  equilibrium for
a threshold of $q=.4$, where nodes are labeled as pink if they have adopted and light blue if they have not.

\begin{figure}[h!]
  \centering
    \includegraphics[width=0.5\textwidth]{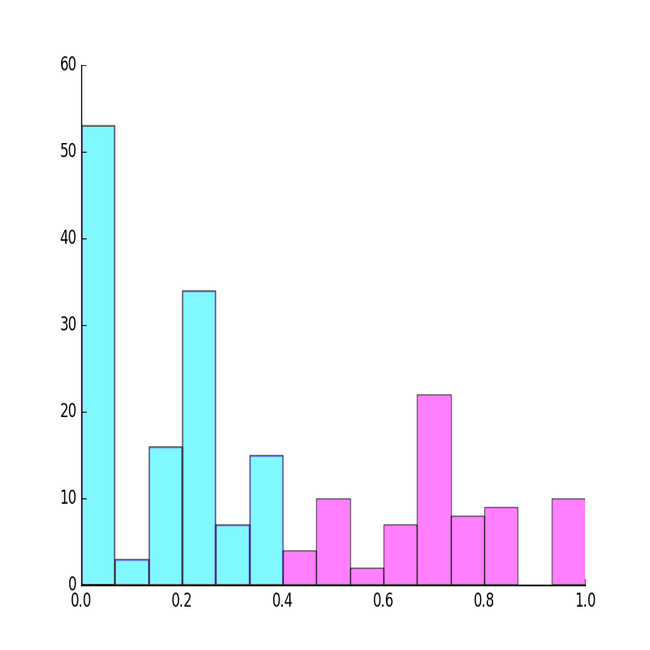}
    \caption{\label{Example_Random_Conventions2}
Distributions of on-neighbor shares for nodes adopting (pink) and not adopting (light blue) from the
network pictured in Figure \ref{Example_Random_Conventions}.  The distributions sandwich the equilibrium threshold of $q=.4$.}
\end{figure}

The equilibrium in Figure \ref{Example_Random_Conventions}
has the corresponding frequency distributions $S_{on}$ and $S_{off}$ pictured in
Figure \ref{Example_Random_Conventions2}.
Observe that $q=.4$ perfectly separates the distributions.

Of course, in most empirical applications there is likely to be heterogeneity in preferences as well as
noisy behavior, so that the observed set of adopters may not form a convention for any single $q$, and the frequency distributions of
the $s_i$'s for the adopters and non-adopters can overlap.
We adapt the model to account this in three different ways---one in this section and two others in Supplementary Appendix \ref{testqs}.

A simple way to allow for noise is to introduce a probability $\varepsilon>0$ with which each agent makes an ``error'' and chooses their adoption decision
independently from what is predicted the threshold $q$ and by the behaviors of their friends.\footnote{This is similar to a quantal-response equilibrium, but without the need to introduce beliefs and Bayesian reasoning in our setting.  Alternatively,  one could also define an ``error'' to be that agents simply reverse what they should do according to $q$ and their friends' behaviors.  Those techniques would result in similar results.  }
We illustrate this process by perturbing the convention on the network from Figure \ref{Example_Random_Conventions} with $\varepsilon=.2$ and $q=.4$.

\begin{figure}[h!]
  \centering
    \includegraphics[width=0.5\textwidth]{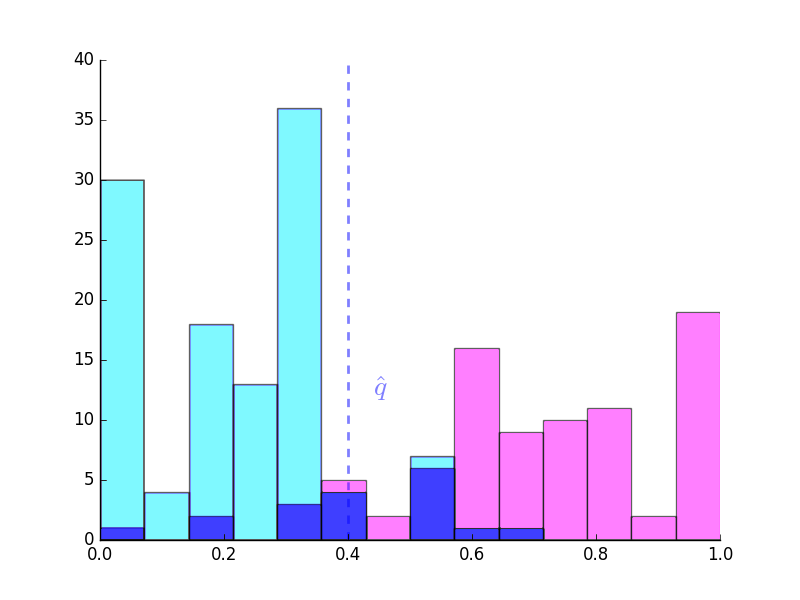}
    \caption{\label{Example_Random_Conventions4}
Frequency distributions of `on'-neighbor shares for nodes adopting in pink and those not adopting in light blue.  The dark blue/purple bars are the overlapping portions of the frequency distributions.}
\end{figure}

This perturbation, wherein $\varepsilon=.2$ of the agents choose randomly, generates the modified frequency distributions $S_{on}$ and $S_{off}$
that overlap as pictured in Figure \ref{Example_Random_Conventions4}.  The true behavioral threshold $q=.4$ still `approximately' separates the distributions in that most of $S_{off}$ lies below $q$ and most of $S_{on}$ lies above it.

Given this sort of noisy behavior, the straightforward way to estimate $q$
is to choose the $q$ for which the largest number of agents are behaving consistently with a coordination threshold of $q$ and the fewest numbers of
agents who are making `errors'.
In particular, consider a statistic that for each $q$ counts how many nodes
behave {\sl inconsistently} with a $q$ threshold:
$$T(q) =  |N_{on} \cap \lbrace i: s_i < q \rbrace | + | N_{off} \cap \lbrace I: s_i\geq q \rbrace |.$$
A $\widehat{q}$ that minimizes $T(q)$ minimizes the
number of deviations from equilibrium behavior, and so finds a $q$ that best fits the behavior with the smallest number of `errors' in the way that people
choose their behavior.\footnote{
One could alternatively jointly estimate $\varepsilon$ and $q$ by 
Bayesian methods, and then presuming a prior with most weight on small errors (well below 1/2),  the
resulting $\widehat{q}$ will also asymptotically be an approximate minimizer of $T$.
If one believes that $\varepsilon$ is large, then one could instead jointly estimate $q,\varepsilon$ via maximum likelihood, which could then produce larger $\varepsilon$'s than the minimizer.  }

Though we have only discussed estimation in the relative threshold $q$ case, the approach extends directly to the absolute threshold ($t$) case by substituting the number of $i$'s friends taking the action rather than the share (replace $s_i$ with $s_i \times d_i$) and then substituting $t$ for $q$:
$$T(t) =  |N_{on} \cap \lbrace i: s_i\times d_i < t\rbrace | + | N_{off} \cap \lbrace I: s_i \times d_i\geq t \rbrace |.$$

\subsubsection{An Illustration of Estimating $q$}

\begin{figure}[h!]
  \centering
    \includegraphics[width=0.45\textwidth]{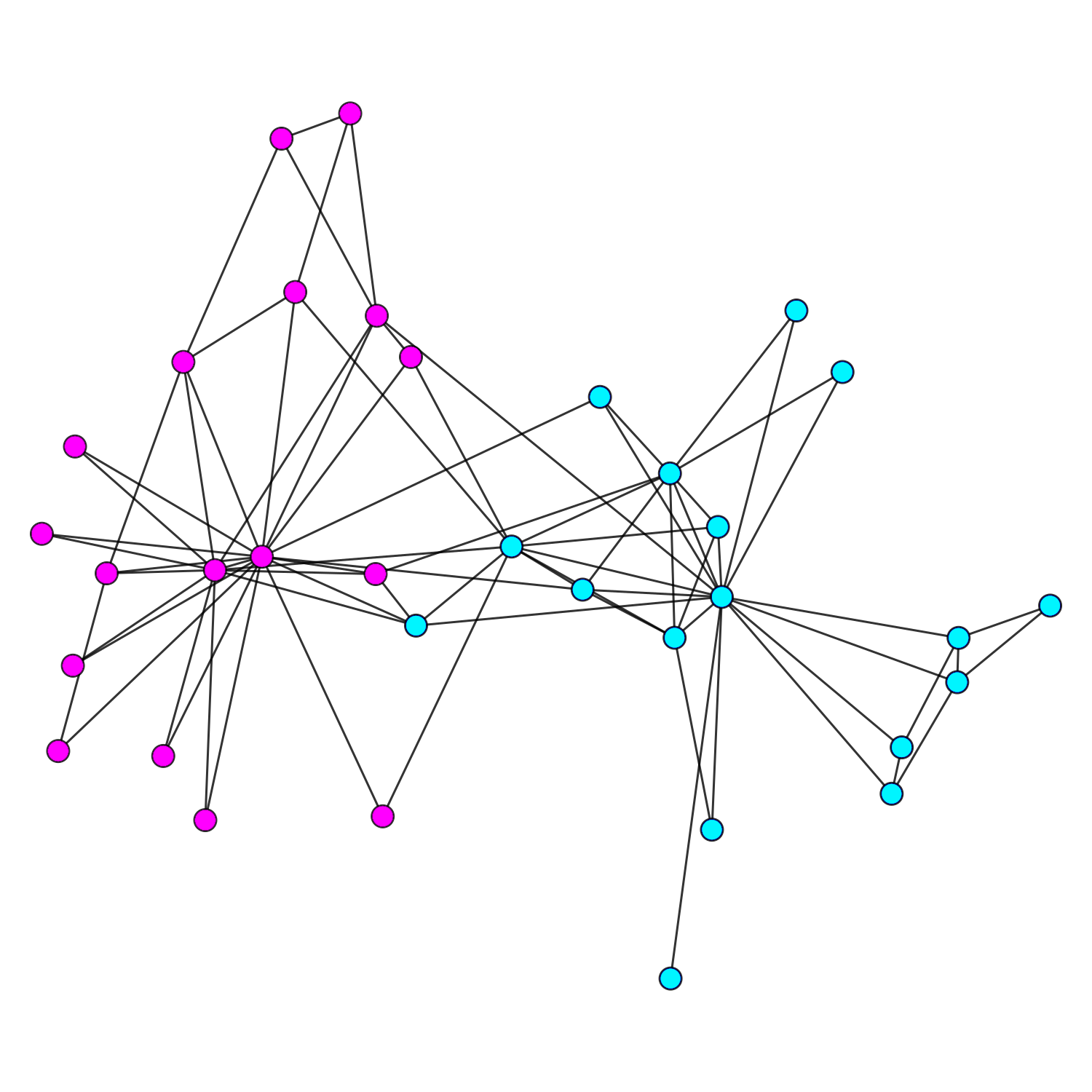}
    \caption{\label{karate1}
Zachary's Karate Club: the pink nodes are those who split off from the club while the light blue nodes stayed.}
\end{figure}

We illustrate the above technique by applying it to the most famous example from the community detection
literature:  Zachary's Karate club.
This club spit into two pieces, with some of its members breaking off to form a new club.   The split is pictured in Figure \ref{karate1}.

\begin{figure}[H]
  \centering
    \includegraphics[width=0.5\textwidth]{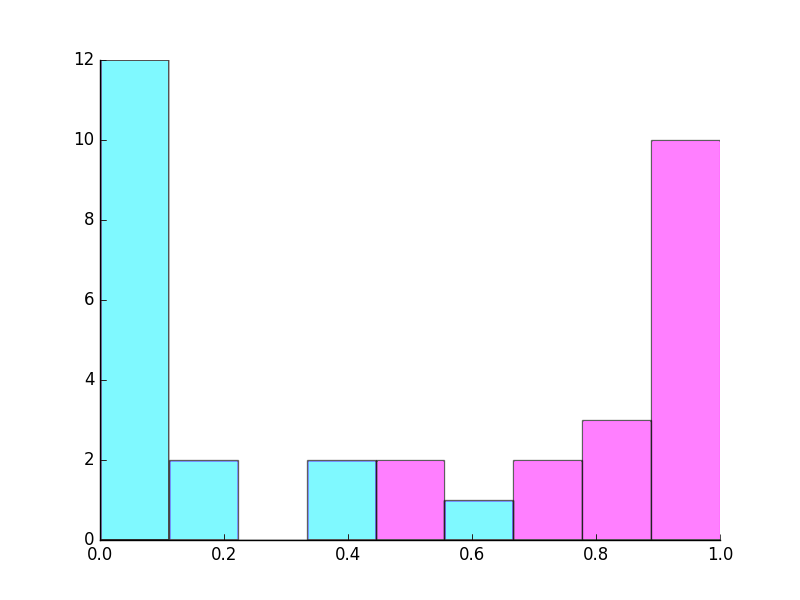}
    \caption{\label{karate2}
The distribution of neighbors who split off from the club:   Those who split off had most of their friends also split, while those who stayed had fewer friends who split off.}
\end{figure}

We can consider the coordination game in which people prefer to split off if and only if at least a share
$q$ of their friends do.  We can then estimate $q$ using our technique from above
by comparing $S_{\text{on}}$ and $S_{\text{off}}$, as pictured in Figure \ref{karate2}.

To estimate $q$, we calculate the value of our ``errors'' statistic $T$ at various potential thresholds:

\begin{figure}[H]
  \centering
    \includegraphics[width=0.5\textwidth]{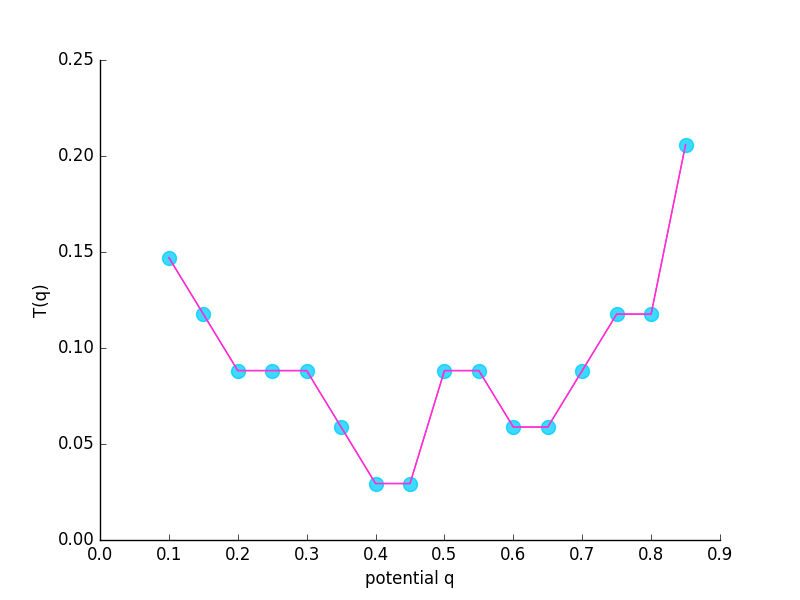}
    \caption{\label{karate3}
The fraction of nodes that are deviating from coordination if the true threshold were $q$, for various values of $q$. }
\end{figure}

We see that fraction of people making errors $T(q)$ is minimized for $q$ between $0.4$ and $0.45$, and
involves less than three percent errors.  The set of $q$ that minimize $T$ will generally be an interval, since ``on''-neighbor shares must vary by discrete multiples of the inverse of the least common product of the nodes' degrees; but the size of the interval will shrink as degree heterogeneity and the size of the network grow. This interval serves as an estimate.

In Supplementary Appendix \ref{relabs} we extend the technique to allow thresholds to depend on individual characteristics, and  apply the technique to analyze peer effects on smoking in the Add Health data set, and we also show that a fractional threshold fits those data
better than an absolute threshold.

\subsection{Seeding with First-Stage Estimation}

Since the threshold $q$ is unknown in many seeding applications, one first has to estimate $q$ from some previous adoption
of the same behavior in a different network or part of the network, or a similar behavior, and then use that to estimate the $q$-atomic structure to inform seed selection.
In this section we examine this joint procedure via simulations where we can vary the true unobserved $q$ and the noise in behavior.

The gap between seeding with the true $q$ and an estimate $\hat{q}$ gives a metric for evaluating how well the estimation works.
We find that the performance of our seeding algorithm is not significantly additionally degraded due to the noisy estimation,
provided that the share of agents whose adoption is completely random is not too large.

To be able to compare the results to our earlier seeding results, we test our joint estimation and seeding performance on the Indian Village networks. We introduce the noise by picking a fraction $\alpha$ of the nodes to be random adopters, who adopt with probability $1/2$ completely independently of all other nodes' adoption decisions. The simulation procedure for a given village network and true threshold $q$ is as follows.

\begin{enumerate}

\item{} Simulate a convention with some given true $q$ and noise level $\alpha$:

\begin{itemize}
\item Generate the noise: a fraction $\alpha$ of the $n$ nodes adopt the behavior independently with probability $1/2$.

\item Generate a convention: randomly select 16 nodes in the remaining $1-\alpha$ share as initial adopters; then iteratively update nodes' behaviors via the process described in subsection \ref{dynamics} with threshold $q$, leaving the adoption decisions of the noisy adopters fixed. Thus, noisy agents are independent of others, but the remaining agents adopt in response to their peers (including the noisy agents).
\end{itemize}

\item{} Estimate $\hat{q}$ by applying the procedure developed in Section \ref{estq} to the simulated convention from Step 1.

\item{} Select $8$ seeds using the estimated $\hat{q}$ and associated estimated atomic structure, as defined in Section \ref{seedalgo}.

\item{} Compare the seeding based on the estimated $\hat{q}$ and our algorithm to that would be obtained under the fully optimal seeding for the true $q$, as well as the seeds that would be chosen by the algorithm if $q$ were known, where the behavior is based on the true $q$ (as defined in Section \ref{dynamics}).
\end{enumerate}

\begin{figure}[H]
  \centering
    \includegraphics[width=0.9\textwidth]{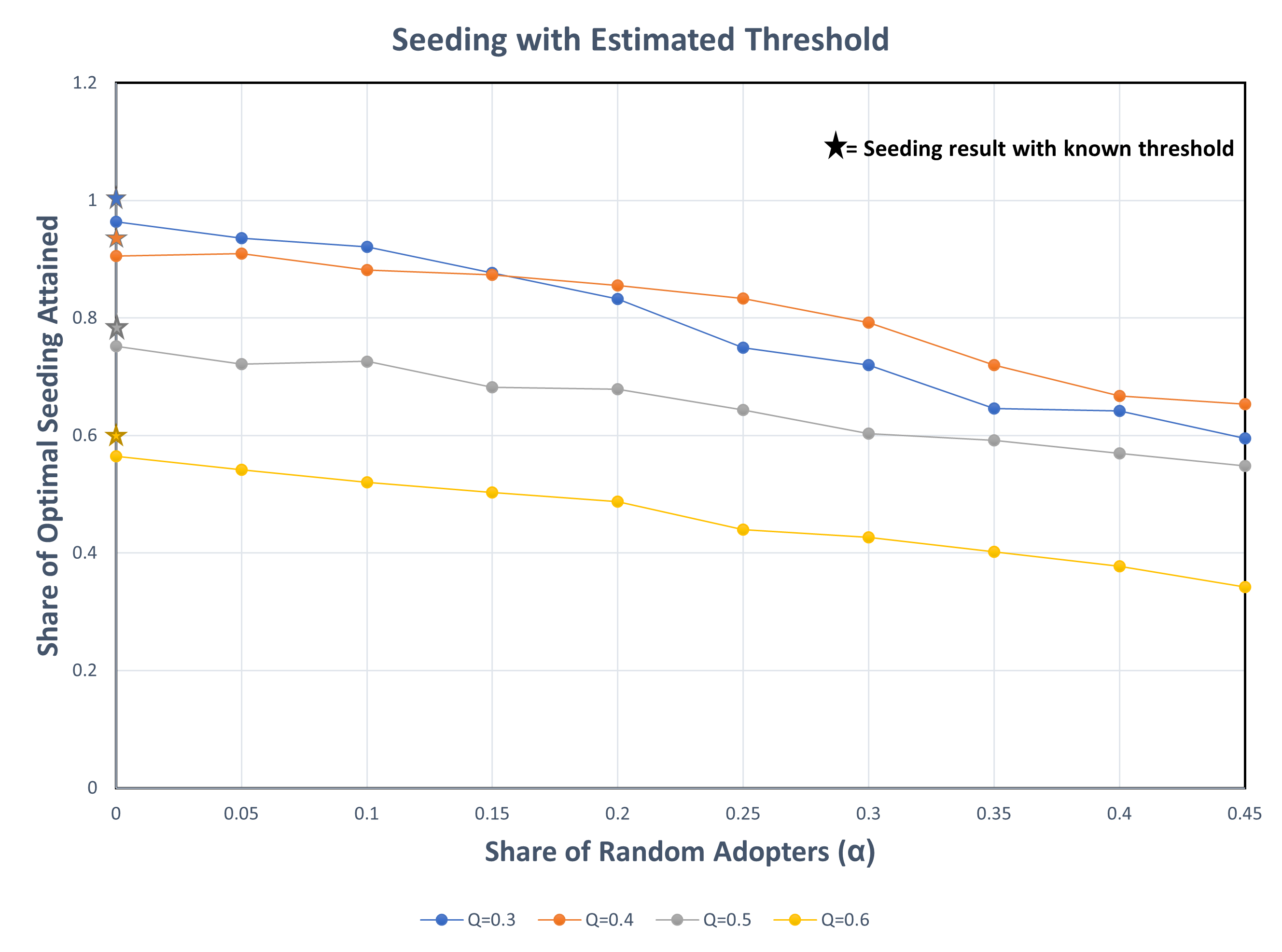}
    \caption{\label{seed_w_estimate}
The performance of the atom-based seeding algorithm using an estimated $\hat{q}$ compared to fully optimal seeding knowing the true $q$.
The `stars' indicate the share of the fully optimal seeding that are obtained by using our seeding algorithm with
the true $q$, and then the curves show the further deterioration that comes from estimating the $q$ as a function of the amount of noisy behavior $\alpha$.    }
\end{figure}

Figure \ref{seed_w_estimate} shows how well the combined algorithm of estimating $q$ and then using our
seeding algorithm compares to what would obtain if one know $q$ and used the fully optimal algorithm.

With no random adopters ($\alpha=0$), our estimate approximates the true threshold well and then our
seeding results match those with known $q$, indicated by the stars.  Those are below 100 percent of optimal since the atom-based algorithm is not fully optimal.  We then see a small but increasing difference between the results for atom-based seeding with $q$ known and $q$ estimated. As the noisiness of simulated data increases, the difference between the estimated and true $q$ degrades the performance of the atom-based seeding. Up to $\alpha=1/5$, the decrease in performance is modest: in this range, on average, the share of the optimum result attained by the algorithm drops by $0.02$ for every $0.05$ increase in the share of random adopters.

The degradations (negative slope) as a function of noise is greater the further the true $q$ is from $0.5$.  This occurs because noise in the estimation procedure tends to push the estimated $\hat{q}$ towards $0.5$, the average estimate that would obtain if all adoption decisions were an independent coin flip.

\subsection{Link Persistence within Atoms}

Our focus has been on how network structure influences behavior, and how using that understanding can
define communities or an atomic structure of a network.
Of course, behaviors can also play a role in network formation  (e.g., see \cite{kandel1978}).
As a last observation, we point out that identifying atoms can shed new light on
the co-evolution of networks and behaviors over time.

In particular, we examine two data sets that have time series of networks with good coverage of the population.
We show that, in both contexts,
pairs of linked agents within an atom are more likely to maintain that link across time than pairs of linked agents who sit in different atoms.
We work with  $Q=1/3  \pm \varepsilon$-robust atoms since these tend to provide distinctive and not overly fragmented atoms.

The first data set that we examine consists of borrowing and lending relationships among households in Indian villages measured in 2007 and 2012 in a study by \cite*{banerjeecdj2016}.
We find that  61 percent  of the links present in 2007 that lie within the $Q=1/3  \pm \varepsilon$-robust atoms
survive are also present in 2012, while only 34 percent of the links that connect across different atoms in 2007 survive to be present in 2012. By comparison, for the partition given by the Louvain community detection algorithm, 54 percent of the links within communities persist, and 39 percent of the links across communities persist.

The second data set is from \cite*{gillensy2018,jacksonnsy2018} and consists of the friendship network among Caltech undergraduates from the class of 2017, including more than 92 percent of the students.
We find that 
53 percent of the links that lie within the $Q=1/3  \pm \varepsilon$-robust atoms survive from sophomore to junior year, while only 38 percent of the links
that connect across different atoms survive.
By comparison, for the partition given by the Louvain community detection algorithm, 44 perfect of the links within communities persist, and 40 percent of the links across communities persist.

In both cases, not only are links within atoms are significantly more likely to remain than links across atoms, but the atomic structure of the network is more useful than modularity-based methods for predicting which links persist and which disappear.\footnote{There are hundreds of links measured in Caltech network and tens of thousands in the Indian village networks, and so both of these differences are significant beyond the 99 percent confidence level if one assumes independent link changes.  Of course, links are rarely independent, and so we do not report significance on these observations.  An appropriate
model for analyzing the evolution statistically involves correlated link formation (e.g.,
see \cite{chandrasekharj2016}), but takes us well beyond the scope of this paper. }
Our model provides a potential explanation for this observation: that the behaviors of agents in an atom always coincide,
while across atoms behaviors can differ.  Thus, since coordinating on a common behavior leads to a higher payoff,
 links within atoms would lead to higher payoffs than links across atoms and therefore
have a higher likelihood of surviving over time.

The difference in survival of relationships within and across atoms another reason for paying attention to atoms and the communities that they define.
This also provides an interesting hypothesis for the evolution of networks over time, and complements other findings that link persistence is also related to their presence in triads (e.g., \cite{banerjeecdj2016,lyu2022investigating}.
As links internal to an atom are more likely to survive, this could lead to the network becoming more fragmented over time.
Given the lack of models of the coevolution of networks and behaviors, this could provide the basis for a new class of models that blend behaviors and network evolution.

\section{Concluding Remarks}

Traditional approaches to detecting communities in a network are based on graph properties---for instance, spectral, modular, or structural equivalence---but without any foundation in the consequences of network structure for human behavior.
Instead, we have taken a novel approach to defining communities in a network based on the behaviors that the communities induce when agents are influenced by their peers.
This perspective reflects the fact that much of a social scientist's interest in networks comes from the roles that networks play in determining behaviors and outcomes.
We have shown that this approach identifies an atomic structure that differs significantly from those identified by standard community-detection techniques.
We also have shown how this behavioral approach provides new methods of estimating network structure, seeding diffusions and estimating peer effects.

Although we  introduced the idea of `behavioral communities', and the associated atomic structure of a network, in the context of coordination on a behavior, it is clear
that our approach can be extended to provide a more general method of identifying communities in networks.
For instance, one could extend our approach to settings in which many actions are possible rather than two, or in which behavior takes on a continuum of values.   One then needs to define when it is that a group should be considered a convention and how that relates to network structure.
There are many metrics that could be used, and even definitions that allow for overlapping conventions, which opens many avenues for future research.\footnote{There are also relationships between core structures of a graph and equilibrium structure (e.g., \cite{manshadij2009}), as well as centrality measures such as Katz-Bonacich centrality and equilibrium structure (e.g., \cite*{ballestercz2006}) that could be exploited and analyzed.  }

There are also further questions associated with our approach that we have not explored here, but are important for further research.
For example,  our technique could give new ways of defining influential people in a network.  This could be done by
asking which people, when removed, lead to a significant change in the atomic structure.
We have also provided a basic algorithm for detecting atoms, but given the complexity of this problem, there are unexplored details of how to develop heuristic and fast algorithms to work on large graphs.
The atomic structure of networks could also be used to develop new understandings of the relationships of different (multiplexed) networks occupied by the same set of people.   For instance, in the various networks studied by \cite*{banerjeecdj2013} have very different structures over same villagers.
Do the villagers have different atomic structures for different types of interactions and how are those dependent on the type of interaction?

\bibliographystyle{ecta}
\bibliography{behcomm}

\section*{Appendix}

\subsection*{Proofs}

\medskip
\noindent{\bf Proof of Theorem \ref{random}}:
We prove the theorem via three lemmas.

The first lemma shows that number of links from a node to a block is within a bound of its expected value, and that this holds simultaneously for all nodes with high probability.


Let $b_i$ denote node $i$'s block and $d_i(b',n)$ node $i$'s degree to block $b'$ on a randomly realized network with $n$ nodes.

\begin{lemma}\label{chernoff}
The probability that all nodes have their degrees to all blocks within a factor of $(1\pm \delta)$ times the expected degree goes to 1, for any
$\delta>0$.  That is,  for every $\delta>0$:
$$Prob \left( n= \# \{ i:     (1-\delta) d_{b_i b'}(n)  <  d_i(b',n)  <   (1+\delta) d_{b_i b'}(n) \forall  b'\in B(n) \} \right) \rightarrow^P 1.
$$
\end{lemma}

\noindent {\bf Proof of Lemma \ref{chernoff}:}
Consider some node $i$ from block $b_i\in B(n)$.     

From Chernoff bounds it follows that
\[
\Pr  \left(  (1-\delta) d_{b_i b'}(n)  <  d_i(b',n)  <   (1+\delta) d_{b_i b'}(n) \right) >   1-  2e^{-\delta^2 d_{b_i b'}(n)/3}.
\]
Thus, looking across all $b'$:
\[
\Pr  \left( \forall b':  [ (1-\delta) d_{b_i b'}(n)  <  d_i(b',n)  <   (1+\delta) d_{b_i b'}(n)] \right) >   (1-  2 e^{-\delta^2 \min_{b'} d_{b_i b'}(n)/3})^{|B(n)|}.
\]
It follows that
\[
\Pr  \left( \forall b':  [ (1-\delta) d_{b_i b'}(n)  <  d_i(b',n)  <   (1+\delta) d_{b_i b'}(n)] \right) >   1- 2 | B(n) | e^{-\delta^2 \min_{b'} d_{b_i b'}(n)/3}.
\]
Therefore,
\[
\Pr  \left( \exists  b':  [  d_i(b',n) < (1-\delta) d_{b_i b'}(n)  {\rm \ \ or \ \ }  d_i(b',n) >  (1+\delta) d_{b_i b'}(n)] \right)  <  2 |B(n)| e^{-  \delta^2 \min_{b'} d_{b_i b'}(n)/3}.
\]

Thus the expected number of nodes for which their degree to some block lies outside  of a $\delta$ bound around the expected
degree that they should have with that block satisfies:
\[
E  \left( \# \{ i: \exists  b',  [  d_i(b',n) < (1-\delta) d_{b_i b'}(n)  {\rm \ \ or \ \ }  d_i(b',n) >  (1+\delta) d_{b_i b'}(n)] \} \right)  <   n 2 |B(n)| e^{- \delta^2 \min_{b b'} d_{bb'}(n)/3}
\]
Examining the right hand side of the inequality, and the assumption on $f(n)$, it follows that
\[
n 2 |B(n)| e^{-\delta^2 \min_{bb'} d_{bb'}(n)/3} <   n  2|B(n)| e^{- \delta^2 f(n) \log(n)/3} =  n 2 |B(n)| n^{- \delta^2 f(n)/3 }  \rightarrow 0.
\]
The last implication follows from the fact that  $|B(n)|\leq n$
which implies that
$$n|B(n)|/ n^{\delta^2 f(n)/3}< n^{2-\delta^2 f(n)/3}  \rightarrow 0$$
since $f(n)\rightarrow \infty$.
Thus, the probability that all nodes have all their degrees to all blocks within a factor of $(1\pm \delta)$ times the expected degree goes to 1, for any
$\delta>0$.\eproof

\medskip

\begin{lemma}
\label{modularity1}
Fix any $\varepsilon>0$.  Consider any sequence of $b_n,b_n'\in B(n)$ (including $b_n=b_n'$).  The probability that there exists
some $S_n\subset b_n\cup b_n'$ such that the average fraction of friends that nodes in $S_n\cap b_n$
have to nodes to $S_n\cap b_n'$ is more than $\varepsilon$ different
than the average fraction of friends that nodes in $S_n^c\cap b_n$
have to nodes to $S_n\cap b_n'$
tends to 0.
That is,
\footnotesize
$$Prob \left( \# \{ S_n\subset b_n\cup b_n'  :  \left| \frac{1}{\# S\cap b_n} \sum_{i\in S\cap b_n}  d_i(S\cap b_n', n)/d_i(n)  -
\frac{1}{\# S^c\cap b_n} \sum_{i\in S^c\cap b_n}  d_i(S\cap b_n', n)/d_i(n)\right| >\varepsilon   \} = 0 \right)\rightarrow^P 1. $$
\normalsize
\end{lemma}

Lemma \ref{modularity1}
follows from an application of Theorem 1.2 (c) of  \cite{mcdiarmid2018ER_modularity}.
In particular, if one considers a sequence of E-R random graphs on nodes in $b_n\cup b_n'$ with probabilities of links $p_{b_n,b_n'}$, then the modularity tends to 0 with probability going to 1.\footnote{Note that
both $b_n$ and $b_n'$ are growing in $n$ given that cross degrees are growing, and also that $p_{b_nb_n'}n > f(n)\log(n)\rightarrow \infty$.}
Thus, for large n the probability is tending to 1 that, for all $S_n\subset b_n\cup b_n' $, the average fraction of degrees of nodes in $S_n$ to those in $S_n^c$ and vice versa are within $\varepsilon$ of
the expected fraction.   Applying the same theorem to nodes in $S_n\cap b_n$ and $S_n^c\cap b_n$, as well as
$S_n\cap b_n'$ and $S_n^c\cap b_n'$, we can deduce that relative fractions of expected degrees each are within $\varepsilon/3$ of each other, with probability going to 1.

\medskip

\begin{lemma}
\label{modularity2}
Fix any $\varepsilon>0$.  In a stochastic block model with a finite set of blocks, the probability that
a sequence of robust conventions with an interval of $q$s of at least width $\varepsilon$
 fails to be a superset of the blocks converges to 0.
\end{lemma}

Lemma \ref{modularity2} follows from Lemma \ref{modularity1}, noting that if some sequence of robust conventions
are not a superset of the blocks,  then there exists some subsequence
of $S_n$ that cut across some $b_n$ and form robust conventions.
Then since nodes in $S_n \cap b_n$ have a higher fraction of their links in $S_n$ by at least $\varepsilon$ than
those in  $S_n^c \cap b_n$, there must exist at least one $b_n'$ such that $S_n \cap b_n$ has a relative fraction of friends in
$S_n\cap b_n'$ that is greater than
the relative fraction that $S_n^c \cap b_n$
has in $S_n\cap b_n'$  by at least $\varepsilon$.
Then we apply Lemma \ref{modularity1} to reach a contradiction.

The first part of Theorem \ref{random} then follows from Lemma \ref{modularity2}.
The second part of the theorem then follows from Lemmas \ref{modularity2} and \ref{chernoff}, as does the third part.
This concludes the proof of Theorem \ref{random}.\eproof

\bigskip
\noindent{\bf Proof of Theorem \ref{optseed}:}

Consider a sequence of weakly homophilous convergent block models satisfying the conditions of Theorem \ref{random}.
Suppose that it has $|B(n)|=K$ approximately equal sized blocks for large enough $n$.
Suppose as well that there exists $\varepsilon>0$ and some number of $S(n)$ seeds for each $n$ such that
 $ (1+\varepsilon) \frac{qn}{K\overline{q}}  < S(n) < (1-\varepsilon)q n$.
 Note that this implies that $K>1$.
Under  Theorem \ref{random} it follows that the atoms are supersets of the blocks with a probability going to 1.

We find a lower bound on the number of nodes that adopt the behavior due to the atom-based seeding strategy by considering
  a seeding in which seeds are evenly divided and put into $\lfloor \frac{S(n)\overline{q}}{(1+\varepsilon) q m(n)} \rfloor$ blocks.
  In such a seeding at least $\frac{qm(n)(1+\varepsilon)}{\overline{q}}$ seeds appear in a block.
  Thus, at least a fraction $\frac{q(1+\varepsilon)}{\overline{q}}$ of the nodes in each of the seeded blocks are seeded.
   By an argument analogous to that from the proof of Theorem \ref{random}, we can use Chernoff bounds (using $\varepsilon/2$ in the role of $\delta$) to show that the probability that all  nodes (including the seeds) in each seeded block have a fraction of more than $\overline{q} \left(\frac{q }{\overline{q}}\right) = q$  of their neighbors among the seeds goes to one.
   Thus, with a probability going to one, all nodes in all the seeded blocks adopt the behavior and will maintain the behavior in whatever the resulting convention is.
   This is then a lower bound on the impact of the atom-based seeding, and so all nodes in $\lfloor \frac{S(n)\overline{q}}{(1+\varepsilon) q m(n)} \rfloor$ blocks adopt the behavior, leading
   to the claim.

Next, note that with a completely random seeding, the expected ratio of any node's neighbors that are infected is at most $S(n)/(n-1)$ (it is less for seeds), which is less than $(1-\varepsilon) q \frac{n}{n-1}$, by assumption.
Again, by a Chernoff-bound argument similar to that in Theorem \ref{random}, the probability that all nodes have fewer than $q$ of their neighbors infected under the seeding goes to one.
Thus, with a probability going to one, no nodes other than seeds ever adopt the behavior, and if seeds can choose to revert, then no nodes at all adopt the behavior.\eproof

\newpage

\centerline{\bf Supplementary Appendix: }
\centerline{Behavioral Communities and the Atomic Structure of Networks}
\centerline{by Matthew O. Jackson and Evan C. Storms
}

\section{Inferring Network Information---Atoms and Blocks---from Observed Conventions: A Variation of the Chocolate Bar Theorem}

Network data can be costly to acquire, and so it can be useful to
infer an unobserved network (e.g., see \cite*{brezaetal2017}).
In this section we discuss how seeing
conventions  allows us to infer the atoms, which then provide valuable network information (e.g., the densities of links within and across different groups of agents can be inferred from the atomic structure).

For example, suppose that $q=1/2$ and that one can observe a variety of choices by people between pairs of  technologies for which they prefer to use the technology that most of their friends use (e.g., video games, operating systems, software, apps, etc.).  If one observes
enough of those behaviors, then one can recover the atoms, and then predict the possible conventions for some new choice between two technologies,
 and even implement our atom-based seeding algorithm---as under Theorem \ref{optseed} one does not need to know the network within the atom to achieve
the bound.

Consider an (unobserved) network $g$ and a behavioral threshold $q$.
Let $K$ denote the number of atoms associated with the threshold $q$, and
consider a network such that any union of atoms forms a convention.\footnote{This condition is satisfied, for instance, if there is a high enough density of friendships within atoms and low enough across atoms.  The condition is not essential for the analysis here, it simply provides a symmetry to the analysis.  More generally, one can analyze how many conventions split each pair of atoms and develop
analogous results from those numbers, but the calculations are more cumbersome.}
What is the minimal number of conventions on past behaviors that is required to recover full knowledge of the $K$ atoms?
It turns out that the minimal number of conventions needed to
recover the $K$ atoms is $\lceil log_2 (K) \rceil$.

This is a tight bound and is much smaller than the $K-1$ obtained under the well-known ``Chocolate Bar Theorem,'' which one might have superficially conjectured to be the answer.
In our application, instead, the conventions can be overlapping - each break can end up splitting many of the previous pieces.\footnote{The Chocolate Bar Theorem asks how many linear breaks it takes to split a rectangular chocolate bar into an even number, $K$, of equal-sized rectangles, when one
can only break one given piece at a time.   It is easy to
see that the answer is $K-1$ by induction (each break only gives one more piece than one had before).  In our situation a new break can split many of the previous pieces at the same time - a new convention can
cut across all the previous conventions - so it would be as if one could stack the pieces of the chocolate bar and break them all at once - so it breaks as a power of two rather than linearly, one at a time. }
For instance, if one has $K=8$ atoms - labeled $\{1,2, \ldots, 8\}$,  one can list a convention simply by listing which atoms are in the convention (adopting the behavior) - for instance $\{1,2,5,6\}$ indicates a convention where the people in atoms $1, 2, 5$ and $6$ all adopt the behavior and the people in the other atoms do not.
In particular,  the three conventions $\{1,2,3,4\}$, $\{1,2,5,6\}$, $\{1,3,5,7\}$ identify all of the atoms: for any two atoms there is at least one of the conventions
such that one of the atoms is in the convention and the other atom is out.
The proof that the minimal number is $\lceil log_2 (K) \rceil$ is an extension of this example and left to the reader.

We can also ask how many conventions one would typically need to see in a case where one cannot expect to see the minimum number, but instead
expects the conventions to be randomly generated.   In particular, consider a set of conventions that are independently formed by randomly having each atom adopt the behavior with prob 1/2.
If $k$ such randomly drawn conventions are observed, then the chance that any two atoms are not split by any observed convention is $2^{-k}$.
The expected number of pairs that are not split is
\[
\frac{K(K-1)}{2} \frac{1}{2^k}.
\]
By Markov's inequality the probability that all pairs are split is at least
\[
1- \frac{K(K-1)}{2^{k+1} }.
\]
In the limit, this bound becomes tight in that the probability that exactly one pair is not split tends to  $\frac{K(K-1)}{2^{k+1} }$, as the chance that this happens to more than one pair vanishes as $k$ grows.
If $k>2 log_2(K)-1$ then this is more than 1/2, and converges to 0 exponentially in  $k+1-2 log_2(K)$.
Thus, the number of randomly generated conventions that one needs to see to be able to identify all of the atoms is not much more than double the minimal number needed.

We collect these observations in the following proposition.

\begin{proposition}
\label{chocolate}
Consider a network and $q$ and let $K$ denote the number of atoms associated with $q$.
If 
any union of atoms forms a convention,
then the minimum number of $q$-conventions that one needs to see to be able to recover all of the atoms is $\lceil log_2 (K) \rceil$.
Moreover, if one observes $k$ randomly generated $q$-conventions (with each atom in or out independently with probability 1/2), then the probability that one can infer all of the atoms is at least
$1- \frac{K(K-1)}{2^{k+1} } $.
\end{proposition}

Thus, if one can observe a set of different conventions for a collection of $q$'s that generate the same atoms, then seeing $\lceil log_2 (K) \rceil$ different conventions is enough to  identify all the atoms/blocks even without knowing anything about the network.
The second part of the result implies that if the conventions are randomly generated,
and on sees at least $k$ conventions, where  $1- \frac{K(K-1)}{2^{k+1} } >.95$, then one would be more than 95 percent sure of being able to exactly identify all of the atoms.   This is satisfied, for instance if  $k>3.322 +2 log_2(K)$.

It is worth pointing out that the logic behind the proposition extends
to infer all of the atoms, even without knowing how many atoms there are (i.e., without knowing $K$).
In particular, suppose that one sees a random sequence of conventions (with each atom in or out independently with probability 1/2).
Eventually, new conventions will stop generating new atoms.  Once one sees a number of conventions in a row that do not generate any
new atoms, then one can be increasingly certain that one has recovered all of the atoms.
This happens at a rate that we can make precise, as follows.

Consider seeing a sequence of randomly generated conventions and
let $K(k)$ be how many different atoms have been generated by the first $k$ conventions.  If $K>K(k)$, then the probability that $K(k+\ell)=K(k)$ is no more than $1/2^\ell$, since each new convention has a chance of 1/2 of splitting any two atoms that have not yet been split.  So, for instance, starting from any $k$, under the null hypothesis that $K>K(k)$, the probability of observing seven additional conventions that generate no new atoms is no more than 1/128: less than 1 percent.



\section{Absolute Thresholds }

The above definitions are relative to some fraction of at least $q$ of neighbors taking an action.
This applies naturally to coordination problems.
For some other games of strategic complements, it can be natural to adopt a behavior if at least $t$
neighbors do, for some $t\in \{0,1,2,\ldots, n-1\}$.
For instance, one might benefit from learning to play a particular game that requires $k$ agents if and only if at least $k-1$ friends play the game.

If the network is regular of degree $d$ (i.e., $d_i(g)=d$ for all $i\in N$), then
$S$ is a convention for $q$ in a relative (fraction) setting if and only if $S$ is a convention for $t=qd$ in the absolute setting.
However, for some networks that are not regular, there exist $q$'s that generate conventions that are not conventions
for any $t$,  and vice versa.

\subsection{Absolute Threshold-based Communities in Stochastic Block Models}

The conclusions in Theorem \ref{random} are different from what occurs in the
absolute threshold ($t$) setting.
For any absolute threshold, as $n$ increases, even for very sparse networks,
behavior is contagious both within and across blocks and the whole society becomes the only convention.

\begin{theorem}
\label{random2}
Consider some finite $t>0$ and a growing sequence of stochastic block networks.\footnote{This result does not require
that the stochastic block model be homophilous, nor does it require that  there exists $f(n)\rightarrow \infty$ such that  $d_{bb}(n)> f(n)\log(n)$ for all $b$. }
If there exists $\varepsilon>0$  for which
$p_{bb'}(n)> (1+\varepsilon) \left(\frac{t\log(n)}{n}\right)^{1/t}$ for each $b', b$, then
with a probability going to 1 as $n$ grows the partition corresponding to $C(t,g^n)$ is the degenerate one generated by the atom of $N$.
\end{theorem}

Clearly,  if $p_{bb'}$ falls substantially below the threshold for enough $bb'$ pairs then the graph will fragment.   Provided the probabilities internal to blocks is large enough to generate conventions within the blocks, then the community structure will be non-degenerate.

Theorem \ref{random2} is stated for a fixed $t$.  The result also holds when $t$ grows with $n$.
For example, consider a case in which there exists $f(n)\rightarrow \infty$ such that  $p_{bb'}(n)> f(n)  \frac{\log(n)}{n}$ for each $b', b$, so that Theorem \ref{random} applies.
In that case, if there exists $\varepsilon>0$ such that
$t(n) < \min_b d_b(n) \left[ \min [1-\max_{b} d_{bb}(n)/d_b(n),  \min_{b,b'} d_{bb'}(n)/d_b(n) ]+\varepsilon\right]$,
then the atomic structure is the degenerate one with an atom of $N$ with a probability going to 1.
Effectively, a low $q$ behaves similarly to a growing $t$ if the network is sufficiently connected across blocks.
Theorem \ref{random2} works with a lower threshold on $p_{bb'}(n)$ and
then establishing the precise growth rate of $t$ that is admissible is challenging, given that it enters in complicated ways in the expressions in the proof below.

This provides a more subtle but important distinction between what has been called ``simple'' versus ``complex'' contagion in the literature (e.g., \cite{centolaem2007,centola2010}), noting that contagion that has $t=1$ can behave differently from behavior with $t>1$.   Our results show that the meaningful distinction is between whether $t$ is large enough to be a nontrivial fraction of a person's degree - essentially a large enough proportional $q$ rather than a smaller threshold.   It is not requiring multiple friends rather than just one taking an action that is the key, but whether the fraction of friends required to trigger behavior is large enough to have to concentrate within communities.


We prove Theorem \ref{random2} by proving Theorem \ref{random3} about random graphs, which appears in the appendix and is of independent interest.
We prove that $ (1+\varepsilon) \left(\frac{t\log(n)}{n}\right)^{1/t}$ is in fact a strong threshold (in the sense defined in random graph theory) for
the existence of a nontrivial set of nodes being $t$-closed -- so that all nodes outside the set have fewer than $t$ connections into the set.  We show
how this relates to what are known as $k$-cores of random graphs.
We could not find Theorem \ref{random3} in the graph-theory literature, so we have proven it directly.
Our method of proof appears to be new and may be of independent interest.

\medskip
\noindent{\bf Proof of Theorem \ref{random2}}:
We can decompose the network into an ER random graph with link probability
$p(n)> (1+\varepsilon) \left(\frac{t\log(n)}{n}\right)^{1/t}$, plus extra links.
The result is then a corollary of Theorem \ref{random3}, which we state and prove next.\eproof

The following definitions are needed to state Theorem \ref{random3}, which is used to establish Theorem \ref{random2}.

\paragraph{$k$-cores of Random Graphs}

Following standard definitions, a $k$-core of an undirected graph $g$ is a maximal subgraph that includes fewer than $n$ nodes and such that all nodes
in the subgraph have degree of at least $k$ within the subgraph. When a nonempty $k$-core exists, then it must be that the $k$-core forms a convention for (absolute)
threshold $t=k$.

We define a {\sl weak} $k$-core to be a nonempty subgraph, including fewer than $n$ nodes, for which all nodes in the subgraph have degree at least $k$ within the subgraph and for which no single node could be added and have degree at least $k$. The set of weak $k$-cores are exactly the set of conventions (other than the whole set $N$) for threshold $t=k$.

A $k$-closed set is a nonempty subgraph, which has at least $k$ nodes and fewer than $n$ nodes, for which all nodes outside of the subgraph have fewer than $k$
connections to the nodes in the subgraph.

Every (weak) k-core is $k$-closed, but the converse is not true as being $k$-closed does not require that nodes in the subgraph have degree at least $k$.

We use the standard notation $G(n,p)$ to indicate an Erdos-Renyi random graph on $n$ nodes with a i.i.d. probability $p(n)$ of any link existing.

\begin{theorem}
\label{random3}
Consider a growing sequence of Erdos-Renyi random graphs $G(n,p)$.
\begin{itemize}
\item If $p(n)>(1+\varepsilon) \left(\frac{k\log(n)}{n}\right)^{1/k}$ for any $\varepsilon>0$, then the probability that a $k$-closed set exists goes to 0 (and thus so does the probability that there exist any weak $k$-cores or $k$-cores, or (tight) conventions that have $k=t$ and involve less than all nodes).
\item Conversely, if $p(n)<(1-\varepsilon) \left(\frac{k\log(n)}{n}\right)^{1/k}$ for any $\varepsilon>0$, then the probability that a $k$-closed set exists goes to one.
\end{itemize}
\end{theorem}
\

The term $\left(\frac{k\log(n)}{n}\right)^{1/k}$ is what is known as a {\sl sharp threshold} in random graph theory.

When $k=1$ it reduces to the threshold for connection in an Erdos-Renyi random graph.

Note that even though Theorem \ref{random3} is about Erdos-Renyi random graphs, and Theorem \ref{random2} is about stochastic block models, stochastic block models can be constructed by starting from an Erdos-Renyi random graph, and then adding additional links within blocks (and possibly across
some pairs of blocks).   Thus, Theorem \ref{random2} is a corollary.

Once $p(n)<(1-\varepsilon) \left(\frac{k\log(n)}{n}\right)^{1/k}$ the number of $k$ closed sets of sized $k$, as well as $k+1$...,  goes to infinity.   In that case, the existence of a (weak) $k$-core just requires that the probability is large enough so that out of that infinite sequence of such sets at least one forms a clique.  Thus, as long as the link probability does not drop so low that such cliques disappear, there will exist a (weak) $k$-core.

The proof technique that we use is based on showing that the probability of having a $k$ closed set of $k$ nodes at the threshold of $ \left(\frac{k\log(n)}{n}\right)^{1/k}$
can be bounded by the probability that there exists an isolated node at the threshold of $\frac{\log(n)}{n}$.   This is useful given that the behavior of this other event is well-known, while the first event is not and involves more intricate correlations.  These bounds turn out to be tight and so are useful in proving this theorem.
We have not seen this technique, of bounding the probability of one class of events in one random graph model by bounding it by the probability of a different class of events in a different random graph model, used before.

\medskip
\noindent{\bf Proof of Theorem \ref{random3}}:

First we prove that if $p(n)\geq (1+\varepsilon) \left(\frac{k\log(n)}{n}\right)^{1/k}$ then there is no $k$-closed set with probability going to 1.

We prove this at the threshold of $(1+\varepsilon) \left(\frac{k\log(n)}{n}\right)^{1/k}$ and thus it also holds for any larger $p$ since
{\sl not } having a  $k$-closed set is a monotone property (e.g., see \cite{bollobas2001}:  if a graph has the property then adding more links
maintains the property).   Take $\varepsilon>0$ to be small.   Since $k$ is fixed and $\varepsilon$ is arbitrary, we equivalently work with
$ \left(\frac{(1+\varepsilon)k\log(n)}{n}\right)^{1/k}$.

First, note that the probability that some node has fewer than $k$ connections to some set of  $k$ nodes is
\[
1-p(n)^k.
\]
The probability of having some set $B$ of cardinality $n_B\geq k$  be $k$-closed is at
most
\begin{equation}
\label{pnb}
\left[\left(1-p(n)^k\right)^{n-n_B}\right]^{\lfloor \frac{n_b}{k} \rfloor }.
\end{equation}
This is because a necessary condition for $B$ to be closed is having all of its sets of $k$ nodes closed to nodes outside of $B$, and there are at least
$\lfloor \frac{n_b}{k} \rfloor $ {\sl disjoint} sets of  $k$ nodes in $B$ and for which each of them being closed to nodes outside of $B$ is an independent event
So this only counts those disjoint sets being closed to nodes outside of $B$, and hence is a loose upper bound on the probability of $B$ being closed.\footnote{
The possibility of overlap in the sized-$k$ subsets induces correlation between the events that the subsets are closed to nodes outside of $B$ and produces a much
more complicated expression for the exact probability, but
this loose upper bound is easier to calculate and suffices for our proof.}

Next, note that the probability of having some set $B$ of cardinality $n_B\geq k$  be $1$-closed when the formation of a
link happens with probability $p'(n)$ is
\[
\left(\left(1-p'(n)\right)^{n_B}\right)^{n-n_B}.
\]

We rewrite this as
\begin{equation}
\label{pnb2}
\left[\left(\left(1-p'(n)\right)^k\right)^{n-n_B}\right]^{\frac{n_B}{k} }.
\end{equation}

We show that
when
 $p(n)=  \left(\frac{(1+\varepsilon) k \log(n)}{n}\right)^{1/k}$ and   $p'(n)=  (1+\varepsilon)\frac{\log(n)}{n}$
 then the expression in
 (\ref{pnb}) is less than the expression in  (\ref{pnb2}).

 It is enough to show that
 \begin{equation}
\label{pnb3}
\left(1-p(n)^k\right) < \left[\left(1-p'(n)\right)^k\right]^{\frac{n_B}{k} / \lfloor \frac{n_B}{k} \rfloor}.
\end{equation}
Noting that $\frac{n_B}{k} / \lfloor \frac{n_B}{k} \rfloor \leq 2$, we verify that
\[
1-p(n)^k < \left(1-p'(n)\right)^{2k}.
\]
At the designated values of $p(n)$ and $p'(n)$ this becomes
\[
1-(1+\varepsilon)\frac{k \log(n)}{n} <\left( 1- \frac{(1+\varepsilon)\log(n)}{n}\right)^2,
\]
for which it is enough  that
\[
\frac{k (1+\varepsilon)\log(n)}{n} >2\frac{(1+\varepsilon)\log(n)}{n}.
\]
This establishes that the probability that any set of at least $k$ nodes is $k$-closed under this $p(n)$
 is less than the corresponding probability that the same set is $1$ closed under this $p'(n)$.

 This implies that the expected number of sets of at least $k$ nodes that are $k$-closed under this $p(n)$ is less than
 the expected number of sets of nodes least $k$ nodes that are $1$-closed under this $p'(n)$.

 The second number is known to converge to 0   (see   \cite{bollobas2001}\footnote{Usual statements of
 the threshold of connectivity of $G(n,p)$ are in terms of the probability of connectedness.   Here we are using the stronger
 statement that the expected number of components is going to 0 above the threshold.  This can be pieced together from results in \cite{bollobas2001},
 but one can also find direct treatments, for instance, see Section 4.5.2 in \cite*{blumhk2016}.}  )
 Thus, the expected number of sets of at least $k$ nodes that are $k$-closed under this $p(n)$ converges to 0.
 This implies that the probability of having any such sets converges to 0, as claimed.

\medskip

To complete the proof, we show that
if
$p(n)=  \left(\frac{(1-\varepsilon) k \log(n)}{n}\right)^{1/k}$
then the probability of having at least one $k$-closed set of size $k$ goes to one.

The probability that any particular set of $k$ nodes is closed is
\[
(1-p^k)^{n-k}.
\]

Thus, the
expected number of closed sets of size $k$ is
\[
\sum_{B\subset N: \# B=k} (1-p^k)^{n-k} =
\binom{n}{k}(1-p^k)^{n-k}.
\]
At the threshold this is
\[
\binom{n}{k}
\left(1- \frac{(1-\varepsilon)k\log(n)}{n}\right)^{n-k}.
\]

For fixed $k$, this is of the order of
\[
n^k  n^{ - (1-\varepsilon) k}    = n^{k \varepsilon}\rightarrow \infty.
\]

Thus, the expected number of $k$-closed sets of $k$ nodes goes to infinity.

To complete the proof, we show that the variance of the number of $k$-closed sets of $k$ nodes
compared to the mean is bounded.   Once that is established, the fact that the expected number of  $k$-closed sets of $k$ nodes goes to infinity implies that
the probability that there exists at least one $k$-closed set of $k$ nodes exists goes to one can be proven using Chebychev's inequality by bounding the variance compared to the mean
(for omitted details, see this technique used in a proof on page 95 of \cite{jackson2008}).

Therefore, letting $X_{n,k}$ be the number of $k$ closed sets of $k$ nodes, we show that the variance of $X_{n,k}$ is a bounded multiple of $E[X_{n,k}]$.

First, note that we can write
\[
Var(X_{n,k})=
E[X_{n,k}(X_{n,k}-1)] + E[X_{n,k}] -  E[X_{n,k}]^2.
\]
Next, note that $E[X_{n,k}(X_{n,k}-1)]$ is the expected number of ordered pairs of $k$ closed sets of $k$ nodes,
and that
\[
E[X_{n,k}(X_{n,k}-1)] \leq  \binom{n}{k}  (1-p^k)^{n-k}  \sum_{k'=1}^{ k}\binom{n-k}{k'}(1-p^{k'})^{n-k-k'}
\]
Here $k'$ is the number of nodes of the second set of nodes that does not overlap with the first, and
$(1-p^{k'})^{n-k-k'}$ ignores any possible overlapping links between the two sets of $k$ nodes, and so is a loose upper bound on the probability that a second set is closed conditional upon the first one being closed.
Thus,
\[
Var(X_{n,k})\leq E[X_{n,k}] +
 \binom{n}{k} (1-p^k)^{n-k}\left(\sum_{k'=1}^{ k}\binom{n-k}{k'} (1-p^{k'})^{n-k-k'}\right)  - \binom{n}{k} ^2 (1-p^k)^{2n-2k}
 \]
and so
\[
Var(X_{n,k})\leq E[X_{n,k}] +
 E[X_{n,k}]   \left(  \sum_{k'=1}^{ k}\binom{n-k}{k'} (1-p^{k'})^{n-k-k'}   - \binom{n}{k}  (1-p^k)^{n-k}     \right) .
\]
This implies that\footnote{ Note that $\binom{n-k}{k'} (1-p^{k'})^{n-k-k'} < E[X_{n-k,k'}]$ since it is $p=  \left(\frac{(1-\varepsilon) k \log(n)}{n}\right)^{1/k}$ rather than $
 \left(\frac{(1-\varepsilon) k' \log(n-k)}{n-k}\right)^{1/k'}$ that is
used in the expression above, and that for large $n$, small $\varepsilon$, and $k'<k$,  $ \left(\frac{(1-\varepsilon) k' \log(n-k)}{n-k}\right)^{1/k'}< \left(\frac{(1-\varepsilon) k \log(n)}{n}\right)^{1/k}$ .}
\[
Var(X_{n,k})\leq E[X_{n,k}] +
 E[X_{n,k}]   \left( \sum_{k'=1}^{ k} E[X_{n-k,k'} ]- E[X_{n,k}]\right).
\]
Along the lines of our proof above, $\sum_{k'=1}^{ k} E[X_{n-k,k'} ]- E[X_{n,k}]$ is of the order of
\[
\sum_{k'=1}^{ k} (n-k)^{k'\varepsilon} - n^{k\varepsilon},
\]
and
\[
\sum_{k'=1}^{ k} (n-k)^{k'\varepsilon} - n^{k\varepsilon} \leq \frac{(n-k)^{k\varepsilon}}{1-\frac{1}{(n-k)^{\varepsilon}}} - n^{k\varepsilon} =
(n-k)^{k\varepsilon} \left(\frac{(n-k)^{k\varepsilon}}{(n-k)^{\varepsilon}-1} - \frac{n^{k\varepsilon}}{(n-k)^{\varepsilon}}\right) <0,
\]
which completes the proof.\eproof

\subsection{Testing Relative vs Absolute Thresholds}\label{relabs}

We  next use the statistic ${T}(\cdot)$ defined above to evaluate whether a relative, $q$, or absolute, $t$, threshold model better fits the observed adoption decisions.\footnote{Previous work on testing for different models of peer influence have looked at whether it is information or influence (e.g., \cite*{banerjeecdj2013}) or whether it is a simple or complex contagion (e.g., $t=1$ versus $t>1$ \cite{centola2011,monsted2017}).  Here we are testing different types of influence against each other.}

If every person has the same degree, then the two models are interchangeable -- every $q$ has an equivalent $t$.   However,  networks in which degree varies across people, then the two models diverge.  In the relative model a higher degree person requires more friends taking the action than a lower degree person in order to want to take the action.  In the absolute threshold model, the number of friends needed to induce a person to take the action is independent of the degree.
This is the key to identifying which model better fits the data.

To illustrate how the two models lead to different predictions and can be distinguished, we apply the procedure to data on smoking decisions in a high school social network from the Add Health data set.\footnote{For a more detailed analysis of smoking and friendships in networks, see \cite{badev2017}.}

\begin{figure}[h!]
\centering
\subfloat[Whole High School]{
\includegraphics[width=0.4\textwidth]{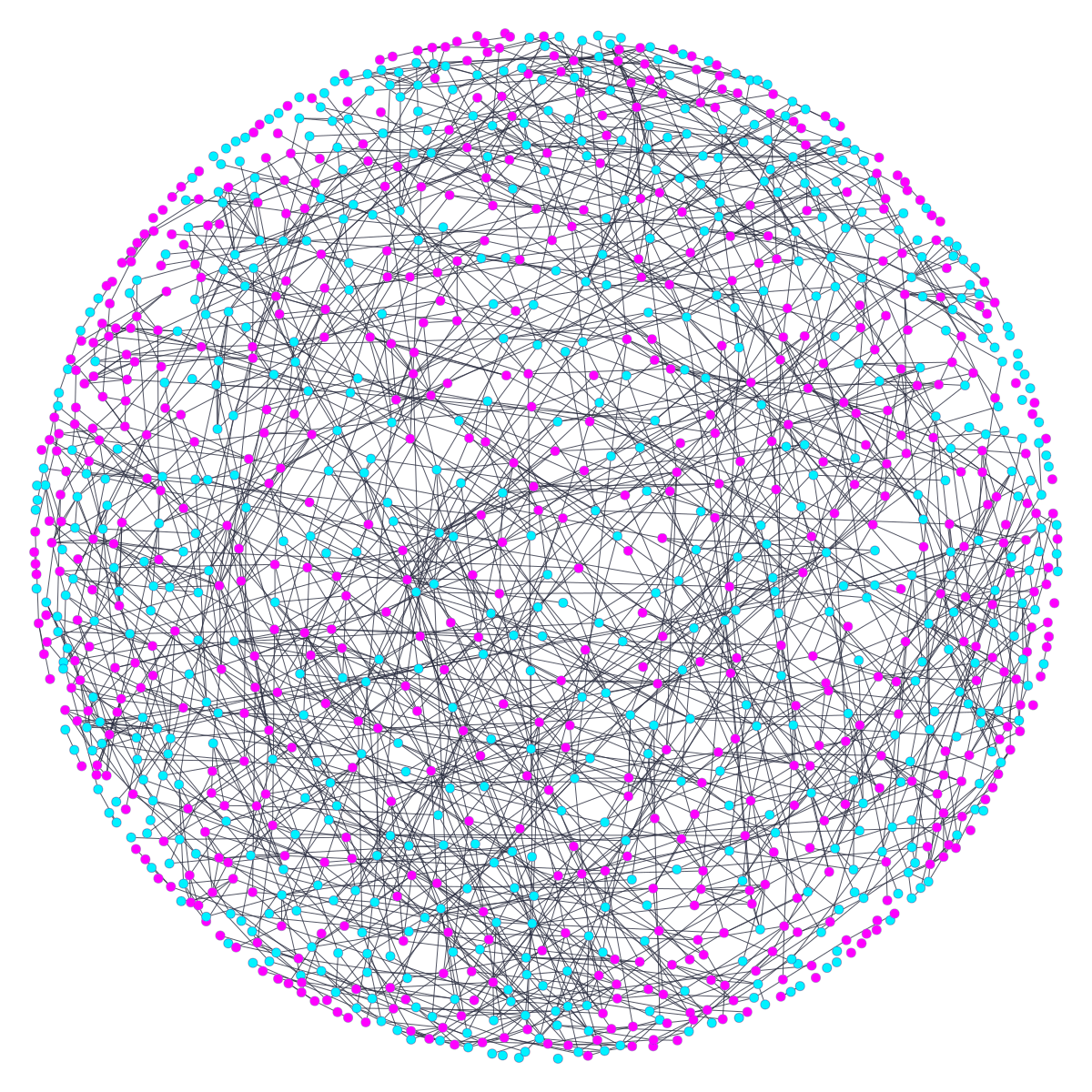}
}
\includegraphics[width=0.02\textwidth]{blank.jpg}
\subfloat[12th grade networks among boys]{
\includegraphics[width=0.4\textwidth]{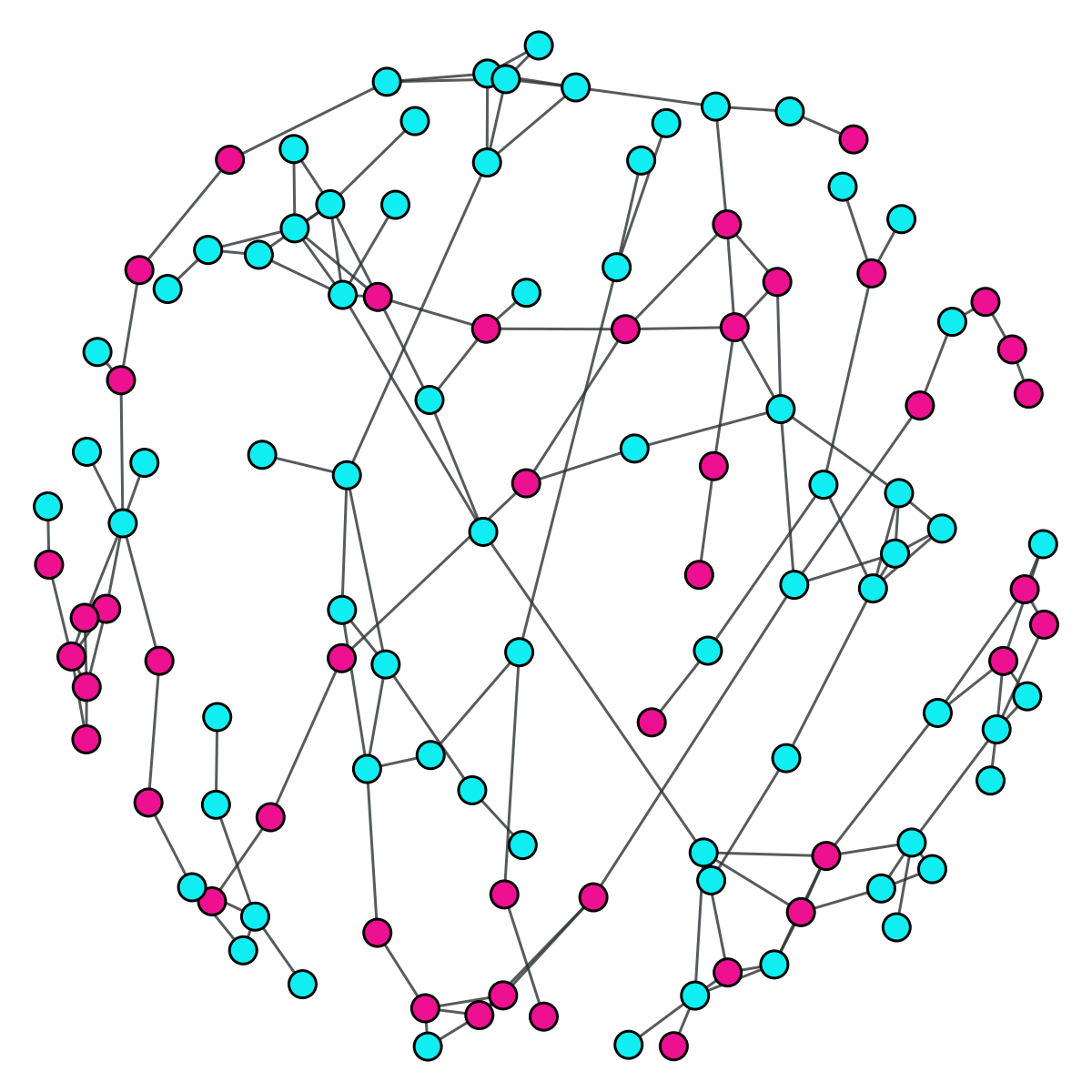}
}
\caption{\label{smokes} Smoking Adoption in a high school social network.  In (b) we see a selection of the smallest group, where the network patterns become clearer and one can see that smokers have more smoking friends, and non-smokers have more non-smoking friends. }
\end{figure}

Figure \ref{smokes} pictures smoking in a high school social network: pink nodes represent students who said that they had smoked a cigarette in the past twelve months,
while blue nodes represents student who had not.
The excerpt of a subset of students makes the patterns of adjacency among smokers and among non-smokers clearer.

We can compare how well the best-fitting absolute and relative thresholds divide the distributions of on-neighbor shares and numbers, respectively.

\begin{figure}[h!]
\centering
\subfloat[Best relative threshold split]{
\includegraphics[width=0.4\textwidth]{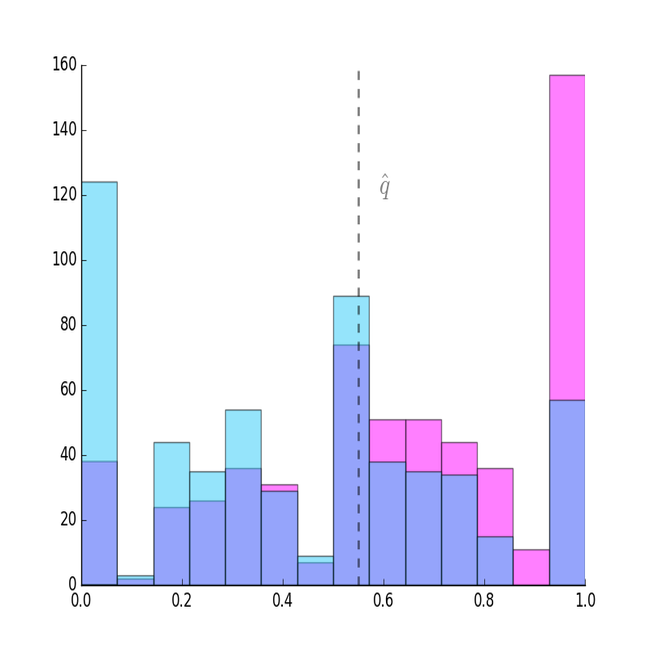}
}
\includegraphics[width=0.02\textwidth]{blank.jpg}
\subfloat[Best absolute threshold split]{
\includegraphics[width=0.4\textwidth]{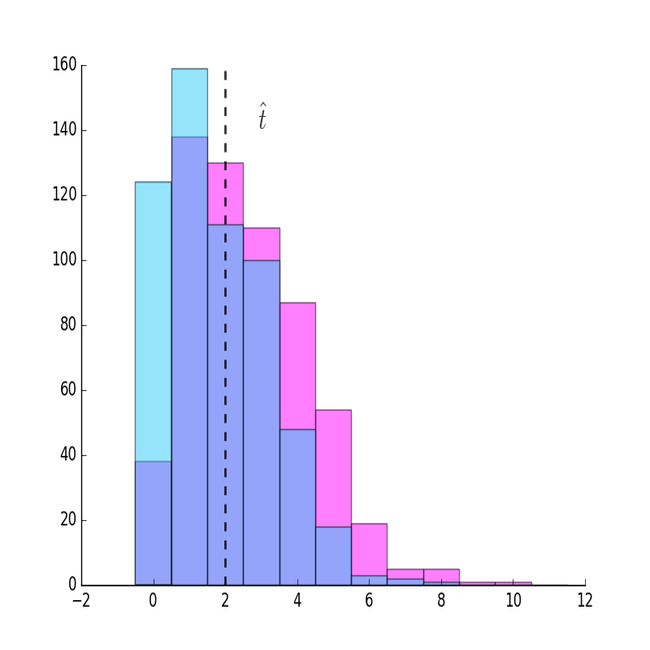}
}

\caption{\label{addhealth_splits}
{\bf } A comparison of how well a relative vs. an absolute threshold splits the on-neighbor distributions
}
\end{figure}

The best-fit estimate of a relative threshold is $\hat{q} =0.55$, which gives  ${T}(\hat{q})/n=0.36$ - so 36 percent of the students acting in `error'; and
the best-fit absolute threshold is$\hat{t}=2$, which gives $\hat{T}(\hat{t})/n=0.4$.

In this network there are $n=1221$ students.  The relative threshold $\hat{q} =0.55$ results in 438 students who are not acting according to the predicted action based on the best fitting relative threshold, while the absolute threshold $\hat{t}=2$ has 486 students who are not acting according to the predicted action based on the best fitting absolute threshold - or an extra 48 students, which is four percent of the population.
The relative threshold model thus better predicts behaviors than the absolute threshold.

Distinguishing the two models statistically can be thought of as follows.    Letting $z$ be the number of students who are
not acting in accordance with the threshold, the chance of matching the observed behaviors is\footnote{ $1-\epsilon + \frac{\epsilon}{2}= 1-\frac{\epsilon}{2}$ is the chance that an agent acting in accordance with the threshold's behavior is correct (the agent acts according to the model, or is random and happens to pick the right action with probability one half), and then $\frac{\epsilon}{2}$ is the chance that an agent who is not acting in accordance with the threshold is correctly matched.}
\[
\left(1-\frac{\epsilon}{2}\right)^{n-z} \left(\frac{\epsilon}{2}\right)^{z} =   \left(\frac{2-\epsilon}{2}\right)^{n} \left(\frac{\epsilon}{2-\epsilon}\right)^{z}.
\]

This probability decreases exponentially in the number of errors $z$.  Thus, the likelihood is exponentially higher, by a factor
$\left(\frac{2-\epsilon}{\epsilon}\right)^{z-z'}$ when the number of errors is $z'<z$.    When $z-z'=48$, then even for $\epsilon$ that is very high, the likelihood is much higher under $q$ than $t$.    Then a log likelihood ratio test leads to a $p$-value of effectively zero.

If one did not use either model, then the best one could do would be to predict everyone to be a smoker.
That would lead to an error rate of $0.49$ (the fraction of non-smokers) or 600 students.    That is 162 worse than the relative threshold and 114 worse than the absolute model.   Again,  a log likelihood ratio test leads to a $p$-value of effectively zero.

The fact that these strict $q$ and $t$ models are still mis-predicting more than a third of the students' actions is partly due to the fact that we are only using the network to predict actions and not including information from demographics.  It is natural to expect the payoff to a behavior and coordinating with one's friends to
depend on demographic characteristics.     If we allow the threshold to vary with demographics (e.g., race, income, gender, etc.), then one can significantly increase the predictiveness of the model.

The technique we have outlined extends directly to allow for demographics, as follows.    If $X$ is the demographic information, then one can fit a function $q(X)$ to predict peoples thresholds and behaviors and then use
$$T(q(\cdot)) =  |N_{on} \cap \lbrace i: s_i < q(X_i) \rbrace | + | N_{off} \cap \lbrace I: s_i\geq q(X_i) \rbrace |,$$
as the objective and select $q(\cdot)$ to minimize the function (and similarly for an absolute threshold function $t(\cdot)$).

\begin{figure}[h!]
\centering
\subfloat[Grade 9]{
\includegraphics[width=0.4\textwidth]{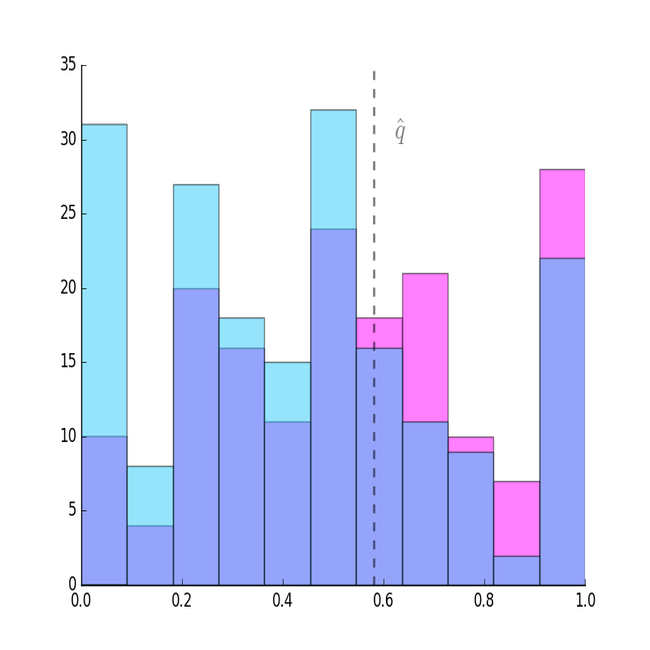}
}
\includegraphics[width=0.02\textwidth]{blank.jpg}
\subfloat[Grade 10]{
\includegraphics[width=0.4\textwidth]{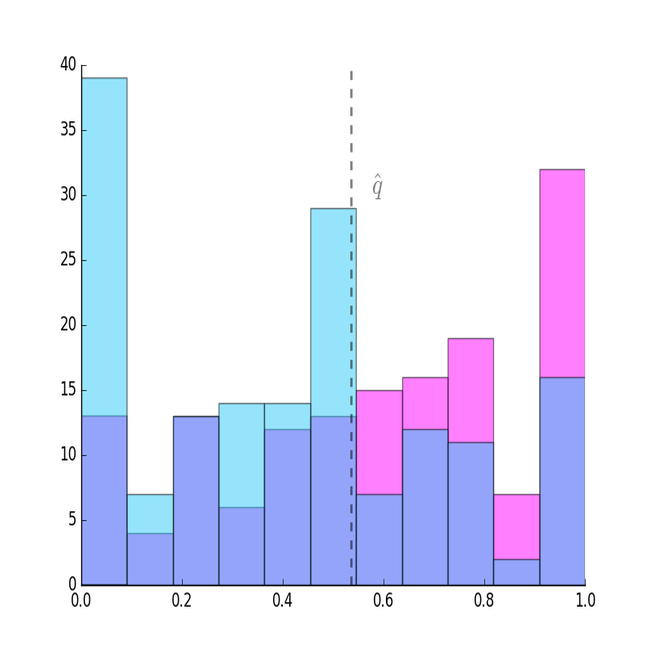}
}
\includegraphics[width=0.02\textwidth]{blank.jpg}
\subfloat[Grade 11]{
\includegraphics[width=0.4\textwidth]{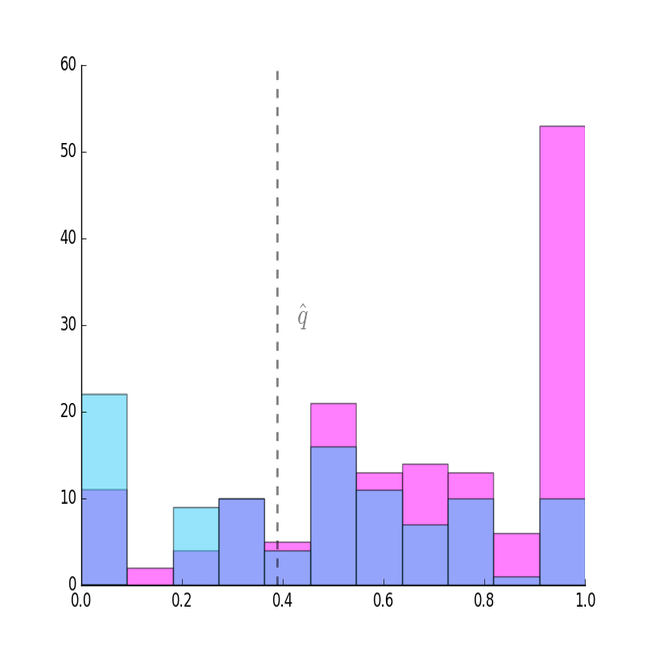}
}
\includegraphics[width=0.02\textwidth]{blank.jpg}
\subfloat[Grade 12]{
\includegraphics[width=0.4\textwidth]{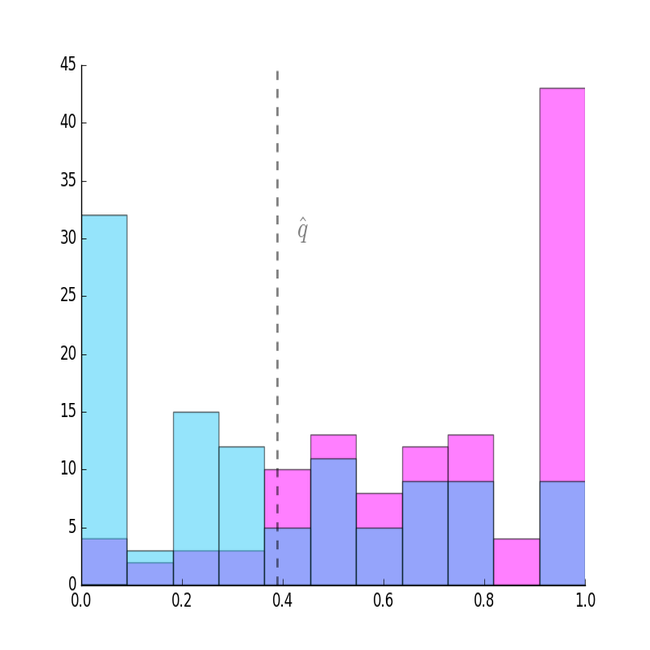}
}
\caption{\label{allboys_q_by_grade}
{\bf :} Estimating $\hat{q}$ for each of the grades 9,10,11, and 12 separately.
}
\end{figure}

To illustrate, we estimate $q(X)$ for the high school smoking example, where we take $X$ to be a node's year in school. Figure \ref{allboys_q_by_grade} below shows the on-node neighbor shares for smokers vs. non-smokers for each grade separately.  (We still include students in other years when calculating the share of a node's neighbors who smoke.)

We estimate $\hat{q}$ values of 0.58, 0.54, 0.39, and 0.39 for grade 9, 10, 11, and 12. So in this example, younger students are less easily influenced to smoke by the smoking behaviors of their friends.
Using the grade specific thresholds, the share of agents of any gender whose behavior the relative threshold model mis-predicts falls marginally, from $0.36$ under the universal threshold above to $0.34$ under the grade-specific thresholds, which corresponds to mis-predicting 24 fewer students -- which again leads to an exponential increase in likelihood, but is less dramatic.    Still, it has a log-likelihood improvement $p$ value of effectively 0.\footnote{The test statistic is
$2\times ln( \left(\frac{2-\epsilon}{\epsilon}\right)^{24} )$, which for values of $\epsilon$ all the way up to being well above $2/3$, leads to a statistic above a $\chi^2$ at a .999 level with four degrees of freedom.}
There may be additional demographics that would lead to additional improvements in fit, the point here is simply to illustrate the potential approach.

In the Appendix, Section \ref{testqs}, we describe two other methods of estimating $q$ from an observed network.
In the above approach, the heterogeneity or error terms are introduced via the ($\epsilon$) probability that
agents choose behavior independently of their neighbors.   In the appendix the two other approaches are to introduce other sorts of noise:  either mismeasurement of the network, or individual person-by-person variations in the thresholds.

In the online appendix (Section \ref{testg}), we  also  discuss another extension of the above technique.
In settings where multiple networks are observed, for instance borrowing and lending as well as kinship as well as pure socializing (e.g., \cite*{banerjeecdj2013,banerjeecdj2015,banerjeecdj2016}),
it might be that certain networks are important in determining certain behaviors.  For instance, in determining which statistical software to use a person might
coordinate with his or her co-workers, while deciding on a game or social app he or she might look to family and friends.
By observing the behaviors, one can then look to see which combination of network and threshold results in the fewest errors in predicted behaviors.

\section{Computability of the Atomic Partition}
\label{Compute}

Finding a $Q$-convention that separates two nodes in an arbitrary network is a combinatorial set problem that, because of the complementarity of nodes' influence in spreading behavior, cannot be reduced to a submodular maximization problem. It is unsurprising, then, that the problem is NP hard---indeed, even just determining whether a network has a proper $Q$ convention (one other than the entire network) is NP hard. We therefore cannot hope for a method that identifies all of the $Q$ atoms for an arbitrary network in polynomial time unless $P=NP$.

In this section, we present two approaches to this challenge. First, we describe a method for reducing the atom-finding problem to an integer linear program (ILP), which is known to perform well in many NP-hard problems on real-world data. Second, we discuss a local-richness condition on a network that arises naturally in many settings of interest and show that under this condition the $Q$ atoms can be determined efficiently.

In addition, we use a variety of real-world networks to demonstrate that standard ILP solvers are able to solve the associated ILP and that the real-world networks exhibit our local-richness condition.

\subsection{The $i,j \: Q-SPLIT$ Problem}

Given a network $(N,g)$, two nodes $i\neq j \in N$, and a compact interval of thresholds $Q=[\underbar{q},\overline{q}]$, we term the problem of determining whether there exists a $Q$-convention in $g$ which contains exactly one of $i$ and $j$ the $i,j \: Q-SPLIT$ problem, and refer to such a convention as a \textbf{separating} $Q$-convention for $i$ and $j$.
Note that finding the $Q$-atomic structure of $g$ is equivalent to solving $i,j \: Q-SPLIT$ for each ordered pair of nodes $i,j$.
Since there are $(n^2-n)/2$ such pairs, the problem of finding the $Q$-atomic structure is efficiently reducible to $i,j \: Q-SPLIT$.

\subsection{ILP Formulation}

The $i,j \: Q-SPLIT$ problem admits a natural representation as an ILP.  Given a network $g$ with associated adjacency matrix $G$, let $\widetilde{G}$ be the degree-regularized adjacency matrix obtained by dividing each row by the degree of the corresponding node. For $S \subseteq V$, let $x_S \in \{ 0,1 \}^n$ be a binary vector with $x_S(i)=1$ if node $i$ is in $S$.  Then $i,j \: Q-SPLIT$ is exactly the problem of finding a feasible solution $x_S$ to the following system of equations:
\begin{empheq}[box = {\Garybox[ILP Formulation]}]{align*}
x_S(i)+x_S(j)&={\bf 1} \\
(\widetilde{G}-\overline{q}I)x_S &\geq 0 \\
(\widetilde{G} - (1-\underbar{q})I)x_{S^c} &\geq 0.
\end{empheq}
The first condition requires that the subset $S$ contain exactly one of $i$ and $j$.  The second condition requires that $S$ be $\overline{q}$-cohesive.  The third condition requires that $S$ be $\underbar{q}$-closed, expressed in the equivalent formulation that the complement of $S$ be $1-\underbar{q}$-cohesive (where $\underbar{q}$ is a rational number, $\underbar{q}$ can be replaced with $\underbar{q}-\eta$ for $\eta<1/n$ to deal with potential ties).

Although finding the solution to an ILP is NP-hard, state-of-the-art ILP engines have consistently found feasible (and near-optimal) solutions in many real-world applications in runtimes that are orders of magnitude below the theoretical worst-case. Moreover, since each solution to the $i,j \: Q-SPLIT$ problem separates many pairs of nodes, in practice we can determine the $Q$-atomic partition by solving far fewer than $n^2-n$ instances of $i,j \: Q-SPLIT$. At the end of this section we demonstrate that with a standard ILP solver we can correctly identify the $Q$-atoms in the
Indian Village, Add Health, and Caltech networks in times that grow linearly with the number of nodes.

Before showing the efficiency in the data, we describe a condition that guarantees that an ILP works efficiently.

\subsection{$k$-seedable Conventions}

Social networks exhibit clustering and sparsity, both of which help promote the existence of conventions that grow from a set of seeds much smaller than the total number of nodes. If the collection of such conventions is rich enough to generate the atomic structure of the network, then computing the atomic structure becomes tractable.

Formally, we say that a $Q$-convention $C$ is \textbf{generated} by a set of nodes $S \subset C$ if it is
the one that is obtained the process we described for our seeding analysis in Section \ref{seeding} when using threshold $\underbar{q}$.  Note that since it is a $Q$-convention, it remains a convention for $\overline{q}$.
A $Q$-convention $C$ is \textbf{$k$-seedable} if $C$ is generated by some $S\subset C$ with $|S|\leq k$.

Our local structural richness condition requires that {$k$-seedable} conventions generate the $Q$-atomic structure of $g$:  The $Q$-atomic structure of $g$ is \textbf{$k$-generatable} if, whenever there exists an $i,j$-separating $Q$-convention, there exists a $k$-seedable $i,j$-separating $Q$-convention.  Note that this does not require that $g$ have \textit{only} $k$-seedable conventions, but instead that
if two nodes can be separated by a $Q$-convention, then they can be separated by some $k$-seedable $Q$-convention.

Since there are polynomially many subsets of size less than or equal to a given $k$, and since the convention generated by a particular subset can be computed in $O(n^2)$ time, $k$-generatability of the $Q$-atomic structure is a sufficient condition for efficient computation.  This is formally expressed as follows:

\begin{claim}
In the class of networks with $k$-generatable $Q$-atomic structures for $k \geq 4$, $i,j\: Q-SPLIT$ can be solved in $O(n^{k+4})$ time, and the $Q$ atomic structure of a network can be determined in $O(n^{k+6})$ time.
\end{claim}

To see the proof of this claim, first observe that the $\underbar{q}$-closure of a subset $S$ can be computed by multiplying $x_S$ by $\widetilde{G}$ to determine which nodes have a share of at least $\underbar{q}$ of their neighbors in $S$, replacing $x_S$ by $x_S'$ where $S'=\{i \in N| i\in S \: \: \text{or } \widetilde{G}x_S(i)\geq \underbar{q} \}$, and iterating until the set of nodes does
not change.  This process terminates in at most $n$ steps, and since matrix multiplication can be computed in $O(n^3)$ time, the closure can be obtained in $O(n^4)$ time.
Finding the convention generated by a subset involves two closure operations, and so can also be completed in $O(n^4)$ time.
Next, note that for any nodes $i,j$ of $g$, if the $Q$ atomic structure of $G$ is $k$-generatable, then there exists a subset $S$ of size at most $k$ generating a $Q$-convention that separates and separable pair of nodes $i$ and $j$.  There are $O(n^k)$ such subsets, and so determining whether there exists an $i,j$-separating $Q$ convention can be performed in $O(n^{k+4})$ time.
Finally, solving $i,j \: Q-SPLIT$ for each of the $(n^2-n)/2$ pairs of nodes determines the $Q$-atomic partition of $g$, and so this partitioning can be determined in $O(n^{k+6})$ time.

The property of a network having a $k$-generatable $Q$-atomic partition is an intuitive and natural one, but clearly
determining whether a network $g$ has a $k$-generatable $Q$-atomic partition is necessarily NP-hard.  Providing sufficient conditions on a network structure for this to hold is something that we leave for future research.
Still, we can verify that this property is satisfied on several data sets that have been used extensively in the literature
and have nontrivial sizes to the network.
In particular, our next section provides evidence that in many real-world social networks, the $Q$-atomic structure is $k$-generatable for $k$ orders of magnitude smaller than the number of nodes, so that $k$-generatability for small $k$ may be reasonable a-priori in at least applications using common data sets.

\subsection{Real-World Testing}

The local richness of social networks and the power of modern ILP solvers suggest a practical two-stage approach for determining the $Q$-atomic structure of a network: first generate the maximal $\overline{q}$-cohesive subset of the $\underline{q}$-closure of each subset of size up to some value $k$, then store the resulting conventions as initial solutions for the ILP-solver to use in solving $Q$-split for each pair of nodes in the second stage. The details of this approach are covered further below.

Here we report the run-time and accuracy of our method for three network datasets typical of the empirical networks literature: the AdHealth high school social networks, the Indian village social reciprocity networks, and the Caltech undergrad social networks. Runs were carried out on a 64-core, 256 GB RAM server. Accuracy is measured in terms of the share of networks for which our algorithm completely correctly identifies the atomic structure, as tested against brute force search guaranteed to find the correct atoms. We note that even when the atomic structure is miscomputed, the difference between the computed and actual atomic partitions, as measured by the number of additional conventions needed to generate the correct partition, is at most three additional conventions in all but two cases.

\begin{table}[]
\begin{tabular}{llllll}
                                                             &                                                                     & \multicolumn{3}{c}{\textbf{Dataset}}                                                         &                           \\
                                                             &                                                                     & Indian Villages               & AdHealth                      & CalTech                      &                           \\ \cline{2-5}
\multicolumn{1}{l|}{\multirow{12}{*}{\textbf{Network Size}}} & \multicolumn{1}{l|}{\multirow{4}{*}{50\textless{}n\textless{}175}}  & \multicolumn{1}{l|}{1min 13s} & \multicolumn{1}{l|}{1min24s}  & \multicolumn{1}{l|}{-}       & Avg. Runtime              \\ \cline{3-5}
\multicolumn{1}{l|}{}                                        & \multicolumn{1}{l|}{}                                               & \multicolumn{1}{l|}{5}        & \multicolumn{1}{l|}{6}        & \multicolumn{1}{l|}{-}       & Avg. k for k-generability \\ \cline{3-5}
\multicolumn{1}{l|}{}                                        & \multicolumn{1}{l|}{}                                               & \multicolumn{1}{l|}{1}        & \multicolumn{1}{l|}{1}        & \multicolumn{1}{l|}{N/A}     & Accuracy Rate             \\ \cline{3-5}
\multicolumn{1}{l|}{}                                        & \multicolumn{1}{l|}{}                                               & \multicolumn{1}{l|}{21}       & \multicolumn{1}{l|}{20}       & \multicolumn{1}{l|}{0}       & Sample Size               \\ \cline{2-5}
\multicolumn{1}{l|}{}                                        & \multicolumn{1}{l|}{\multirow{4}{*}{175\textless{}n\textless{}275}} & \multicolumn{1}{l|}{2min13s}  & \multicolumn{1}{l|}{3min 8s}  & \multicolumn{1}{l|}{2min22s} & Avg. Runtime              \\ \cline{3-5}
\multicolumn{1}{l|}{}                                        & \multicolumn{1}{l|}{}                                               & \multicolumn{1}{l|}{8}        & \multicolumn{1}{l|}{7}        & \multicolumn{1}{l|}{7}       & Avg. k for k-generability \\ \cline{3-5}
\multicolumn{1}{l|}{}                                        & \multicolumn{1}{l|}{}                                               & \multicolumn{1}{l|}{0.95}     & \multicolumn{1}{l|}{0.93}     & \multicolumn{1}{l|}{1}       & Accuracy Rate             \\ \cline{3-5}
\multicolumn{1}{l|}{}                                        & \multicolumn{1}{l|}{}                                               & \multicolumn{1}{l|}{38}       & \multicolumn{1}{l|}{15}       & \multicolumn{1}{l|}{2}       & Sample Size               \\ \cline{2-5}
\multicolumn{1}{l|}{}                                        & \multicolumn{1}{l|}{\multirow{4}{*}{275\textless{}n\textless{}500}} & \multicolumn{1}{l|}{7min 12s} & \multicolumn{1}{l|}{14min12s} & \multicolumn{1}{l|}{-}       & Avg. Runtime              \\ \cline{3-5}
\multicolumn{1}{l|}{}                                        & \multicolumn{1}{l|}{}                                               & \multicolumn{1}{l|}{9}        & \multicolumn{1}{l|}{10}       & \multicolumn{1}{l|}{-}       & Avg. k for k-generability \\ \cline{3-5}
\multicolumn{1}{l|}{}                                        & \multicolumn{1}{l|}{}                                               & \multicolumn{1}{l|}{0.92}     & \multicolumn{1}{l|}{0.91}     & \multicolumn{1}{l|}{N/A}     & Accuracy Rate             \\ \cline{3-5}
\multicolumn{1}{l|}{}                                        & \multicolumn{1}{l|}{}                                               & \multicolumn{1}{l|}{16}       & \multicolumn{1}{l|}{40}       & \multicolumn{1}{l|}{0}       & Sample Size               \\ \cline{2-5}
\end{tabular}
\end{table}
\newpage

\section{Finding a combination of networks and $q$ from Observed Behaviors and Conditioning on Nodal Characteristics }\label{testg}

A straightforward extension of the technique of estimating $q$ to allow for multiplexed networks, is to not only consider which $q$ best rationalizes
behavior, but also which network out of some set of possible networks does.  For instance, if one is considering students' decisions to adopt some software, and one had lists of both their friends and their study partners, one could see which of the two networks combined with some $q$ best predicts the behavior.
The extension is straightforward, as we simply define $s_i(g)$ to be the fraction of $i$'s neighbors who have adopted the behavior if we consider the network to be $g$.
We can also allow the $q$ to depend on some vector of (finite range) characteristics $X$, so that we allow $q(X_i)$ to be node $i$'s threshold as a function of $i$'s characteristics.
Then we can define the total number of errors as a function of the network and threshold to be:
$$T(g,q(\cdot) ) =  |N_{on} \cap \lbrace i: s_i(g) < q(X_i) \rbrace | + | N_{off} \cap \lbrace I: s_i(g)\geq q(X_i) \rbrace |.$$
Then we can minimize $T$ across both $g$ and $q(\cdot)$ rather than just some fixed $q$.

\section{Alternative Methods of Estimating $q$ from Observed Behaviors }\label{testqs}

\subsubsection{Estimating $q$ by Minimizing the Number of Additional Links Needed to Rationalize Behavior}

\

The preceding approach estimates the behavioral threshold $q$ by minimizing the number of agents whose behavior cannot be rationalized by a given threshold $q(\cdot)$ (again letting this be characteristic dependent.  This presumes that there either behaviors are seen in error, or else that agents randomly choose a different behavior with some probability.
A different perspective is to think of the network being observed with errors.
Given that the vast majority of real-world network data feature measurement error, another
reasonable (corresponding) approach is to estimate $q(\cdot)$ by minimizing the number of links that need to
be changed in order to rationalize the observed behavior.

For some observed set of agents $S\subset N$ adopting the behavior,
let $G(S,q(\cdot))$ be the set of all networks for which
$S$ is a convention relative to $q(\cdot)$.   Then the estimate(s) of $q(\cdot)$ are
those in
\[
\argmin_{q}   \left[ \min_{g'\in G(S,q(\cdot))} |g'-g| \right]
\]
where $|g'-g|$ is the distance between $g'$ and $g$ in terms of the size of the symmetric difference of their lists of links.

If we apply this approach to the high school smoking network of
Figure \ref{smokes}, we find $\hat{q}=0.49$, corresponding to a change
in $196$ links required, or 8 percent of the 2,443 links in the data.

In some applications measurement error comes in the form of omitted links rather than including links that are not present.   For instance, people may forget to name others in a survey, or may tire from the survey, and measured social networks via emails, social media, phone calls, or other data may not detect all relationships that are present.   Thus, a variation on this method is instead to only consider who many
links need to be added to rationalize a given $q$ and then to pick the $q$ which requires adding the fewest links.
If we do that on this data set, then
we find $\hat{q}=0.51$, corresponding to an additional $468$ links required, or 19 percent more links than the 2,443 in the data.\footnote{The algorithm for doing this analysis is quite straightforward.  For any given $q$, one can look at each individual who has more than $q$ of their friends taking the action and who is not taking it themselves, or has fewer than $q$ of their friends taking the action but is taking the action.
For each such individual, one can count how many links need to be added to non-adopters or to adopters, respectively, to rationalize that person's behavior.
The total number of links that would then need to be added for such a $q$ is then half as many as adding this number up across all agents (modulo the number being odd for one or both group).
In some cases with a very asymmetric convention or with an extreme $q$, it might be that there would not be enough people in one of these groups to add all of the links required, in which case there would be no way to rationalize the observed behavior simply by adding links.  In that case, one could discard such $q$'s.  If no $q$ could be rationalized, then one would need to turn to a different technique to obtain an estimate.
This could only happen when even forming a complete clique among all adopters and among all non-adopters could not rationalize the behavior, which would seem to be a rare case.   }

\subsection{Estimating Heterogeneous $q$'s}

The above two approaches are built on counting errors in terms of the number of individual behaviors that do not conform, or in terms of how different a network would have to be to rationalize the observed behavior.

A third approach is to suppose that the individual $q_i$'s are heterogeneous, distributed about some population mean threshold $q$. Under this assumption, a natural estimator $\hat{q}$ of $q$ is the value which minimizes the degree of heterogeneity needed to rationalize the agents' behavior, that is, the variance in the node-wise differences $q_i-q$ where $q_i$ is the closest value to $q$ which rationalizes agent $i$'s behavior.\footnote{Yet another approach would be to look at a set of conventions of behaviors, and then find the $q$ whose atoms best fit those conventions.  Best fitting would involve penalties for splitting atoms, and one can also penalize for the number of atoms.}
Specifically, for a given value of $q$, this nearest fitted $q_i$, call it $\hat{q}_i(q)$, is given by:
$$\hat{q_i}(q)=\twopartdef{\min(q,\frac{|N_{on} \cap N_i|}{|N_i]})}{a_i=1}{\max(q,\frac{|N_{on} \cap N_i|}{|N_i]})}{a_i=0}, $$
and we take $\hat{q}=\min_{\tilde{q}} \sum_i (\tilde{q} - \hat{q}_i(\tilde{q}))^2$.
When we apply this approach to the high school smoking network above, we get $\hat{q}=0.51$, close to both of the estimates above.

An obvious extension of this would be to allow the $q$ and $\widehat{q}_i$ to depend on observable (or estimated latent) characteristics, so have a different one for twelfth grade males, a different one for twelfth grade females, and for eleventh grade boys, etc.

\section{Distinguishing Homophily from Coordination}


When estimating a threshold from an observed network and convention, there is an alternative hypothesis that could also potentially explain the data: homophily.
This is a well-known confound of correlated behaviors (e.g., see \cite*{aralms2009,shalizi2011homophily,jacksonrz2017}).
People with similar backgrounds and tastes tend to be linked to each other.   Thus, there could be strong correlation patterns in neighbors' behaviors simply because they have the same tastes without any influence from one to another.
With controlled experiments (e.g., random assignment of roommates as in \cite{sacerdote2001}) or instruments under some conditions (\cite*{bramoulledf2009}, \cite{araln2017}) one can test for influence directly as we discussed in the paper.
In the absence of such identification, faced with just observations of networks and behavior, it is more challenging to distinguish homophily from influence.

To fix terms, we use `homophily' to describe a world in which the probability that two agents are linked to each other depends on some distance function of a vector of their characteristics and their adoption decisions are a stochastic function of that vector, but in which the adoption decision does not depend on the realized adoption decisions' of one's neighbors.  In contrast, by `coordination', we mean a model in which each agent in a network (one possibly formed under the influence of homophily) makes his adoption decision as a function only of whether or not the share of his neighbors adopting the behavior is above some common threshold $q$.   Of course, there could be both homophily and coordination, but to fix ideas we discuss testing for one versus the other.

Admitting homophily in our null hypothesis yields a more demanding test of coordination, since homophily will generate positive coordination between neighbors' adoption decisions just as coordination does. Testing against homophily thus requires a more subtle analysis of the structure that correlation takes.

It is impossible to entirely rule arbitrary models of homophily.  For instance, suppose we observe a convention $S$ that matches perfectly with some threshold $q$.
Another model that could explain the data is that people who adopted are those who prefer to adopt, and the other people preferred not to adopt.
The associated `pure homophily' model of network formation is that
all people who prefer to adopt also prefer to form links in such a way as to end up with a network in which a fraction of at least $q$ (respectively, more than $t$) of their friends have the same preference; and conversely, all
those who prefer not to adopt want to be in a network in which less than a fraction $q$ (respectively, fewer than $t$) of their friends prefer to adopt.  Beyond that, agents have no preferences over network structure.

This leads to the following straightforward observation, which is a variation on a result of \cite{shalizi2011homophily} in our setting:
\begin{claim}
Given any convention $S$ associated with some network $g$ and adoption threshold $q$ (or $t$), there is a homophily model of network formation with no complementarities in behavior that generates
the same data.
\end{claim}

The claim means that ironclad testing for causation of behavior requires controlled experiments or valid instruments.
Of course, the homophily model underlying the extreme version of the claim above is pathological.
If one is willing to restrict to a class of models of homophily and network formation, clearly a strong assumption, then
one can distinguish coordination behavior from homophily from an observed convention.
The logic is as follows.

Consider a model of homophily-driven behavior of the following form.   Nodes have some characteristics, denoted by $\theta_i$ that lie in some
(possibly multi-dimensional) Euclidean space, that affect $i$'s behavior and
also the probability that two nodes $i,j$ are linked.    Label the characteristics so that the probability of adoption is an increasing function of $\theta_i$.
The probability that $i,j$ are linked is strictly decreasing in some measure of the distance between $\theta_i $ and $\theta_j$,
holding the rest of the network fixed.\footnote{This does not preclude that other considerations affect network formation - such as indirect connections.
Working on the margin is all that is necessary for our suggested approach.}

This model of homophily is empirically distinguishable from behavior based on coordination according to some $q$.
Under this model of homophily, nodes' adoption decisions are correlated with their neighbors' adoption decisions {\sl conditional on a share of at least $q$ of those neighbors adopting} for any choice of $q$.
This follows since increasing the number of neighbors adopting increases the distribution that one has over neighbors' $\theta_j$'s.   This increased inference leads one to expect higher values of $\theta_i$, which then implies a higher expected behavior.
Thus, the behavior of linked individuals is correlated under homophily, {\sl and}  that correlation remains after conditioning on fractions of neighbors adopting
exceeding some level as types are correlated via network formation and seeing additional behavior is still informative even conditional upon exceeding some threshold.
In contrast, when agents' behaviors are instead driven
by complementarities and some $q$, then this correlation disappears if we condition on $q$:  once one exceeds the threshold,  additional behavior leads to no increase in correlation.

We outline a formal test of threshold-based coordination versus such homophily-driven behavior.

Given a network $G$ and a set of "on" nodes $S$, let $s_i=|N_i(G) \cap S|/N_i(G)$.Define the statistic ${R}$ as a function of a threshold $q$ as:
$${R}(q) = {\textrm{Corr}}(s_i,{1}_{i \in S} \: \: | \: \: s_i > q) +  \cor(s_i,{1}_{i \in S} \: \: | \: \: s_i < q)$$
In words, the statistic $R$ is the sum of the correlations between adoption and on-neighbor shares conditional on nodes' on-neighbor shares being above and below $q$.

The idea of the formal test is to examine the joint statistic $({T},{R})$ (where $T$ is as defined in the paper). If there is no coordination, and also no homophily in characteristics correlated with adoption, then for all $q$ ${T}(q)$ should be large while ${R}(q)$ should be small.  If there is no coordination, but there is homophily in characteristics correlated with adoption, then for all $q$ ${R}(q)$ should be large.  Only coordination leads to a $q$ such that ${T}(q)$ and ${R}(q)$ are both small. This suggests the following procedure:

First estimate $q$ by minimizing ${H}(q)=F({T}(q),{R}(q))$  for a function $H$ which is increasing in ${T}$ and ${R}$.  Second, assuming some model for the linking probabilities (as a decreasing function of some measure of the distance between $\theta_i $ and $\theta_j$) and adoption probabilities (as an increasing function of $\theta_i$), use maximum likelihood estimation to estimate the parameters of this model along with the values $\theta_i$.  Third, compare the minimized value ${H}(\hat{q})$ to the distribution of ${H}$ generated by the maximum likelihood estimates (this may have to be bootstrapped depending on the functional form assumptions) to obtain a p-value.

One could also fit a hybrid model in which one simultaneously estimates the latent space of $\theta_i$'s and the $q$'s as a function of that space.     The idea would be that instead of simply choosing a best fit $q(X_i)$ function as we discussed in the paper, one could fit
a function of the form $q(X_i,\theta_i)$ together with the assignment of $\theta_i$'s based on the network and observed behaviors.
This would be a variation on the approach of \cite{goldsmith-pinkhami2013} (see also \cite{jackson2013,goldsmith-pinkhami2013b})
adapted to our setting.

\section{Seed Sets and Fragile Conventions}

In this section we provide some alternative definitions of conventions and behavioral communities.

Conventions may have subsets that still form a convention.
So, let us define various sorts of minimal conventions.

\subsection{Self-Sustaining Seed Sets}

First, let us define a $q$-{\sl self-sustaining seed set} to be a subset $S\subset N$ that is $q$-cohesive and has no nonempty strict subset that is $q$-cohesive.

So, a self-sustaining seed set is a minimal set of nodes such that if they all adopt the behavior then their best responses will continue to be to adopt the behavior.  They may also spread the behavior beyond their set, but that is not necessary as part of the definition.

The reason to have the ``self-sustaining'' part, is that it might be possible to seed a convention with a smaller set if one can pay or force nodes to adopt a behavior and not change back.    Self-sustaining means that none of the nodes would change back if they are all turned on, and so that requires that the group be $q$-cohesive.

Note that $q$-tight sets and self-sustaining seed sets can be distinct:  there are clearly self-sustaining seed sets that are not tight since closure is not required of a self-sustaining seed set, and there are $q$-tight sets that are not self-sustaining seed sets since there can be a subset that is cohesive (just not closed).
For example, a triad with $q<1/2$ is tight, but any two nodes form a self-sustaining set.

Nonetheless, every $q$-tight set contains at least one self-sustaining seed set.

\subsection{Fragile Conventions}

Let us call a convention {\sl fragile} if there exists some node in the convention, such that changing that node to not adopting the behavior and then iterating on best responses would lead {\sl all} agents to change to not adopting the behavior - so changing just one node's behavior can completely eradicate the convention.

There are fragile conventions that are not tight, and vice versa.

For instance, the two conventions on the right side of Figure \ref{ExampleConvention2} are fragile but not tight.   There is a single node in each of those conventions that when changed leads the convention to collapse, but there is still a subset of each convention that forms a convention (those on the left of the figure) and so the conventions are not tight.

To see a tight convention that is not fragile, again consider a triangle with $q<1/2$.  Changing any node's behavior will not alter the other two, and so it is not fragile.

A self-sustaining seed set that is a convention is necessarily fragile.    All subsets of a self-sustaining seed set are not cohesive and so changing {\sl any} node's behavior will unravel the convention.

Moreover, every fragile convention contains a self-sustaining seed set.

Any convention that contains at least two disjoint tight subsets is not fragile.

There are conventions that contain overlapping but distinct tight subsets that are fragile.

\begin{claim}
Consider the set of nodes that can unravel a fragile convention:  all of them need to be part of any self-sustaining seed set that generates the convention.
\end{claim}

This follows from the observation that any node that is taken out and can unravel a fragile convention is such that there is no cohesive subset left that excludes that node, so any cohesive set must include all such nodes.

\subsection{Fragile and Seeding Community Structures}

Let ${C}^F(Q,g)$ denote the finest $\sigma$-algebra whose elements are all coarser than all the fragile conventions that are robust relative to $Q$.

Here, instead of grouping two nodes together if they act the same way in every convention, they are grouped together if they act the same way in at least one fragile convention.  Thus, their behavior is sometimes tied together by the network structure, but does not have to always be tied together.

The fragility of the convention is important as otherwise taking the union of two completely disjoint conventions would put people together without their behaviors really being tied down by each other.
The finest elements of ${C}^F(Q,g)$ are referred to as {\sl fragile atoms}.

One can also define a version which also requires that incorporates a self-seeding requirement. For instance, let ${C}^{FS}(Q,g)$ denote the finest $\sigma$-algebra whose elements are all coarser than all the fragile conventions that are robust relative to $Q$
and which are $q$-self-sustaining seed sets for all $q\in Q$.

\section{Atom-Finding Algorithm: Details}
\label{algorithm}

Here we provide a more detailed description of our approximation algorithm for finding the $Q=[\underbar{q},\overline{q}]$ atomic structure of a network $G$.  As discussed in section \ref{Compute}, the high-level approach is to first generate conventions by finding the $Q$-conventions generated by small (up to size $k$) subsets of nodes, and then using an ILP solver to test whether as of yet unseparated nodes have separating conventions not found in the first step.

If we choose $k$ such that the $Q$-atomic partition if $k$-generatable, the second step is unnecessary: the first step will always find the correct partition.  However, since in practice we do not know for what $k$ the partition is $k$-generatable, we can modify the first step so that we find conventions even when they are not necessarily $k$-seedable, by adding additional nodes to a subset when the convention it generates is empty. We select which nodes to add by whichever most increase the cohesion of the subsets $\underline{q}$-closure. Note that if we can generate a non-trivial convention in this way, we can generate that convention starting from a connected subsets of nodes; hence with this modification we can restrict the subsets we attempt to grow into conventions to connected size-$k$ subsets.

\begin{algo}[Generating the $Q$-Atomic Partition]
\mbox{}
\begin{enumerate}
\raggedright
\item{} Generate all connected subsets of $G$ up to size $k$.
\item{} For each subset $S$ generated above, compute the $Q$-convention generated by $S$.  If this is the empty convention, add to $S$ the node which most increases the cohesion of the $\underline{q}$-closure of $S$, and repeat with the augmented subset in place of $S$. Let $\mathcal{C}$ be the collection of $Q$-conventions generated this way, and let $P$ be the set of pairs of nodes not separated by some convention in $\mathcal{C}$.
\item{} While $P$ is nonempty, select a pair of nodes $(i,j)\in P$, and solve the ILP $i,j$ $Q-SPLIT$.  If a separating convention is found, add this convention to $\mathcal{C}$ and remove from $P$ all pairs which the convention separates.  If no separating convention is found, remove $(i,j)$ from $P$.
\item{} Return the partition generated by the conventions in $\mathcal{C}$.
\end{enumerate}
\end{algo}

Because step 3 involves solving an ILP, there is no polynomial-time method for ensuring a correct solution (i.e. identifying a $Q$-convention if it exists).  We rely on ILP solvers\footnote{Reported results use Gurobi, one of the most widely-used combinatorial solvers. Accuracy is similar with the open source solver Cbc, though the time taken to compute a solution is on average thirty percent longer.} to find a correct solutions with high probability; the accuracy results reported in Section \ref{Compute} suggest that the solver is in fact finding the correct solution with high enough probability to generate the correct atomic structure in the vast majority of real-world networks in our test set.

Code implementing the above algorithm can be found in the supplementary online appendix.

\end{document}